\documentclass[final,3p,times]{elsarticle}
\usepackage{graphics}
\usepackage{graphicx}
\usepackage{epsfig}

\usepackage{amssymb}
\usepackage{amsthm}

\usepackage{lineno}
\usepackage{color}

\begin{document}

\begin{frontmatter}

%% Title, authors and addresses
 
%% use the tnoteref command within \title for footnotes;
%% use the tnotetext command for the associated footnote;
%% use the fnref command within \author or \address for footnotes;
%% use the fntext command for the associated footnote;
%% use the corref command within \author for corresponding author footnotes;
%% use the cortext command for the associated footnote;
%% use the ead command for the email address,
%% and the form \ead[url] for the home page:
%%
%% \title{Title\tnoteref{label1}}
%% \tnotetext[label1]{}
%% \author{Name\corref{cor1}\fnref{label2}}
%% \ead{email address}
%% \ead[url]{home page}
%% \fntext[label2]{}
%% \cortext[cor1]{}
%% \address{Address\fnref{label3}}
%% \fntext[label3]{}

\title{
Statistical Lyapunov theory based on bifurcation analysis of energy cascade in 
isotropic homogeneous turbulence: a physical--mathematical review 
%Bifurcations modeling for studying the turbulent energy cascade
}
 
%% use optional labels to link authors explicitly to addresses:
%% \author[label1,label2]{<author name>}
%% \address[label1]{<address>}
%% \address[label2]{<address>} 

\author{Nicola de Divitiis}

\address{"La Sapienza" University, Dipartimento di Ingegneria Meccanica e 
Aerospaziale, Via Eudossiana, 18, 00184 Rome, Italy, \\
phone: +39--0644585268, \ \ fax: +39--0644585750, \\ 
e-mail: n.dedivitiis@gmail.com, \ \  nicola.dedivitiis@uniroma1.it}

\begin{abstract} 
This work presents a review of previous articles dealing with an original turbulence theory proposed by the author, and provides new theoretical insights into some related issues. 
The new theoretical procedures and methodological approaches confirm and corroborate the previous results. 
{%\color{blue} 
These articles study the regime of homogeneous isotropic turbulence for incompressible fluids
and propose theoretical approaches based on a specific Lyapunov theory for determining the closures of the von K\'arm\'an--Howarth and Corrsin equations, and the statistics of velocity and temperature difference. 
While numerous works are present in the literature which concern the closures of the autocorrelation equations in the Fourier domain (i.e. Lin equation closure), few articles deal with the closures of the autocorrelation equations in the physical space. These latter, being based on the eddy--viscosity concept, describe diffusive closure models.
On the other hand, the proposed Lyapunov theory leads to nondiffusive closures based on the property that, in turbulence, contiguous fluid particles trajectories continuously diverge.
Therefore, the main motivation of this review is to present a theoretical formulation which does not adopt the eddy--viscosity paradigm, and summarizes the results of the previous works.
Next, 
%unlike the other theories, 
this analysis assumes that the current fluid placements, together with velocity and temperature fields, are fluid state variables. This leads to the closures of the autocorrelation equations and helps to interpret the mechanism of energy cascade as due to the continuous divergence of the contiguous trajectories.
Furthermore, novel theoretical issues are here presented among which we can mention the following ones.
The bifurcation rate of the velocity gradient, calculated along fluid particles trajectories, is shown to be much larger than the corresponding maximal Lyapunov exponent. On that basis, an  interpretation of the energy cascade phenomenon is given and the statistics of finite time Lyapunov exponent of the velocity gradient is shown to be represented by normal distribution functions. Next, the self--similarity produced by the proposed closures is analyzed, and a proper bifurcation analysis of the closed von K\'arm\'an--Howarth equation is performed. This latter investigates the route from developed turbulence toward the non--chaotic regimes, leading to an estimate of the critical Taylor scale Reynolds number.
A proper statistical decomposition  based on extended distribution functions and on the Navier--Stokes equations is presented, which leads to the statistics of velocity and temperature difference.
} 
\end{abstract}

\begin{keyword}
%% keywords here, in the form: keyword \sep keyword

%% MSC codes here, in the form: \MSC code \sep code
%% or \MSC[2008] code \sep code (2000 is the default)
Energy cascade, Bifurcations, Lyapunov theory.
\end{keyword}

\end{frontmatter}

\newcommand{\no}{\noindent}
\newcommand{\be}{\begin{equation}}
\newcommand{\ee}{\end{equation}}
\newcommand{\bea}{\begin{eqnarray}}
\newcommand{\eea}{\end{eqnarray}}
\newcommand{\bc}{\begin{center}}
\newcommand{\ec}{\end{center}}

\newcommand{\calr}{{\cal R}}
\newcommand{\calv}{{\cal V}}

\newcommand{\bff}{\mbox{\boldmath $f$}}
\newcommand{\bfg}{\mbox{\boldmath $g$}}
\newcommand{\bfh}{\mbox{\boldmath $h$}}
\newcommand{\bfi}{\mbox{\boldmath $i$}}
\newcommand{\bfm}{\mbox{\boldmath $m$}}
\newcommand{\bfp}{\mbox{\boldmath $p$}}
\newcommand{\bfr}{\mbox{\boldmath $r$}}
\newcommand{\bfu}{\mbox{\boldmath $u$}}
\newcommand{\bfv}{\mbox{\boldmath $v$}}
\newcommand{\bfx}{\mbox{\boldmath $x$}}
\newcommand{\bfy}{\mbox{\boldmath $y$}}
\newcommand{\bfw}{\mbox{\boldmath $w$}}
\newcommand{\bfk}{\mbox{\boldmath $\kappa$}}

\newcommand{\bfA}{\mbox{\boldmath $A$}}
\newcommand{\bfD}{\mbox{\boldmath $D$}}
\newcommand{\bfI}{\mbox{\boldmath $I$}}
\newcommand{\bfL}{\mbox{\boldmath $L$}}
\newcommand{\bfM}{\mbox{\boldmath $M$}}
\newcommand{\bfS}{\mbox{\boldmath $S$}}
\newcommand{\bfT}{\mbox{\boldmath $T$}}
\newcommand{\bfU}{\mbox{\boldmath $U$}}
\newcommand{\bfX}{\mbox{\boldmath $X$}}
\newcommand{\bfY}{\mbox{\boldmath $Y$}}
\newcommand{\bfK}{\mbox{\boldmath $K$}}

\newcommand{\bfeta}{\mbox{\boldmath $\eta$}}
\newcommand{\bfrho}{\mbox{\boldmath $\rho$}}
\newcommand{\bfchi}{\mbox{\boldmath $\chi$}}
\newcommand{\bfphi}{\mbox{\boldmath $\phi$}}
\newcommand{\bfPhi}{\mbox{\boldmath $\Phi$}}
\newcommand{\bflambda}{\mbox{\boldmath $\lambda$}}
\newcommand{\bfxi}{\mbox{\boldmath $\xi$}}
\newcommand{\bfLambda}{\mbox{\boldmath $\Lambda$}}
\newcommand{\bfPsi}{\mbox{\boldmath $\Psi$}}
\newcommand{\bfomega}{\mbox{\boldmath $\omega$}}
\newcommand{\bfOmega}{\mbox{\boldmath $\Omega$}}
\newcommand{\bfeps}{\mbox{\boldmath $\varepsilon$}}
\newcommand{\bfepsn}{\mbox{\boldmath $\epsilon$}}
\newcommand{\bfzeta}{\mbox{\boldmath $\zeta$}}
\newcommand{\bfkappa}{\mbox{\boldmath $\kappa$}}
\newcommand{\bfsigma}{\mbox{\boldmath $\sigma$}}
\newcommand{\itPsi}{\mbox{\it $\Psi$}}
\newcommand{\itPhi}{\mbox{\it $\Phi$}}

\newcommand{\bint}{\mbox{ \int{a}{b}} }
\newcommand{\ds}{\displaystyle}
\newcommand{\Sum}{\Large \sum}

% \linenumbers

% main text

\bigskip

\section{Introduction \label{intro}}

This article presents a review of previous works of the author regarding an original Lyapunov analysis of the developed turbulence which leads to the closures of the von K\'arm\'an--Howarth and Corrsin equations and to the statistics of both velocity and temperature difference
\cite{deDivitiis_1, deDivitiis_2, deDivitiis_3, deDivitiis_4, deDivitiis_5, deDivitiis_6, deDivitiis_7}.
This theory studies the fully developed homogeneous isotropic turbulence through the bifurcations of the incompressible Navier--Stokes equations using a specific statistical Lyapunov analysis of the fluid kinematic field. 
In addition, now it is introduced the energy cascade interpretation and explained some of the mathematical properties of the proposed closures. This work is organized into two parts. One is the reasoned review of previous results, but with new demonstrations and theoretical procedures.
The other one, presented in sections marked with asterisk symbol "*", concerns new theoretical issues of the proposed turbulence theory.

%------------------> Inizio modifica

{%\color{blue}

Although numerous articles were written which concern the closures of the Lin equation in the Fourier domain \cite{Obukhov41, Heisenberg48, Kovasznay48, Ellison61, Pao65, Leith67, Clark98, Nazarenko04, Clark09}, few works address the closures of the autocorrelation equations in the physical space. These last ones, being based on the eddy--viscosity concept, describe diffusive closure models. Unlike the latter, the proposed Lyapunov theory provides nondiffusive closures in the physical space based on the property that, in developed turbulence, contiguous fluid particles trajectories continuously diverge.
Thus, the main purpose of this review is to summarize the results of the previous works based on a theory which does not use the eddy--viscosity paradigm, and to give new theoretical insights into some related issues.

The homogeneous isotropic turbulence is an ideal flow regime characterized by the energy cascade
phenomenon where the diverse parts of fluid exhibit the same statistics and isotropy. On the other hand, the turbulent flows occurring in nature and in the various fields of engineering are generally much more complex than homogeneous isotropic turbulence. In such  flows, spatial variations of average velocity and of other statistical flow properties can happen causing very complex simultaneous effects that add to the turbulent energy cascade and interact with the latter in a nontrivial fashion. 
Hence, the study of the energy cascade separately from the other phenomena requires the analysis of isotropic homogenous turbulence.

The von K\'arm\'an--Howarth and Corrsin equations are the evolution equations of longitudinal velocity and temperature correlations in homogeneous isotropic turbulence, respectively.
Both the equations, being unclosed, need the adoption of proper closures \cite{Karman38, Batchelor53, Corrsin_1, Corrsin_2}.
In detail, the von K\'arm\'an--Howarth equation includes $K$, the term due to the inertia forces and directly related to the longitudinal triple velocity correlation $k$, which has to be properly modelled.
The modeling of such term must take into account that, due to the inertia forces, 
$K$ does not modify the kinetic energy and satisfies the detailed conservation of energy \cite{Batchelor53}. This latter states that the exchange of energy between wave--numbers is only linked to the amplitudes of such wave--numbers and of their difference \cite{Eyink06}. 
Different works propose for the von K\'arm\'an--Howarth equation the diffusion approximation \cite{Hasselmann58, Millionshtchikov69, Oberlack93}
\bea
\ds k = 2 \frac{D}{u} \frac{\partial f}{\partial r}
\label{diff closure}
\eea
where $r$ and $D=D(r)$ are separation distance and turbulent diffusion parameter, respectively, and $u^2 = \langle u_i u_i \rangle/3$ corresponds to the longitudinal velocity standard deviation. 
Following Eq. (\ref{diff closure}), the turbulence can be viewed as a diffusivity phenomenon depending upon $r$, where $K$ will include a term proportional to $\partial^2 f / \partial r^2$.
In the framework of Eq. (\ref{diff closure}), Hasselmann \cite{Hasselmann58} proposed, in 1958, a closure suggesting a link between $k$ and $f$ which expresses $k$ in function of the momentum convected through a spherical surface. His model, which incorporates a free parameter, expresses $D(r)$ by means of a complex expression.
Thereafter, Millionshtchikov developed a closure of the form $D(r) = k_1 u r$, where $k_1$ represents an empirical constant \cite{Millionshtchikov69}.
Although both the models describe two possible mechanisms of energy cascade, in general, do not satisfy some physical conditions. For instance, the Hasselmann model does not verify the continuity equation for all the initial conditions, whereas the Millionshtchikov equation gives, following Eq. (\ref{diff closure}), values of velocity difference skewness in contrast with experiments and energy cascade \cite{Batchelor53}.
More recently, Oberlack and Peters \cite{Oberlack93} suggested a closure where 
$D(r) = k_2 r u \sqrt{1-f}$, being $k_2$ a constant parameter. The authors show that such closure reproduces the energy cascade and, for a proper choice of $k_2$, provides results  in agreement with the experiments \cite{Oberlack93}.

For what concerns the Corrsin equation, this exhibits $G$, the term responsible for the thermal energy cascade. This quantity, directly related to the triple velocity--temperature correlation $m^*$,
also needs adequate modellation.
As $G$ depends also on the velocity correlation, the Corrsin equation requires the knowledge of $f$, thus it must be solved together to the von K\'arm\'an-–Howarth equation.
Different works can be found in the literature which deal with the closure of Corrsin equation.
Some of them study the self--similarity of the temperature correlation in order to analyze properties and possible expressions for $G$. Such studies are supported by the idea that the simultaneous effect of energy cascade, conductivity and viscosity, makes the temperature correlation similar in the time. This question was theoretically addressed by George (see \cite{George1, George2} and references therein) which showed that the decaying isotropic turbulence reaches the self--similarity, while the temperature correlation is scaled by the Taylor microscale whose current value depends on the initial condition. More recently, Antonia et al. \cite{Antonia} studied the temperature structure functions in decaying homogeneous isotropic turbulence and found that the standard deviation of the temperature, as well as the turbulent kinetic energy, follows approximately the similarity over a wide range of length scales. There, the authors used this approximate similarity to estimate the third--order correlations and found satisfactory agreement between measured and calculated functions.
On the other hand, the temperature correlation can be obtained using proper closures of von K\'arm\'an-–Howarth and Corrsin equations suitable for the energy cascade phenomenon. 
On this argument, several articles has been written. 
For instance, Baev and Chernykh \cite{Baev} (and references therein) analyzed velocity and temperature correlations by means of a closure model based on the gradient hypothesis which relates pair longitudinal second and third order correlations, by means of  empirical coefficients.

Although other works regarding the von K\'arm\'an--Howarth equation were written 
\cite{Mellor84, Onufriev94, Grebenev05, Grebenev09, Antonia2013}, to the author's
knowledge a physical--mathematical analysis based on basic principles which provides 
analytical closures of von K\'arm\'an--Howarth and Corrsin equations has not received 
due attention. Therefore, the aim of the this work is to present a review of the Lyapunov analysis presented in \cite{deDivitiis_1, deDivitiis_2, deDivitiis_3, deDivitiis_4, deDivitiis_5, deDivitiis_6, deDivitiis_7} and new theoretical insights into some related issues.

In the present formulation, based on the Navier--Stokes bifurcations, the current fluid placements, together with velocity and temperature fields, are considered to be fluid state variables. This leads to the closures of the autocorrelation equations and helps to interpret the mechanism of energy cascade as due to the continuous divergence of the contiguous trajectories.
}
%------------------> fine modifica

In line with Ref. \cite{deDivitiis_3}, the present work first addresses the problem for defining the bifurcations for incompressible Navier--Stokes equations, considering that these latter can be reduced to an opportune symbolic form of operators for which the classical bifurcation theory of differential equations can be applied \cite{Ruelle71}. 
In such framework, this analysis remarks that a single Navier--Stokes bifurcation will generate a doubling of the velocity field and of all its several properties, with particular reference to the characteristic length scales. If on one side the lengths are doubled due to bifurcations, on the other hand the characteristic scale for homogeneous flows in infinite domains is not defined. Hence, the problem to define the characteristic length --and therefore the flow Reynolds number-- in such situation is also discussed.
Such characteristic scale is here defined in terms of spatial variations of initial or current velocity field in such a way that, in fully developed homogeneous isotropic turbulence, this length coincides with the Taylor microscale. As far as the characteristic velocity is concerned, this is also defined in terms of velocity field so that, in developed turbulence, identifies the velocity standard deviation.

The trajectories bifurcations in the phase space of the velocity field are here formally dealt with using a proper Volterra integral formulation of the Navier--Stokes equations, whereas the turbulence transition is qualitatively analyzed through general properties of the bifurcations and of the route toward the fully developed chaos. This background, regarding the general bifurcations properties and the route toward the chaos, will be useful for this analysis.

The adopted statistical Lyapunov theory shows how the fluid relative kinematics can be much more rapid than velocity and temperature fields in developed turbulence, so that fluid strain and velocity fields are statistically independent with each other.
{%\color{blue}
Moreover, in addition to Refs. \cite{deDivitiis_1, deDivitiis_2, deDivitiis_3, deDivitiis_4, deDivitiis_5, deDivitiis_6,  deDivitiis_7}, this analysis introduces the bifurcation rate of the velocity gradient, a quantity providing the frequency at which the velocity gradient determinant vanishes along fluid particles trajectories. The bifurcation rate, in fully developed turbulence, is shown to be much greater than the corresponding maximal Lyapunov exponent. This allows to explain the energy cascade through the relation between material vorticity, Lyapunov vectors and bifurcation rate
using the Lyapunov theory. 
In detail, the energy cascade 
can be viewed as a continuous and intensive stretching and folding process of fluid particles which involves smaller and smaller length scales during the fluid motion, where the folding frequency equals
the bifurcation rate.}

Next, the statistics of the Lyapunov exponents is reviewed. 
In agreement with Ref. \cite{deDivitiis_6}, we show that the local Lyapunov exponents are uniformely unsymmetrically distributed in their interval of variation. Unlike to Ref. \cite{deDivitiis_6} which uses the criterion of maximum entropy associated with the fluid particles placements, the isotropy and homogeneity hypotheses are here adopted.
A further result with respect to the previous issues pertains the finite time Lyapunov exponents statistics: through the bifurcation analysis and the central limit theorem, we show that the finite time Lyapunov exponent tends to a fluctuating variable distributed following a normal distribution function. 

Thereafter, the closure formulas of von K\'arm\'an--Howarth and Corrsin equations are derived through Liouville equation and finite scale Lyapunov exponent statistics. These closures do not correspond to a diffusive model, being the result of the trajectories divergence in the continuum fluid. Such formulas coincide with those just obtained in Refs. \cite{deDivitiis_1, deDivitiis_4} and \cite{deDivitiis_5} where it is shown that such closures adequately describe the energy cascade phenomenon, reproducing, negative skewness of velocity difference,  the Kolmogorov law and temperature spectra in line with the theoretical argumentation of Kolmogorov, Obukhov--Corrsin and Batchelor \cite{Batchelor_2, Batchelor_3, Obukhov}, with experimental results \cite{Gibson, Mydlarski}, and with numerical data \cite{Rogallo, Donzis}.
These closures are here achieved by using different mathematical procedures with respect to the other articles \cite{deDivitiis_1, deDivitiis_4} and \cite{deDivitiis_5}. 
While the previous works derive such closures studying the local fluid act of motion in the finite scale Lyapunov basis \cite{deDivitiis_1, deDivitiis_4} and adopting maximum and average finite scale Lyapunov exponents \cite{deDivitiis_5}, here these closures are obtained by means of the local finite scale Lyapunov exponents PDF, showing that the assumptions of Refs. \cite{deDivitiis_1, deDivitiis_4} and \cite{deDivitiis_5} agree with this analysis, corroborating the previous results. 
Some of the properties of the proposed closures are then studied, with particular reference to the evolution times of the developed correlations and their self--similarity.
In detail, as new result with respect the previous articles, this analysis shows that the proposed closures generate correlations self--similarity in proper ranges of separation distance, which is directly linked to the particles trajectories divergence.

Furthermore, a novel bifurcation analysis of the closed von K\'arm\'an--Howarth equation is proposed, which considers the route starting from the fully developed turbulence toward the non--chaotic regimes. 
This extends the discussion of the previous works and represents an alternative point of view for studying the turbulent transition.
According to this analysis, the closed von K\'arm\'an--Howarth equation is decomposed in several ordinary differential equations through the Taylor series expansion of the longitudinal velocity correlation. This procedure, which also accounts for the aforementioned self--similarity, leads to estimate the Taylor scale Reynolds number at the transition. This latter is found to be $10$, a value in good agreement with several experiments which give values around to $10$, and in particular with the bifurcations analysis of the  energy cascade of Ref. \cite{deDivitiis_3} which provides a critical Reynolds number of $10.13$ if the route toward the turbulence follows the Feigenbaum scenario \cite{Feigenbaum78, Eckmann81}.

Finally, the statistics of velocity and temperature difference, of paramount importance for estimating the energy cascade, is reviewed.
While Refs. \cite{deDivitiis_1, deDivitiis_2, deDivitiis_4} and \cite{deDivitiis_7} 
determine such statistics through a concise heuristic method,
this analysis uses a specific statistical decomposition of velocity and temperature which adopts appropriate stochastic variables related to the Navier--Stokes bifurcations.
The novelty of the present approach with respect to the previous articles is that the random variables of such decomposition are opportunely chosen to reproduce the Navier--Stokes bifurcation effects and the isotropy: these are highly nonsymmetrically distributed stochastic variables following opportune extended distribution functions which can assume negative values.
Such decomposition, able to reproduce negative skewness of longitudinal velocity difference, provides a  statistics of both velocity and temperature difference in agreement with theoretical and experimental data known from the literature \cite{Kolmogorov41, Kolmogorov62, She94, Tabeling96, Tabeling97}. Here, in addition to Refs.  \cite{deDivitiis_1, deDivitiis_2, deDivitiis_4, deDivitiis_7}, a detailed mathematical analysis is presented which concerns the statistical properties of the aforementioned extended distribution functions in relation to the Navier--Stokes bifurcations.

\bigskip

In brief, the original contributions of the present work can be summarized as:

{%\color{blue}
$i$) The bifurcation rate associated with the velocity gradient is shown to be much larger than the maximal Lyapunov exponent of the velocity gradient.
}

$ii$) As the consequence of $i$), the energy cascade can be viewed as a succession of stretching and folding of fluid particles which involves smaller and smaller length scales, where the particle folding happens at the frequency of the bifurcation rate.

$iii$) As the consequence of $i$), the central limit theorem provides reasonable argumentation that the finite time Lyapunov exponent is distributed following a gaussian distribution function. 

$iv$) The proposed closures generate correlations self--similarity in proper ranges of variation of the separation distance which is directly caused by the continuous fluid particles trajectories divergence.

$v$) A specific bifurcation analysis of the closed von K\'arm\'an--Howarth equation is proposed which allows to estimate the critical Taylor scale Reynolds number in isotropic turbulence.

$vi$) A statistical decomposition of velocity and temperature is presented which is based on stochastic variables distributed following extended distribution functions. 
Such decomposition leads to the statistics of velocity and temperature difference, where the intermittency of these latter increases as Reynolds number and P\'eclet number rise.

\bigskip

\section{Background \label{Background}}

In the framework of the link between bifurcations and turbulence, this section deals with some of the fundamental elements of the Navier--Stokes equations and heat equation, useful for the present analysis. In particular, we will address the problem to define an adequate bifurcation analysis for the Navier--Stokes equations, and will analyze the meaning of the characteristic length scales when a homogeneous flow is in an infinite domain. All the considerations regarding the fluid temperature can be applied also to any passive scalar which exhibits diffusivity.
A statistically homogeneous and isotropic flow with null average velocity is considered.

In order to formulate the bifurcation analysis, we start from the Navier--Stokes equations and the temperature equation 
\bea
\begin{array}{l@{\hspace{-0.cm}}l}
\ds \nabla_{\bf x} \cdot {\bf u} =0, \\\\
\ds \frac{\partial {\bf u}}{\partial t} =
-  \nabla_{\bf x} {\bf u} \ {\bf u} - \frac{\nabla_{\bf x}p}{\rho}  + \nu \nabla_{\bf x}^2 {\bf u}
\label{NS_eq}
\end{array}
\eea
\bea
\begin{array}{l@{\hspace{-0.cm}}l}
\ds \frac{\partial \vartheta }{\partial t} =
-{\bf u} \cdot \nabla_{\bf x} \vartheta + \chi \nabla_{\bf x}^2 \vartheta
\label{T_eq}
\end{array}
\eea
where $\bf u$=${\bf u}(t, {\bf x})$, $p$=$p(t, {\bf x})$ and $\vartheta$=$\vartheta(t, {\bf x})$
are velocity, pressure and temperature fields, $\nu$ and $\chi$=$k \rho/C_p$ are fluid kinematic viscosity and thermal diffusivity, being $\rho=$const, $k$ and $C_p$ density, fluid thermal conductivity and specific heat at constant pressure, respectively. 
In this study $\nu$ and $\chi$ are supposed to be independent from the temperature, thus Eqs. (\ref{NS_eq}) is autonomous with respect to Eq. (\ref{T_eq}), whereas Eq. (\ref{T_eq}) will depend on Eqs. (\ref{NS_eq}).

To define the bifurcations of Eqs. (\ref{NS_eq}) and (\ref{T_eq}), such equations are first expressed in the symbolic form of operators. To this end, in the momentum Navier--Stokes equations, the pressure field is eliminated by means of the continuity equation, thus Eqs. (\ref{NS_eq}) and (\ref{T_eq}) are formally written as
\bea
\begin{array}{l@{\hspace{-0.cm}}l}
 \dot{\bf u} =  {\bf N}({\bf u} ; \nu),
\end{array}
\label{NS_op}
\eea
\bea
\begin{array}{l@{\hspace{-0.cm}}l}
\dot{\vartheta} =  {\bf M}({\bf u}, \vartheta ; \chi)
\end{array}
\label{T_op}
\eea
in which ${\bf N}$ is a nonlinear quadratic operator incorporating $-{\bf u} \cdot \nabla_{\bf x} {\bf u}$, $-\nu \nabla_{\bf x}^2 {\bf u}$ and the integral nonlinear operator which expresses the pressure gradient as a functional of the velocity field, being 
\bea
\ds  p(t, {\bf x}) =  \frac{\rho}{4 \pi}
\int \frac{\partial^2 u_i' u_j'}{\partial x_i' \partial x_j'} \ \frac{d V({\bf x}')}{\vert {\bf x}' - {\bf x} \vert} 
\label{pressure}
\eea
Therefore, $p$ provides nonlocal effects of the velocity field \cite{Tsinober2009}, and the Navier--Stokes equations are reduced to be an integro--differential equation formally expressed by Eq. (\ref{NS_op}). 
For what concerns Eq. (\ref{T_op}), it is the evolution equation of $\vartheta$, where $\bf M$ is a linear operator of $\vartheta$. Accordingly, transition and turbulence are caused by the bifurcations of Eq. (\ref{NS_op}), where $\nu^{-1}$ plays the role of the control parameter. 
At this stage of the analysis, it is worth to remark the following two items: 
a) there is no explicit methods of bifurcation analysis for integro--differential equations such as Eq. (\ref{NS_op}). 
b) since the flow is statistically homogeneous in an infinite domain, characteristic scales of the problem are not defined.

The item a) can be solved according to the analysis method proposed by Ruelle and Takens in Ref. \cite{Ruelle71}: it is supposed that the infinite dimensional space of velocity field $\left\lbrace \bf u\right\rbrace$ can be replaced by a finite--dimensional manifold, then Eq. (\ref{NS_op}) can be reduced to be the equation of the kind studied by Ruelle and Takens in Ref. \cite{Ruelle71}. 
Therefore, the classical bifurcation theory of ordinary differential equations \cite{Guckenheimer90, Ruelle71, Eckmann81} can be formally applied to Eq. (\ref{NS_op}), and the present analysis can be considered to be valid in the limits of the formulation proposed in Ref. \cite{Ruelle71}.

For what concerns the characteristic length, a homogeneous flow in infinite domain is free from boundary conditions, thus the characteristic scale, being not defined, is here chosen in function of the spatial variations of the current velocity field. Thus, for all flow regimes in infinite regions, (i.e. non--chaotic, turbulent and transition flows), characteristic length and velocity, $L$ and $U$ respectively, are here chosen in terms of volume integrals of $\bf u$ in the following manner 
\bea
\begin{array}{l@{\hspace{-0.cm}}l}
\ds U^2 = \lim_{{\cal V} \rightarrow \infty} \frac{1}{{\cal V}} \int_{\cal V} {\bf u}(t, {\bf x})\cdot{\bf u}(t, {\bf x}) d{\cal V}({\bf x}), \\\\
\ds G^2 =  \lim_{{\cal V} \rightarrow \infty} \frac{1}{{\cal V}} \int_{\cal V} 
%\Sum_{i=1}^3 
\nabla_{\bf x} {\bf u} : \nabla_{\bf x} {\bf u} \  d{\cal V}({\bf x}), \\\\
\ds L^2 = c \  \frac{U^2}{G^2}
\end{array}
\label{UL}
\eea
where ${\cal V}$ is the fluid domain volume, ":" denotes the Frobenius inner product and $c$=$O(1)$ is a dimensionless constant which will be properly chosen.
The flow Reynolds number is then defined in terms of $U$ and $L$ as
\bea
\ds  Re =\frac{U L}{\nu}. 
\label{Re}
\eea
{%\color{blue}
 Equation (\ref{Re}) provides an extension of the Taylor scale Reynolds number
which applies for every flow regime. In particular, such definition holds also for non turbulent flows,
where $U$ and $L$, although not velocity standard deviation and statistical correlation scale, provide a generalization of the latter.}
In fully developed homogeneous turbulence, the volume integrals appearing in Eqs. (\ref{UL}) equal  statistical averages calculated over the velocity field ensemble, such as velocity standard deviation and dissipation rate.
%Assuming $c$=$5$, 
Accordingly, in isotropic homogeneous turbulence $L$ and $U$ identify, respectively, Taylor scale $\lambda_T$ and standard deviation $u$ of one of the velocity components, and $Re=U \ L/\nu$ coincides with the Taylor scale Reynolds number $R_T$.
Such definitions (\ref{UL}) extend the concept of velocity variance and correlation scale and will be used for the bifurcation analysis proposed in this work.

\bigskip

\section{Navier--Stokes bifurcations.}

Before introducing the bifurcations analysis of the Navier--Stokes equations in the operatorial form (\ref{NS_op}), it is worth remarking that a given point in the space of velocity fields set $\bar{\bf u} \in \left\lbrace {\bf u} \right\rbrace$ -- or temperature field $\bar{\vartheta} \in \left\lbrace
 \vartheta \right\rbrace$-- corresponds to a spatial distribution including all its characteristics, in particular the length scales associated with $\bar{\bf u}$.

%\bigskip

%\subsection{\bf Navier--Stokes bifurcations}

The bifurcations of Eq.(\ref{NS_op}) happen when the Jacobian ${\nabla_{\bf u}
{\bf N}}$ exhibits at least an eigenvalue with zero real part (NS--bifurcations), 
and this occurs when
\bea
\ds \det (\nabla_{\bf u} {\bf N}) =0.
\eea
Such bifurcations are responsible for multiple velocity fields $\hat {\bf u}$ which provides
the same field $\dot{\bf u}$.
In fact, during the fluid motion, multiple solutions $\hat {\bf u}$ and $\hat{\vartheta}$ can be determined, at each instant, through inversion of Eq. (\ref{NS_op})
\bea
\begin{array}{l@{\hspace{-0.cm}}l}
\dot{\bf u} = {\bf N}({\bf u}; \nu)  \\\\
\hat {\bf u}  = {\bf N}^{-1}(\dot{\bf u}; \nu), \\\\
\hat{\vartheta}=  {\bf M}^{-1}(\dot{\vartheta}, \hat {\bf u}; \chi)
\end{array}
\label{invr0}
\eea
In the framework of the trajectories bifurcations in the phase space, the fluid motion can be expressed by means of Eq.(\ref{NS_op}) and initial conditions ${\bf u}(0)$ and $\vartheta(0)$, using the following Volterra integral formulation 
\bea
\begin{array}{l@{\hspace{-0.cm}}l}
\ds {\bf u}(t) - {\bf u}(0) - \int_0^t {\bf N}({\bf u}(\tau); \nu) \ d \tau 
\equiv {\cal N} \left( {\bf u}; \nu \right)=0 , \\\\
\ds {\vartheta}(t) - {\vartheta}(0) - \int_0^t {\bf M}({\bf u}(\tau), \vartheta(\tau); \chi) \ d \tau \equiv {\cal M} ({\bf u}, \vartheta; \chi) =0
\end{array}
\label{Volterra}
\eea
where $\cal N$ and $\cal M$ are proper operators such that 
\bea
\begin{array}{l@{\hspace{-0.cm}}l}
\ds {\cal N}: \left\lbrace {\bf u} \right\rbrace \rightarrow {\cal N} (\left\lbrace {\bf u}
 \right\rbrace), \\\\
\ds {\cal M}: \left\lbrace {\bf u} \right\rbrace \times \left\lbrace  \vartheta \right\rbrace \rightarrow {\cal M} ( \left\lbrace {\bf u}  \right\rbrace \times \left\lbrace  \vartheta \right\rbrace)
\end{array}
\eea
Specifically, $\cal N$ is a nonlinear operator of $\bf u$, where the image ${\cal N} (\left\lbrace {\bf u} \right\rbrace)$ has the same structure of $\left\lbrace {\bf u} \right\rbrace$, whereas ${\cal M}$ is linear with respect to $\vartheta$ and the image 
${\cal M} (\left\lbrace  {\bf u} \right\rbrace \times \left\lbrace \vartheta \right\rbrace)$
is isomorphic with $\left\lbrace \vartheta \right\rbrace$.
Thus, $\cal N$ and $\cal M$ admit in general the following jacobians
\bea
\begin{array}{l@{\hspace{-0.cm}}l}
\ds \nabla_{\bf u}{\cal N}, \\\\
\ds \nabla_{\vartheta}{\cal M}
\end{array}
\eea
According to Eqs. (\ref{Volterra}), a trajectory bifurcation happens when $\nabla_{\bf u}{\cal N}$ is singular, that is when
\bea
\ds \det \left( \nabla_{\bf u}{\cal N} \right) =0
\eea
and the multiple solutions of Eqs. (\ref{Volterra}), say  $\hat {\bf u}$ and $\hat{\vartheta}$, are given in terms of ${\bf u}$ and ${\vartheta}$ through
\bea
\begin{array}{l@{\hspace{-0.cm}}l}
\ds {\cal N} \left( \hat{\bf u}; \nu \right) = {\cal N} \left( {\bf u}; \nu \right)=0 , \\\\
\ds {\cal M} (\hat{\bf u}, \hat{\vartheta}; \chi) ={\cal M} ({\bf u}, {\vartheta}; \chi) =0
\end{array}
\label{invr}
\eea
using the implicit functions theorem. Therefore, if velocity and temperature fields are supposed to be known for $\nu$=$\nu_0$, the fields calculated for $\nu$$\ne$$\nu_0$ are formally expressed as
\bea
\begin{array}{l@{\hspace{-0.cm}}l}
\ds \hat{\bf u}(\nu)  = 
{\bf u}(\nu_0) 
-\int_{\nu_0}^{\nu} 
\left( \nabla_{\bf u} {\cal N} \right)^{-1}
\frac{\partial {\cal N}}{\partial \nu} \ d\nu, \\\\
\ds  \hat{\vartheta} (\nu) = 
{\vartheta}(\nu_0) 
+\int_{\nu_0}^{\nu}
\left( \nabla_{\vartheta}{\cal M} \right)^{-1}
\left( \nabla_{\bf u} {\cal M} \right)
\left( \nabla_{\bf u} {\cal N} \right)^{-1}
\frac{\partial {\cal N}}{\partial \nu} \ d\nu, \\\\
\end{array}
\label{Dini}
\eea

\bigskip

\section {Qualitative analysis of the route toward the chaos}

With reference to Eq. (\ref{invr0}) or (\ref{Dini}), when $\nu^{-1}$ is relatively small, $\bf N$ and 
$\cal N$ behave like linear operators, and 
Eq. (\ref{invr}) returns $\hat {\bf u} \equiv {\bf u} (t, {\bf x})$ as unique
solution. Increasing $\nu^{-1}$, the Navier--Stokes equations encounter the first bifurcation at $\nu=\nu_1$, the jacobian $\nabla_{\bf u}{\cal N}$ is singular there, and
thereafter Eq. (\ref{invr}) determines different velocity fields $\hat {\bf u}$ with the corresponding length scales.
A single bifurcation causes a doubling of $\bf u$, i.e. a doubling of the velocity values and of the length scales. Although the route toward the chaos can be of different kinds 
\cite{Ruelle71, Feigenbaum78, Pomeau80, Eckmann81}, one common element of these latter is that the number of encountered bifurcations at the onset of the chaotic regimes is about greater than three.
Hence, if $\nu^{-1}$ is quite small, the velocity field can be represented by its Fourier series of a given basic scale. The first bifurcation introduces new solutions $\hat{\bf u}$ whose Fourier characteristic lengths are independent from the previous one. Thereafter, each bifurcation adds new independent scales, and, after the third bifurcation ($\nu^{-1}=\nu_*^{-1}$), the transition occurs, the several characteristic lengths and the velocity values appear to be continuously distributed, thus the velocity field is represented by the Fourier transform there.
In such situation, a huge number of such solutions are unstable, ${\bf u} (t, {\bf x})$ tends to sweep the entire velocity field set, and the motion is expected to be chaotic with a high level of mixing. As for  $\hat{\vartheta}$, $\bf M$ and $\cal M$ are both linear operators
of $\vartheta$, thus $\hat{\vartheta}$ follows the variations of $\hat {\bf u}$.
\begin{figure}[t]
\centering
\includegraphics[width=11.0cm, height=14.0cm]{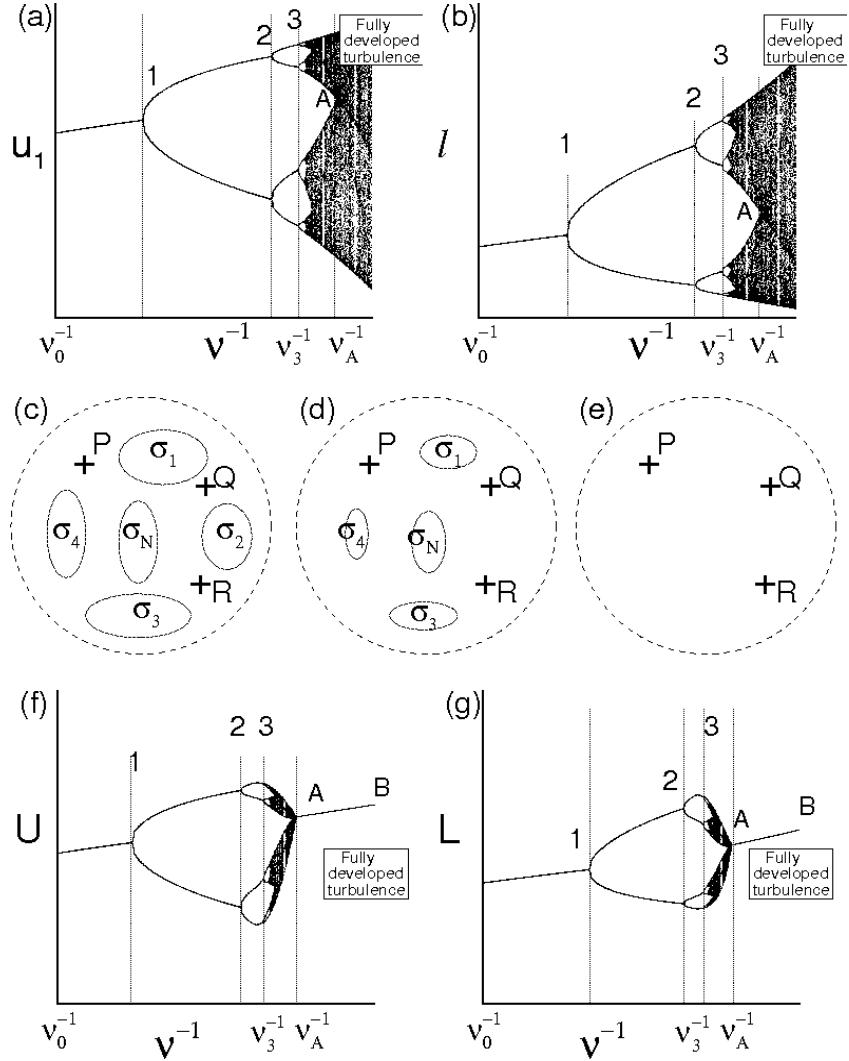}
\caption{Qualitative scheme of the route toward the turbulence. (a) and (b): velocity and length scale in terms of kinematic viscosity. (c), (d) and (e): symbolic representation of solutions in the
velocity fields set. (f) and (g): U and L in terms of kinematic viscosity.}
\label{figura_1}
\end{figure}
If $\nu^{-1}$ does not exceed its critical value value, say $\nu_*^{-1}$, 
the velocity fields satisfying Eq. (\ref{invr}) are limited in number, and this 
corresponds to intermediate stages of the route toward the chaos.
On the contrary, when $\nu^{-1} > \nu_*^{-1}$, the region of developed turbulence where $\lambda_{NS}>$0 is observed, being $\lambda_{NS}$ the average maximal Lyapunov exponent of the Navier--Stokes equations, formally calculated as
\bea
\begin{array}{l@{\hspace{-0.cm}}l}
\ds \lambda_{NS} = \lim_{T \rightarrow \infty} \frac{1}{T} \int_0^T 
\frac{{\bfy} \cdot \nabla_{\bf u} {\bf N} {\bfy} }{{\bfy} \cdot{\bfy}} \ dt,
\\\\
\ds \dot{\bfy} = \nabla_{\bf u} {\bf N}({\bf u}; \nu) {\bfy}, \\\\
\ds \dot{\bf u} = {\bf N} ({\bf u}; \nu),
\end{array}
\eea
and $\bfy$ is the Lyapunov vector associated with the Navier--Stokes equations. 
Then, $\nu_*^{-1}$ depends on ${\bf u}$, and $Re_{*}$, calculated with Eqs. (\ref{UL}), can be roughly estimated as the minimum value of $Re$ for which $\lambda_{NS} \geq$ 0.

Figure \ref{figura_1} qualitatively shows the route from non--chaotic regimes toward
the developed turbulence. Specifically, Figs. \ref{figura_1} (a) and (b) report two bifurcation maps 
at a given instant, providing the velocity component $u_1$ in a point of the space, and one characteristic scale $\ell$ of the velocity field in function of $\nu^{-1}$.
Figures \ref{figura_1} (c), (d) and (e) symbolically represent, for assigned values of $\nu$, the velocity field set (points inside the dashed circle), three different solutions of the Navier--Stokes equations, say P, Q and R, and the several subsets $\sigma_1$, $\sigma_2$,... which correspond to islands that are not swept during the fluid motion.
The figure also depicts $L$=$L(\nu^{-1})$ and $U$=$U(\nu^{-1})$ (Fig. \ref{figura_1} (f) and (g) ), formally calculated with Eqs. (\ref{UL}). 
Following Eq. (\ref{Dini}), these maps are not universal, as $u_1$=$u_1(\nu^{-1})$, $\ell$=$\ell(\nu^{-1})$, $L$=$L(\nu^{-1})$ and $U$=$U(\nu^{-1})$ 
do not represent universal laws and their order of magnitude will depend on velocity field at $\nu_0^{-1}$. 
When $\nu^{-1}>\nu_3^{-1}$, the number of solutions diverges and the bifurcation tree of $u_1$ and $\ell$ drastically changes its structure showing tongue geometries that develop from the different bifurcations. As long as $\nu^{-1}$ does not exceed much $\nu_3^{-1}$, the extension of such tongues is relatively bounded, whereas the measure of the islands $\sigma_k$ is quite large. This means that, although $u_1$ and $\ell$ exhibit chaotic behavior there, these do not sweep completely their variation interval, thus Eqs. (\ref{NS_op}) do not behave like an ergodic dynamic system there. 
This corresponds to Fig. \ref{figura_1} (c), where the velocity fields P, Q and R, being differently placed with respect to $\sigma_k$,  $k=1, 2,..$ will exhibit different values of average kinetic energy and dissipation rate in $\cal V$.
As  $\nu^{-1}$ rises, these tongues gradually increase their extension whereas the measures of $\sigma_k$ diminish (see Fig.  \ref{figura_1} (d)) until to reach a situation where the bifurcation tongues overlap with each others and the islands $\sigma_k$ vanish (Fig.  \ref{figura_1} (e)).
Such developed overlapping corresponds to a chaotic behavior of $u_1$ and $\ell$ where these latter almost entirely describe their variation interval: Eqs. (\ref{NS_op}) behave like an ergodic dynamic system there, whereas all the velocity fields, in particular P, Q, and R, although different with each others, give the same values of average kinetic energy and dissipation rate in $\cal V$.
This is the onset of the fully developed turbulence.

As far as $L$ and $U$ are concerned, these are both functionals of $\bf u$ following Eq. (\ref{UL}), accordingly their variations in terms of $\nu^{-1}$ are peculiar, resulting quite different with respect to $u_1$ and $\ell$.
In particular, the structure of the first three bifurcations do not show important differences
with respect to $u_1$ and $\ell$, whereas, after the third bifurcation ($\nu^{-1}>\nu_3^{-1}$), the  chaotic regime begins, and the bifurcation tree of $U$ and $L$ exhibits completely different shape than the corresponding zone of $u_1$ and $\ell$. In detail, the chaotic region extension of $U$ and $L$ appears to be more limited than that of $u_1$ and $\ell$ until to collaps in the lines A--B when $\nu^{-1} > \nu^{-1}_A$. This is because the several bifurcations in $\nu_3^{-1} < \nu^{-1} < \nu^{-1}_A$ correspond to a large number of solutions that show different levels of average kinetic energy and dissipations rate in ${\cal V}$ which are in some way comparable to each other, respectively. Hence, although the chaotic regime is characterized by myriad of values of $u_1$ and $\ell$ which widely sweep the corresponding ranges, $L$ and $U$, being related to average kinetic energy and dissipation rate, will exhibit smaller variations. For relatively high values of $\nu^{-1}$, when the velocity fluctuations behavior is ergodic, the averages calculated on phase trajectory tends to the spatial averages. The region of chaotic regime collaps into the line A--B there. Along such line, for assigned $\nu$, all the solutions, in particular P, Q and R will exhibit the same level of kinetic energy and dissipation, and this represents the regime of fully developed turbulence. 
\begin{figure}[t]
\centering
\includegraphics[width=10.0cm, height=6.0cm]{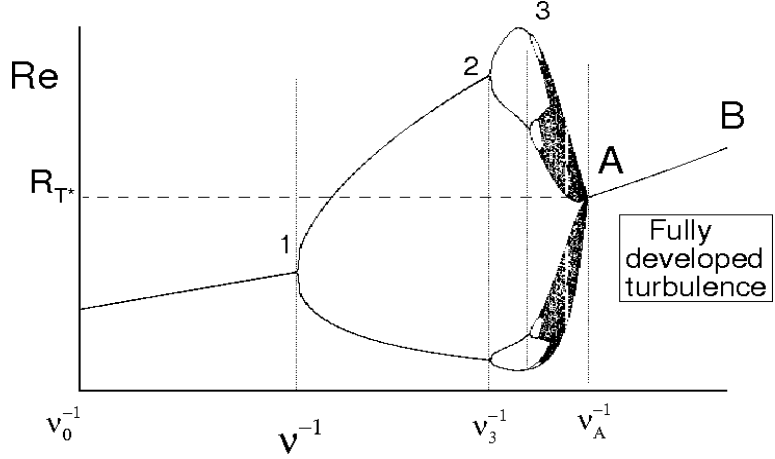}
\caption{Qualitative scheme of the route toward the turbulence: Reynolds number in terms of kinematic viscosity.}
\label{figura_2}
\end{figure}

The Reynolds number $Re=\nu^{-1} U \ L$ is shown in terms of $\nu^{-1}$ in Fig. \ref{figura_2}. 
Also this map is non universal as it depends on $\nu^{-1}_0$.
Nevertheless, such representation allows to identify the critical Reynolds number 
$Re_* = R_{T *}=\nu_*^{-1} U_* \ L_*$, the minimum value of $R_T$ for which the flow maintains statistically homogeneous and isotropic compatible with $\lambda_{N S} \ge 0$. 
Hence, such critical Reynolds number $Re_* = R_{T *}$ will assume an unique value, represented by the point A of Figs. \ref{figura_1} and \ref{figura_2}, which plays the role of an universal limit in homogeneous isotropic turbulence. Then, $\nu_* \equiv \nu_A$, $L_* \equiv L_A$, $U_*=U_A$, and the lines A--B represent regimes of fully developed homogeneous isotropic turbulence where 
\bea
\left.
\begin{array}{l@{\hspace{-0.cm}}l}
\ds L \rightarrow \lambda_T \\\\
\ds U \rightarrow u \\\\
\ds Re \rightarrow R_T
\end{array}
\right\rbrace 
\ \ \ \ \mbox{along A--B}
\eea
We conclude this section by remarking that the characteristic length of the problem is an undefined quantity in infinite domain. Therefore, the length scales of $\bf u$ are used for determining the flow Reynolds number the critical value of which, $Re_*=\nu_*^{-1} U_* \ L_*$ has to be properly estimated. Accordingly, $L_*\equiv \lambda_{T *}$ and $U_*\equiv u_*$, linked with each other, will depend on $R_{T *}$ and $\nu$.

Such qualitative analysis is here used as background to formulate a specific bifurcation analysis of the velocity correlation equation, and to determine an estimate of the critical Reynolds number 
$R_{T *}$.

\bigskip

\section{Kinematic bifurcations. Bifurcation rate} 

The Navier--Stokes bifurcations have significant implications for what concerns the relative kinematics of velocity field. This kinematics is described by the separation vector $\bfxi$ 
(finite scale Lyapunov vector) which satisfies to the following equations
\bea
\begin{array}{l@{\hspace{-0.cm}}l}
\ds \dot{{\bfx} } = {\bf u} (t, {\bfx}),   \\\\
\ds \dot{{\bfxi} } = {\bf u} (t, {\bfx}+{\bfxi}) - {\bf u} (t, {\bfx}),
\end{array}
\label{kin finite}
\eea
being ${\bfx}(t)$ and ${\bf y}(t)={\bfx}(t)+{\bfxi}(t)$ two fluid particles trajectories. In the case of contiguous trajectories, $\vert \bfxi \vert \rightarrow$ 0, 
and Eqs. (\ref{kin finite}) read as
\bea
\begin{array}{l@{\hspace{-0.cm}}l}
\ds \dot{{\bfx} } = {\bf u} (t, {\bfx}),   \\\\
\ds {d \dot{\bfx} } = \nabla_{\bf x} {\bf u} (t, {\bfx})  d {\bfx},
\end{array}
\label{kin}
\eea
where $d {\bfx}$  and ${\nabla_{\bf x} {\bf u}}(t, {\bf x})$ are, respectively, elemental separation vector, and velocity gradient. One point of the physical space is of bifurcation for the velocity field (kinematic bifurcation) if ${\nabla_{\bf x} {\bf u}}(t, {\bf x})$ has at least an eigenvalue with zero real part, and this happens when its determinant vanishes, i.e.
\bea
\ds \det \left( \nabla_{\bf x} {\bf u} (t, {\bf x}) \right) = 0.
\eea

As seen, when $R_{T}$ $>$ $R_{T *}$, due to Navier--Stokes bifurcations, the velocity field evolution will be characterized by continuous distributions of length scales and velocity values. 
{%\color{blue} 
Therefore, for $t>0$, the velocity gradient field will exhibit nonsmooth spatial variations where 
$
\ds \left\langle \nabla_{\bf x} {\bf u}(t, {\bf x}) \right\rangle  = 0
$, 
and its determinant, $\det \left( \nabla_{\bf x} {\bf u}(t, {\bf x})\right)$, is expected to frequently vanish along fluid particles trajectories.
To justify this, one could search a link between such property and the statistics of the eigenvalues of $\nabla_{\bf x} {\bf u}$ which directly arises from the fluid incompressibility \cite{Ashurst}. 
In this regard, observe that an arbitrary particle trajectory $l_t: {\bfx}(t)$ belongs to the surface $\Sigma_1$ 
\bea
\ds \Sigma_1:  \Psi_1(t; x, y, z) \equiv \nabla_{\bf x} \cdot {\bf u} (t, {\bf x})= 0
\label{sigma1}
\eea 
and identically satisfies the equation
\bea
\ds l_t \in \Sigma_1: \frac{\partial \Psi_1}{\partial t} + \nabla_{\bf x} \Psi_1 \cdot \dot{\bfx} = 0
\eea
Thanks to Navier--Stokes bifurcations and fully developed turbulence hypothesis, for $t>$ 0, $\Sigma_1$ and $l_t$ will show abrupt variations in their local placement, orientation and curvatures, and will tend to sweep the entire physical space. On the other hand, the vanishing condition of velocity gradient determinant
\bea
\ds \Sigma_2: {\cal D}(t; x, y, z) \equiv \det \left( \nabla_{\bf x} {\bf u} (t, {\bf x}) \right) = 0.
\label{sigma2}
\eea
defines the surface $\Sigma_2 \ne \Sigma_1$. Thus, the points which satisfy both the conditions (\ref{sigma1}) and (\ref{sigma2}) belong to the line $\ds l_b = \Sigma_1 \cap \Sigma_2$,
and represent all the possible kinematic bifurcations which could happen along $l_t$. 
Because of fully developed turbulence, also $l_b$ will show nonsmooth spatial variations and will tend to describe the entire physical space. Therefore, the kinematic bifurcations which occur along $l_t$ are obtained as $\ds l_t \cap l_b$, being $l_t, l_b \in \Sigma_1$. As $l_t$ and $l_b$ are two different curves of the same surface $\Sigma_1$ that exhibit chaotic behaviors,
their intersections are expected to be very frequent, forming a highly numerous set of points on $\Sigma_1$ according to the qualitative scheme of Fig. \ref{figura_2 b} wherein $l_t$
and $l_b$ are represented by solid and dashed lines. Specifically, for $R_{T}$ $>$ $R_{T *}$,  $t>$0, the Navier--Stokes bifurcations produce the regime of fully developed turbulence, where  length scales and velocity values are continuously doubled, and this causes situations where the number of the intersections between $l_t$ and  $l_b$ (kinematic bifurcations) diverges. To show this, the kinematic bifurcation rate is now introduced. This quantity,  calculated along a fluid particle trajectory, 
is defined as follows 
\bea
\begin{array}{l@{\hspace{-0.cm}}l}
\ds S_b = \lim_{T \rightarrow \infty} \frac{1}{T} \int_{0}^{T} 
\delta ({\cal D}) \ 
\vert \frac{D{\cal D}}{Dt}  \vert \ dt, \\\\
\ds \frac{D{\cal D}}{Dt}  =
\frac{\partial{\cal D}}{ \partial t}  +
\nabla_{\bf x}{\cal D} \cdot {\bf u}
\end{array}
\label{bif rate}
\eea
The rate $S_b$ can be much greater than the eigenvalues modulus of $\nabla_{\bf x}{\bf u}$ and than its maximal Lyapunov exponent.
In fact, due to the Navier--Stokes bifurcations and to the hypothesis of fully developed chaos, the characteristic scales of $\bf u$ are continuously doubled, thus  ${\cal D} \equiv \det  \left( \nabla_{\bf x} {\bf u}  \right)$ is expected to be a function of the kind
\bea
\begin{array}{l@{\hspace{-0.cm}}l}
\ds \det  \left( \nabla_{\bf x} {\bf u}  \right)= {\cal D}\left( {\bf y}_1, {\bf y}_2,..., 
{\bf y}_n\right), \\\\ 
\ds {\bf y}_k= \frac{\bf x}{\ell_k}, \ k = 1, 2,..., n \\\\
\ds \ell_1>\ell_2>...>\ell_n, \\\\
\ds O\vert \frac{\partial {\cal D}}{\partial {\bf y}_1} \vert 
\ds \approx O\vert \frac{\partial {\cal D}}{\partial {\bf y}_2} \vert  ...
\ds \approx O\vert \frac{\partial {\cal D}}{\partial {\bf y}_n} \vert, 
\end{array}
\label{detA}
\eea
\begin{figure}[h]
\centering
\includegraphics[width=8.0cm, height=7.0cm]{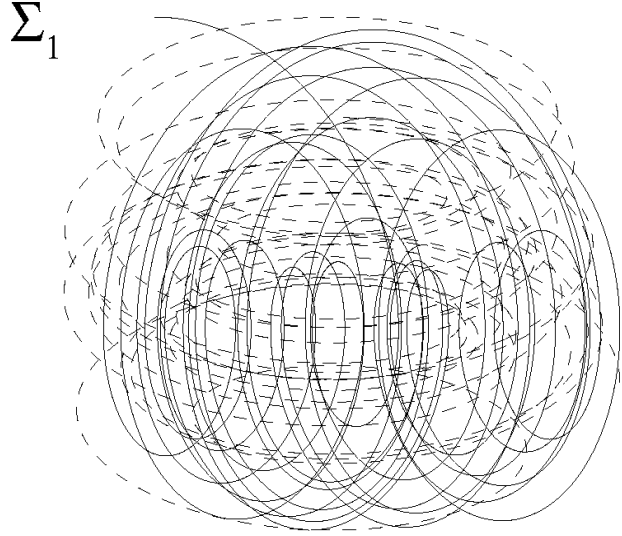}
\caption{%\color{blue}
 Qualitative scheme of fluid particle trajectory $l_t$, bifurcation line $l_b$, and their intersections over $\Sigma_1$.}
\label{figura_2 b}
\end{figure}
where, due to bifurcations, $n$ tends to diverge, and 
\bea
\begin{array}{l@{\hspace{-0.cm}}l}
\ds \nabla_{\bf x} {\cal D} = \Sum_{k=1}^n \frac{\partial {\cal D}}{\partial {\bf y}_k} \frac{1}{\ell_k}, \\\\
\ds O \left( \frac{1}{\ell_n} \ \vert \frac{\partial {\cal D}}{\partial {\bf y}_n} \cdot {\bf u} \vert \right) 
>>> O \left( \vert \frac{\partial {\cal D}}{\partial t} \vert\right) 
\end{array}
\eea
For one assigned velocity field, from Eqs. (\ref{bif rate}) and (\ref{detA}), the simultaneous values of $\bf u$ and $\nabla_{\bf x} (\det (\nabla_{\bf x} {\bf u}))$ can cause very frequent kinematic bifurcations whose rate can be significantly greater than the maximal Lyapunov exponent of Eq. (\ref{kin finite}) or (\ref{kin}).
In fact, following Eqs. (\ref{bif rate}) and (\ref{detA}), the order of magnitude of $S_b$ identifies the ratio (large scale velocity)--(small scale length)
\bea
\ds S_b \approx \frac{u}{\ell_n}
\eea
where the small scale $\ell_n$ represents the minimum distance between to successive kinematic bifurcations encountered along fluid particle trajectory. 
This means that the changing rate of $\nabla_{\bf x} {\bf u}$ along $l_t$ can be much more rapid than the rate of divergence of two contiguous trajectories.

At this stage of the present study, $S_b$ is assumed to be much greater than the maximal
Lyapunov exponent of Eq. (\ref{kin finite}), and its estimation will be performed in the following as soon as $\ell_n$ is identified by means of this analysis.
}
\bigskip

\section{Lyapunov kinematic analysis}

The aim of this section is to discuss how, in fully developed turbulence, the fluctuations of fluid particles displacements and local strain can be much more rapid and statistically independent with respect to the time variations of velocity field.
To analyze this, consider that, in fully developed turbulence, the Navier--Stokes bifurcations cause non smooth spatial variations of ${\bf u}(t, {\bf x})$ which in turn deternine very frequent kinematic bifurcations. 
Due to the fluid incompressibility, two fluid particles will describe chaotic trajectories, 
${\bfx}(t)$ and ${\bf y}(t)={\bfx}(t)+{\bfxi}(t)$, which diverge with each other with a local rate of divergence quantified by the local Lyapunov exponent of finite scale $\xi$
\bea
\ds \tilde{\lambda} = \frac{\dot{\bfxi} \cdot {\bfxi}}{{\bfxi} \cdot {\bfxi}}
\eea
According to such definition of $\tilde{\lambda}$, around to a given instant, $t_0$,  $\bfxi$ and $\dot{\bfxi}$ can be expressed as
\bea
\begin{array}{l@{\hspace{-0.cm}}l}
\ds {\bfxi}={\bf Q}(t){\bfxi}(t_0) \exp\left( \tilde{\lambda} \left( t-t_0 \right) \right), \\\\
\ds \dot{\bfxi}= \tilde{\lambda} {\bfxi} + {\bfomega}_E \times {\bfxi}
\end{array}
\eea
as long as $\vert {\bfxi} \vert \approx \vert {\bfxi}(t_0) \vert=r$,
where $\bf Q$ is an orthogonal matrix giving the orientation of $\bfxi$ with respect to the
inertial frame $\cal R$, and ${\bfomega}_E$ is the angular velocity of ${\bfxi}$ with respect to $\cal R$ whose determination is carried out by means of a proper orthogonalization procedure of the Lyapunov vectors described in Ref. \cite{deDivitiis_6}. The classical local Lyapunov exponent is obtained for  
$\vert \bfxi \vert \rightarrow$0, $\tilde{\lambda} \rightarrow \Lambda$, that is
\bea
\ds \tilde{\Lambda} = \frac{d {\bfx}  \cdot  \nabla_{\bf x} {\bf u}d {\bfx} }{d {\bfx} \cdot d {\bfx}}
\eea
On the other hand, $d {\bfx}$ can be expressed through Eq. (\ref{kin}) as follows  
\bea
\ds d {\bfx} = \exp\left( \int_0^t \nabla_{\bf x} {\bf u} (t', {\bfx}(t')) dt' \right)  d {\bfx}_0
\label{dx}
\eea
where the exponential denotes the series expansion of operators 
\bea
\ds  \exp\left( \int_0^t \nabla_{\bf x} {\bf u} (t', {\bfx}(t')) dt' \right)  = 
 {\bf I} + \int_0^t \nabla_{\bf x} {\bf u} (t', {\bfx}(t')) dt' + ...
\eea
Although in developed turbulence the Navier--Stokes bifurcations cause abrupt spatial variations of velocity and temperature, with $\lambda_{NS}>$0, due to fluid dissipation, $\bf u$ and $\vartheta$ 
are in any case functions of slow growth of $t \in(0, \infty)$, whereas $\bfxi$ and  $d {\bfx}$, being not bounded by the dissipation effects, are functions of exponential growth of $t$. 
Therefore, in line with the analysis of  Ref. \cite{deDivitiis_3}, 
{%\color{blue}
 and taking into account that $S_b >> \sup\left\lbrace \tilde{\lambda}\right\rbrace$,
that $\bfxi$ and $d {\bfx}$ are much more rapid than ${\bf u}(t, {\bf x})$ being $\sup \left\lbrace \tilde \lambda \right\rbrace>>\lambda_{NS}$, it follows that 
}  
$\bfxi$ and $d {\bfx}$ will exhibit power spectra in frequency intervals which are completely separated with respect to those of the power spectum of ${\bf u}$. 
To study this, consider now the Taylor series expansion of $\bf u$ with respect to $t$ of  the trajectories equations, i.e.
\bea
\begin{array}{l@{\hspace{-0.cm}}l}
\ds \dot{{\bfx} } = {\bf u} (0, {\bfx}(t))+ ...,   \\\\
\ds \dot{{\bfxi} } = {\bf u} (0, {\bfx}(t)+{\bfxi}(t)) - {\bf u} (0, {\bfx}(t))+..., \ \ \ \
\mbox{for finite scale $\vert \bfxi \vert$}, \\\\
\ds {d \dot{\bfx} } = \nabla_{\bf x} {\bf u} (0, {\bfx}(t))  d {\bfx}+..., \ \ \ \
\mbox{for contiguous trajectories}
\end{array}
\label{kin1 finite}
\eea
The first terms (terms of 0 order) of such Taylor series do not correspond to time variations in velocity field, thus these do not modify the fluid kinetic energy. Furthermore, as  
$\sup \left\lbrace \tilde \lambda \right\rbrace>>\lambda_{NS}$ (fully developed turbulence),
such terms reproduce the particles trajectories as long as $0 <t< O(1/\lambda_{NS})$, that is
\bea
\begin{array}{l@{\hspace{-0.cm}}l}
\ds \dot{{\bfx} } \simeq {\bf u} (0, {\bfx}(t)),   \\\\
\ds \dot{{\bfxi} } \simeq {\bf u} (0, {\bfx}(t)+{\bfxi}(t)) - {\bf u} (0, {\bfx}(t)), \ \ \ \
\mbox{for  finite scale $\vert \bfxi \vert$}, \\\\
\ds {d \dot{\bfx} } \simeq \nabla_{\bf x} {\bf u} (0, {\bfx}(t))  d {\bfx}, \ \ \ \
\mbox{for contiguous trajectories}
\end{array}
\ds \ \ \ \ \forall  t \in (0, a),  \ \ \ \ a={\cal O}\left( \frac{1}{\lambda_{NS}}\right) 
\label{kin2 finite}
\eea
Following Eq. (\ref{kin2 finite}), the fluctuations of $\bfxi$ and $d {\bfx}$ are statistically independent with respect to the time variations of the velocity field.
Next, $\sup \left\lbrace \tilde \lambda \right\rbrace>>\lambda_{NS}$, thus 
the number of kinematic bifurcation, which happen for $0 <t< O(1/\lambda_{NS})$, is expected to be quite high and can be considered to be significative from the statistical point of view.

Now, according to the mathematical analysis of the continuum media \cite{Truesdell77},
the following map is considered 
\bea
\ds \chi(., t): {\bf x}_0 \rightarrow {\bfx}(t)
\eea
which expresses the placement of material elements at the current time $t$ in function of their referential position, say ${\bf x}_0= {\bfx}(0)$ \cite{Truesdell77}.
From Eq. (\ref{dx}), the local fluid strain $\ds \partial {\bfx}(t)/\partial {\bfx}_0$ 
is then an exponential growth function of $t$ which, thanks to the above mentioned property of independence of $d {\bfx}$ from ${\bf u}(t, {\bf x})$, results to be independent and much faster with respect to the time variations of the velocity field. In fact, from the Lyapunov theory of kinematic field, such strain reads as  
\bea
\ds \frac{\partial {\bfx}}{\partial {\bfx}_0} \equiv  
\exp\left( \int_0^t \nabla_{\bf x} {\bf u} (t', {\bfx}(t')) dt' \right) 
\equiv  
\exp\left( \int_0^t \nabla_{\bf x} {\bf u} (0, {\bfx}(t')) dt' \right)+ ...=
{\bf G} \exp\left( \tilde{\Lambda} \ t\right), 
%\ \ \ \Lambda =  \sup\left\lbrace  \tilde{\Lambda}\right\rbrace 
\label{strain}
\eea 
where $\bf G$ is a proper fluctuating matrix whose elements $G_{i j} = O(1)$ are functions of
of slow growth of $t$. As long as $ t \in (0, a)$ we have 
\bea
\ds \frac{\partial {\bfx}}{\partial {\bfx}_0} \simeq  
%\exp\left( \int_0^t \nabla_{\bf x} {\bf u} (t', {\bfx}(t')) dt' \right) 
%= 
\exp\left( \int_0^t \nabla_{\bf x} {\bf u} (0, {\bfx}(t')) dt' \right)=
{\bf G} \exp\left( \tilde{\Lambda} \ t\right), 
\ds \ \ \ \ \forall  t \in (0, a) 
\label{strain1}
\eea 
that is $\ds \partial {\bfx}(t)/\partial {\bfx}_0$ is independent of the time variations of the velocity field.

In brief, as $\sup\left\lbrace \tilde{\lambda} \right\rbrace>>$ $\lambda_{NS}$, two time scales are here considered: one associated with the velocity field and the other one related to the relative fluid kinematics. Thus, $\bfxi$, $\ds \partial {\bfx}(t)/\partial {\bfx}_0$ and $\tilde{\lambda}$ are statistically independent of $\bf u$. Furthermore, due to very frequent kinematic bifurcations in $(t , t+ 1/\lambda_{NS})$, $\bfxi$, local strain and $\tilde{\lambda}$ are expected to be continuously distributed in their variation ranges.
{%\color{red}
This conclusion is supported by the arguments in Ref. \cite{Ottino89, Ottino90} (and references therein), where the author remarks among other things that
the
fields 
${\bf u}(t, {\bf x})$, (and therefore also ${\bf u}(t, {\bf x+ \bfxi}) - {\bf u}(t, {\bf x})$) produce chaotic trajectories also for relatively simple mathematical structure of 
${\bf u}(t, {\bf x})$ (also for steady fields!). 
}
\bigskip

\section{$^*$Turbulent energy cascade, material vorticity and link with classical kinematic Lyapunov analysis}

By means of theoretical considerations based on the classical Lyapunov theory 
{%\color{red} 
and on the property that the kinematic bifurcation rate is much larger than the maximal Lyapunov exponent of  the velocity gradient}, an interpretation of the kinetic energy cascade phenomenon is given which shows that $\bfeta \equiv d \bfx$ is much more rapid and statistically independent with respect to  $\bf u$. Following such considerations, the vorticity equation of a material element (material vorticity) --directly obtained making the curl of the incompressible Navier--Stokes equations-- is compared with the evolution equation of $\bfeta$ which follows the classical Lyapunov theory. These equations read as
\bea
\begin{array}{l@{\hspace{-0.cm}}l}
\ds  \frac{D {\bfomega}}{D t} \equiv  
\frac{\partial {\bfomega}}{\partial t} + 
\nabla_{\bf x} {\bfomega} \ {\bf u} 
= \nabla_{\bf x} {\bf u} \ {\bfomega} +
\nu \nabla_{\bf x}^2 {\bfomega}, \\\\ 
\mbox{being} \ \ {\bfomega} = \nabla_{\bf x} \times {\bf u}, \\\\
\ds \frac{D {\bfeta}}{D t} \equiv \dot{\bfeta} = \nabla_{\bf x} {\bf u} \ {\bfeta}, \\\\
\ds \frac{D {\bfx}}{D t} \equiv \dot{\bfx} = {\bf u}(t, {\bfx}), \\\\
\ds t \in \left( t_0, t_0+a \right) 
\end{array}
\label{vorticity}
\eea
From such relations, it is apparent that, for inviscid fluids ($\nu=0$), the time variations
of $\bfeta$ and of $\bfomega$ along a fluid particle trajectory $\bfx$=$\bfx(t)$, follow the same equation, thus $\bfomega$ identifies those particular Lyapunov vectors such that 
$\bfeta \propto \nabla_{\bf x} \times {\bf u}$ at the initial time $t_0$. 
On the other hand, regardless of the initial condition $\bfeta(0)$, $\bfeta(t)$ tends to align with the direction of the maximum rising rate of the trajectories distance \cite{Ott2002}. 
If ${\bfomega}(t_0) = k \ {\bfeta}(t_0)$, then 
${\bfomega}(t)= k \ {\bfeta}(t)$, $\forall t>t_0$ (von Helmholtz), where $k$ does not depend on $t$, while $\bfeta$ is a fast growth function of $t$. Thus, following the Lyapunov theory, for inviscid fluids, $\vert \bfomega \vert$, calculated along $\bfx$=$\bfx(t)$, tends to exponentially rise with $t$.
More in general, for inviscid fluids, $\bfomega$ and $\bfeta$ are both fast growth (exponential) functions of the time, where $\bfomega$ tends to align to the direction of maximum growth rate of $\vert \bfeta \vert$ \cite{Ott2002}.

A nonzero viscosity influences the time variations of the material vorticity making this latter a slow growth function of $t \in (t_0, \infty)$, whereas $\bfeta$ and $\bfxi$ remain in any case exponential growth functions of $t$. This implies that, for $\nu \ne 0$, the characteristic time scales of $\bf u$ (and $\vartheta$) and $\bfeta$ are different, and that after the time $t_0+a$,  
the fluctuations of $\bfxi$ result to be statistically independent from $\bf u$. 
This holds also when $\nu \rightarrow 0$ for properly small length scales, except for $\nu=0$.

Based on the previous observations, the combined effect of very frequent bifurcations and stretching term $\nabla_{\bf x} {\bf u} \ {\bfomega}$ produces the kinetic energy cascade. This phenomenon  regards each fluid particle, where $\nabla_{\bf x} {\bf u} \ {\bfomega}$ acts on the material vorticity in the same way in which $\nabla_{\bf x} {\bf u} \ {\bfeta}$ influences $\bfeta$.
In fact, according to Eqs. (\ref{vorticity}), as long as $\vert \nabla_{\bf x} {\bf u} \ {\bfomega} \vert >> \nu \vert \nabla_{\bf x}^2 {\bfomega} \vert$, arbitrary material lines $\bfeta$ --thus arbitrary material volumes built on different Lyapunov vectors $\bfeta$, i.e. $\bfeta_1\times\bfeta_2\cdot\bfeta_3$-- moving along ${\bfx}(t)$, experience the material vorticity growth and deform according to the Lyapunov theory.
According to the analysis of the previous section, such growth phenomenon, due to $\nabla_{\bf x} {\bf u} \ {\bfomega}$, preserves the average kinetic energy and corresponds to the continuous kinetic energy transfer from large to small scales i.e. the kinetic energy cascade phenomenon. 
Due to the arbitrary choice of ${\bfx}(t)$, this pertains all the fluid particles.
For what concerns the thermal energy cascade, $\vartheta$ is a passive scalar, the temperature  follows the velocity fluctuations according to Eqs. (\ref{invr}), thus the cascade of thermal energy is direct consequence of the mechanism of kinetic energy cascade. 

In brief, the energy cascade can be linked to the material vorticity tendency to be proportional to the classical Lyapunov vectors whose modulus changes according to the Lyapunov theory. 
{%\color{red} 
Specifically, according to Eqs. (\ref{vorticity}) and taking into account that 
$S_b >> \sup\left\lbrace \tilde{\lambda} \right\rbrace$,  $\bfeta$ 
is much faster and statistically independent with respect to the velocity field, while the  energy cascade can be viewed as a continuous and intensive stretching and folding process of fluid particles which involves smaller and smaller length scales during their motion, and where the particle folding process happens with a frequency given by the bifurcation rate.}

\bigskip

\section{Distribution functions of ${\bf u}$, $\vartheta$, $\bfx$, $\bfxi$ and $\tilde \lambda$}

{%\color{red} 
Following the present formulation, ${\bf u}$, $\vartheta$, $\bfx$ and $\bfxi$ are
the fluid state variables.}
Therefore, the distribution function of ${\bf u}$, $\vartheta$, $\bfx$ and $\bfxi$, say $P$, varies according to the Liouville theorem associated with (\ref{NS_op})--(\ref{T_op}) and (\ref{kin finite}) \cite{Nicolis95}
\bea
\begin{array}{l@{\hspace{-0.cm}}l}
\ds \frac{\partial P}{\partial t} +
\frac{\delta}{\delta {\bf u}} \cdot \left( P \dot{\bf u} \right) +
\frac{\delta}{\delta \vartheta} \cdot \left( P \dot{\vartheta} \right) + 
\frac{\partial}{\partial {\bfx}} \cdot \left( P \dot{\bfx} \right) +
\frac{\partial}{\partial {\bfxi}} \cdot \left( P \dot{\bfxi} \right) 
=0
\end{array}
\label{Liouville}
\eea
{%\color{red} 
where, according to the notation of Eqs. (\ref{NS_op})--(\ref{T_op}), 
$\delta/ \delta {\bf u}$ and  $\delta/ \delta {\vartheta}$ are functional partial derivatives
with respect to $\bf u$ and $\vartheta$, respectively and }
$\left( \partial/ \partial \circ \right) \cdot$ stands for the divergence with respect to $\circ$.
In line with the previous analysis and  with Refs. \cite{deDivitiis_5, deDivitiis_6}, $P$ can be factorized as follows
\bea 
\begin{array}{l@{\hspace{-0.cm}}l}
\ds P(t, {\bf u}, \vartheta, {\bfx}, {\bfxi}) = F(t, {\bf u}, \vartheta) P_\xi(t, {\bfx}, {\bfxi})
\end{array}
\label{indep stat}
\eea
being $F$ and $P_\xi$ the distribution functions of ($\bf u$, $\vartheta$), and of ($\bfx$, $\bfxi$), respectively. Their evolution equations are formally obtained from Eq. (\ref{Liouville}) and taking into account the aforementioned statistical independence (\ref{indep stat}). This allows to split the Liouville equation (\ref{Liouville}) in the two following equations
\bea
\begin{array}{l@{\hspace{-0.cm}}l}
\ds \frac{\partial F}{\partial t} +
\frac{\delta}{\delta {\bf u}} \cdot \left( F \dot{\bf u} \right) +
\frac{\delta}{\delta \vartheta} \cdot \left( F \dot{\vartheta} \right) 
=0, \\\\
\ds \frac{\partial P_\xi}{\partial t} +
\frac{\partial}{\partial {\bfx}} \cdot \left( P_\xi \dot{\bfx} \right) +
\frac{\partial}{\partial {\bfxi}}\cdot \left( P_\xi \dot{\bfxi} \right) 
=0
\end{array}
\label{Liouville splitted}
\eea
where the boundary conditions of $P_\xi$ read as 
\bea
\begin{array}{l@{\hspace{-0.cm}}l}
\ds P_\xi = 0, \ \forall ({\bfx}, {\bfxi}) \in \partial
\left\lbrace \left\lbrace {\bfx} \right\rbrace \times \left\lbrace {\bfxi} \right\rbrace \right\rbrace 
\end{array}
\label{BC}
\eea
In case of homogeneous and isotropic turbulence, $P_\xi$ does not depend on $\bfx$, and 
can be expressed in function of the finite scale $r$ as follows
\bea
\begin{array}{l@{\hspace{-0.cm}}l}
\ds P_\xi \approx  \sum_k  \delta({\bfxi} - {\bf r}_k ), \\\\
\ds \vert {\bf r}_k  \vert =r, \forall k
\end{array}
\label{Pxi_0} 
\eea
where $\delta$ denotes the Dirac's delta, and ${\bf r}_k$ are uniformely distributed points on a sphere $\cal S$ of radius $r$ due to isotropy hypothesis, 
{%\color{red} 
being $k$ a generic index indicating the several points on $\cal S$.}
This leads to
\bea
\begin{array}{l@{\hspace{-0.cm}}l}
\ds P_\xi= 
 \frac{1}{4 \pi r^2} \ \delta(\vert \bfxi \vert -r)
\end{array}
= 
\left\lbrace 
\begin{array}{l@{\hspace{-0.cm}}l}
\ds 
C \rightarrow \infty \ \ \mbox{if} \  \vert {\bfxi} \vert= r  \\\\
\ds 0 \ \ \mbox{elsewhere} 
\end{array}\right.
\label{Pxi_1} 
\eea
Also $\tilde \lambda$ and $\bfomega_E$ are statistically independent of the velocity field and are continuously distributed in their ranges of variation. In particular, the PDF of $\tilde \lambda$, say $P_\lambda$, can be calculated by means of $P_\xi$ with the Frobenius--Perron equation
\bea
\ds P_\lambda \left( \tilde{\lambda} \right)
= \int_{\bfx} \int_{\bfxi} P_\xi \ \delta\left( \tilde{\lambda}- \frac{\dot{\bfxi}\cdot {\bfxi}}{{\bfxi}\cdot {\bfxi}}\right)  \ d{\bfx} d{\bfxi} 
\label{FrobeniusPerron}
\eea
Now, in isotropic turbulence, the longitudinal component of the velocity
difference $\dot{\bfxi} \cdot{\bfxi}/r$ is uniformely distributed in its variation range as ${\bfxi}$ sweeps $\cal S$, while, due to the fluid incompressibility, $\tilde \lambda$ is expected to vary in the interval $\left( -\lambda_S/2, \lambda_S\right)$, where $\lambda_S =\sup\left\lbrace \tilde{ \lambda}\right\rbrace$.
Therefore, substituting Eq. (\ref{Pxi_0}) in Eq. (\ref{FrobeniusPerron}), we found that 
$\tilde \lambda$ uniformely sweeps $\left( -\lambda_S/2, \lambda_S\right)$, according to 
\bea
\ds P_\lambda = 
\left\lbrace 
\begin{array}{l@{\hspace{-0.cm}}l}
\ds \frac{2}{3}\frac{1}{{\lambda}_S}, \ \ \mbox{if} \  \tilde{\lambda} \in \left( -\frac{{\lambda}_S}{2}, {\lambda}_S\right)  \\\\
\ds 0 \ \ \mbox{elsewhere} 
\end{array}\right. 
\label{Pl}
\eea
Observe that, Eqs. (\ref{Pxi_1}) and (\ref{Pl}) agree with the results of Ref. \cite{deDivitiis_6}, where the author shows that $\bfxi$ and $\tilde{\lambda}$ are both uniformely distributed in their ranges by means of the condition $\cal H$=max compatible with certain constraints, being $\cal H$  the entropy associated with the kinematic state $(\bfx, \bfxi)$. This is because the isotropic homogeneous turbulence hypotheses, here expressed through Eqs. (\ref{Pxi_0})--(\ref{Pxi_1}), correspond to the maximum of $\cal H$.
The causes of the nonsymmetric distribution of $\tilde{\lambda}$ with respect the origin, also analyzed in Ref. \cite{deDivitiis_6}, are fluid incompressibility and alignment property of $\bfxi$ with respect to the maximum rising rate direction. Following such property, regardless of the initial condition $\bfxi(0)$, $\bfxi(t)$ tends to align with the direction of the maximum rising rate of the trajectories distance \cite{Ott2002}. Therefore, such distribution function provides positive average Lyapunov exponents, and gives the link between average and square mean values of the finite scale Lyapunov exponent according to 
\bea
\begin{array}{l@{\hspace{-0.cm}}l}
\ds \left\langle \tilde \lambda \right\rangle_\xi =
\frac{1}{2} \sqrt{\left\langle \tilde \lambda^2 \right\rangle_\xi} > 0.
\end{array}
\label{mPl}
\eea 
where $\left\langle \circ \right\rangle_\xi$ indicates the average of $\circ$ calculated, through
$P_\xi$ or $P_\lambda$.

\bigskip

\section{$^*$Finite time Lyapunov exponents and their distribution in fully developed turbulence}

Altough the local Lyapunov exponent $\tilde{\lambda}$ quantifies the local trajectories divergence in a point of space, in practice, the trajectory stability is evaluated by observing the particle motion in a finite time interval, say $(t_0, t_0+\tau)$. For this reason, it is useful to define the finite time Lyapunov exponent as the average of $\tilde{\lambda}$ in such time interval, that is
\bea
\begin{array}{l@{\hspace{-0.cm}}l}
\ds \tilde{\lambda}_{\tau} = \frac{1}{\tau} \int_{t_0}^{t_0+\tau} \tilde{\lambda} \ dt =
 \frac{1}{\tau} \int_{t_0}^{t_0+\tau} \frac{d}{dt} \ln \varrho \ dt = 
 \frac{1}{\tau} \ln \left( \frac{\varrho(t_0 + \tau)}{\varrho(t_0)}\right). \\\\
\ds \varrho = \vert {\bfxi} \vert.
\end{array}
\eea
This  exponent trivially satisfies
\bea
\begin{array}{l@{\hspace{-0.cm}}l}
\ds \lim_{\tau \rightarrow 0} \tilde{\lambda}_\tau = \tilde{\lambda}.
\end{array}
\eea
If $\tau$ is properly high, a statistically significant number of kinematic bifurcations $n$ can occur for $t \in (t_0, t_0+\tau)$, thus $\tilde{\lambda}_\tau$ is in general a fluctuating variable which exhibits variations whose amplitude diminishes as $\tau$ increases.
Accordingly, $\tilde{\lambda}_\tau$ will be distributed following a gaussian PDF in fully developed turbulence. In fact, due to the bifurcations encountered in $(t_0, t_0+\tau)$, $\tilde{\lambda}_\tau$ can be written as sum of several terms, each of them related to the effects of a single bifurcation, i.e.
\bea
\begin{array}{l@{\hspace{-0.cm}}l}
\ds \tilde{\lambda}_{\tau} =  
\frac{1}{\tau} \ln \left( \frac{\varrho(t_0 + \tau)}{\varrho(t_0)}\right)  =
\frac{1}{\tau} \ln \left( \frac{\varrho(t_0 + \tau)}{\varrho_{n-1}} \ \frac{\varrho_{n-1}}{\varrho_{n-2}} \ ... \
 \frac{\varrho_{1}}{\varrho(t_0)} \right) = 
\frac{1}{\tau} \Sum_{k=1}^n \ln \left( \frac{\varrho_{k}}{\varrho_{k-1}} \right) 
\end{array}
\label{l clt}
\eea
where $\ln ( \varrho_{k}/ \varrho_{k-1} )$ gives the contribution of the $k$th bifurcation starting from $t_0$, being $\varrho_{k-1}$ and $\varrho_k$ the Lyapunov vectors moduli calculated
immediately before and after the $k$th bifurcation. 
On the other hand, due to fully developed chaos, each of such terms is expected to be statistically independent of all other ones, and if $\tau \rightarrow \infty$, the number of encountered bifurcations $n$ diverges. Hence, a proper variant of the central limit theorem can be applied, and this would guarantee that $\tilde{\lambda}_\tau$ tends to a gaussian stochastic variable \cite{Hirokazu83}. 
{%\color{red} 
The novelty of the present section consists in the implication that the property $S_b >> \lambda_\tau$ has on Eq. (\ref{l clt}). Such property should ensure that $\lambda_\tau$ can be approximated to a gaussian stochastic variable also for certain finite values of $\tau$. 
In fact, if $\tau \approx 1/\lambda_\tau$ or $\tau \gtrsim 1/\lambda_\tau$, the time interval $(t_0, t_0+ \tau)$ should include a statistically significant number of kinematic bifurcations, thus the distribution function of $\lambda_\tau$ is expected to be a gaussian PDF, expecially for relatively high values of the Taylor scale Reynolds number.
}

\bigskip

\section{Closure of von K\'arm\'an--Howarth and Corrsin equations}

Starting from the property of statistical independence (\ref{indep stat}) and adopting the Liouville theorem, the closure formulas of von K\'arm\'an-Howarth and Corrsin equations are here determined and the effects of the chaotic trajectories divergence on these closures are discussed.

In fully developed isotropic homogeneous turbulence, the pair correlation
functions of longitudinal velocity components and of temperature, defined as 
\bea
\begin{array}{l@{\hspace{-0.cm}}l}
\ds f(r) = \frac{\langle u_r(t, {\bf x}) u_r(t, {\bf x}+ {\bf r}) \rangle}{u^2} \equiv
\frac{\langle u_r u_r' \rangle}{u^2}, \\\\
\ds f_\theta(r) = \frac{\langle \vartheta (t, {\bf x}) \vartheta(t, {\bf x}+ {\bf r}) \rangle}{\theta^2}
\equiv \frac{\langle \vartheta  \vartheta' \rangle}{\theta^2}.
\end{array}
\eea
satisfy the von K\'arm\'an--Howarth equation \cite{Karman38} and Corrsin equation 
\cite{Corrsin_1, Corrsin_2}, respectively, where
\bea
\begin{array}{l@{\hspace{-0.cm}}l}
\ds u_r = {\bf u}(t, {\bf x})\cdot \frac{\bf r}{r}, \ u_r' = {\bf u}(t, {\bf x}+ {\bf r})\cdot \frac{\bf r}{r} \ 
\end{array}
\eea
von K\'arm\'an--Howarth and Corrsin equations are properly obtained from the Navier--Stokes and heat equations written in two points of space, say $\bf x$ and $\bf x + r$. These correlation equations read as follows 
\bea
\begin{array}{l@{\hspace{-0.cm}}l}
\ds \frac{\partial f}{\partial t} = 
\ds  \frac{K}{u^2} +
\ds 2 \nu  \left(  \frac{\partial^2 f} {\partial r^2} +
\ds \frac{4}{r} \frac{\partial f}{\partial r}  \right) +\frac{10 \nu}{\lambda_T^2} f, \\\\
\ds \frac{\partial f_\theta}{\partial t} = 
\ds  \frac{G}{\theta^2} +
\ds 2 \chi  \left(  \frac{\partial^2 f_\theta} {\partial r^2} +
\ds \frac{2}{r} \frac{\partial f_\theta}{\partial r}  \right) +\frac{12 \chi}{\lambda_\theta^2} f_\theta,
\end{array}
\label{vk-h}
\eea
The boundary conditions associated with such equations are
\bea
\begin{array}{l@{\hspace{+0.2cm}}l}
\ds f(0) = 1,  \ \ \ \ \ds \lim_{r \rightarrow \infty} f (r) = 0, \\\\
\ds f_\theta(0) = 1,  \ \ \ \ \ds \lim_{r \rightarrow \infty} f_\theta (r) = 0, 
\end{array}
\label{bc0}
\eea
being $u \equiv \sqrt{\langle u_r^2 \rangle}$,  $\theta \equiv \sqrt{\langle \vartheta^2 \rangle}$, 
where $\lambda_T \equiv \sqrt{-1/f''(0)}$ and $\lambda_\theta \equiv \sqrt{-2/f_\theta''(0)}$ are  Taylor and Corrsin microscales, respectively. 
The quantities $K$ and $G$, arising from inertia forces and convective terms, give the energy cascade, and are expressed as \cite{Karman38, Corrsin_1, Corrsin_2}
\bea
\begin{array}{l@{\hspace{-0.cm}}l}
\ds \left( 3 + r \frac{\partial}{\partial r} \right) K = 
\frac{\partial}{\partial r_k} \left\langle  u_i u_i'\left( u_k-u_k'\right) \right\rangle, \\\\
\ds G =  
\frac{\partial}{\partial r_k} \left\langle  \vartheta \vartheta' \left( u_k-u_k'\right) \right\rangle,
\end{array}
\label{KG}
\eea
where the repeated index denotes the summation convention.
Following the theory \cite{Karman38, Corrsin_1, Corrsin_2}, $K$ and $G$ are linked to the longitudinal triple velocity correlation function $k$, and to the triple correlation between $u_r$ and $\vartheta$, according to  
\bea
\begin{array}{l@{\hspace{+0.0cm}}l}
\ds K(r) = u^3 \left( \frac{\partial }{\partial r} + \frac{4}{r} \right) k(r), 
\ \ \mbox{where} \ \ 
\ds k(r) = \frac{\langle u_r^2 u_r' \rangle}{u^3}, \\\\
\ds G(r) = 2 u \theta^2 \left( \frac{\partial }{\partial r} + \frac{2}{r} \right) m^*(r), 
\ \ \mbox{where} \ \ 
\ds m^*(r) = \frac{\langle u_r \vartheta \vartheta' \rangle}{\theta^2 u},
\end{array}
\eea
As well known from the literature \cite{Karman38, Corrsin_1, Corrsin_2}, without particular hypotheses about the statistics of $\bf u$ and $\vartheta$,  $K$ and $G$ are unknown quantities which can not be expressed in terms of $f$ and $f_\theta$, thus at this stage of this analysis, both the correlations equations (\ref{vk-h}) are not closed.

In order to obtain analytical forms of $K$ and $G$, observe that these latter, representing the energy flow between length scales in the fluid, do not modify the total amount of kinetic and thermal energies \cite{Batchelor53, Corrsin_1}. Indeed, convective term, inertia and pressure forces determine interactions between Fourier components of velocity and temperature fields providing the transfer of kinetic and thermal energy between volume elements in the wavenumber space, whereas the global effect of such these interactions leaves $u^2$ and $\theta^2$ unaltered \cite{Batchelor53, Corrsin_1}.
On the other hand, the proposed statistical independence property (\ref{indep stat}) allows to write 
the time derivative of $P$ as sum of two terms
\bea
\begin{array}{l@{\hspace{+0.0cm}}l}
\ds \frac{\partial P}{\partial t} = 
P_\xi \frac{\partial F}{\partial t} + 
F \frac{\partial P_\xi}{\partial t} 
\end{array}
\label{f_t}
\eea 
the first one of which, being related to ${\partial F}/{\partial t}$, provides the time variations of velocity and temperature fields. The second one, linked to ${\partial P_\xi}/{\partial t}$, not producing changing of $u^2$ and $\theta^2$, identifies the energy cascade effect. Therefore, $K$ and $G$ arise from the second term of (\ref{f_t}), and can be expressed, by means of the Liouville theorem (\ref{Liouville}) and Eqs. (\ref{Liouville splitted}), in terms of material displacements $\bfxi$, taking into account flow homogeneity and fluid incompressibility.
Specifically, from Eq. (\ref{Liouville})--(\ref{Liouville splitted}), $K$ and $G$, directly arising from $- F \partial(P_\xi \dot{\bfxi})/\partial{\bfxi}$, are calculated as follows
\bea
\begin{array}{l@{\hspace{+0.0cm}}l}
\ds K = - \int_{\cal U} \int_\Xi F \frac{\partial}{\partial {\bfxi}} \cdot \left( P_\xi \dot{\bfxi} \right) 
u_\xi u_\xi^* \ d{\cal U} d\Xi, \\\\
\ds G = - \int_{\cal U} \int_\Xi F \frac{\partial}{\partial {\bfxi}} \cdot \left( P_\xi \dot{\bfxi} \right) 
\ \vartheta \vartheta^* d{\cal U} d\Xi,
\end{array}
\label{KG1}
\eea
where $\ds {\cal U} = \left\lbrace {\bf u}\right\rbrace \times \left\lbrace {\vartheta}\right\rbrace$,  $\ds \Xi =  \left\lbrace {\bfxi}\right\rbrace$ and $d{\cal U}$ and $d\Xi$ are the corresponding elemental volumes, and
\bea
\begin{array}{l@{\hspace{-0.cm}}l}
\ds u_\xi = {\bf u}(t, {\bfx})\cdot \frac{\bfxi}{\xi}, \ \ u_\xi^* = {\bf u}(t, {\bfx}+ {\bfxi})\cdot \frac{\bfxi}{\xi}, \\\\
\ds \vartheta = \vartheta(t, {\bfx}), \ \ \vartheta^* = \vartheta(t, {\bfx}+ {\bfxi}), 
\end{array}
\eea
Integrating Eqs. (\ref{KG1}) with respect to ${\cal U}$, we obtain 
\bea
\begin{array}{l@{\hspace{+0.0cm}}l}
\ds K = - u^2 \int_\Xi \frac{\partial}{\partial {\bfxi}} \cdot \left( P_\xi \dot{\bfxi} \right) 
f(\xi) \ d\Xi, \\\\
\ds G = - \theta^2 \int_\Xi \frac{\partial}{\partial {\bfxi}} \cdot \left( P_\xi \dot{\bfxi} \right) 
 f_\theta(\xi) \ d\Xi,
\end{array}
\label{KG2}
\eea
Again, integrating by parts Eq. (\ref{KG2}) with respect to $\Xi$, taking into account the boundary conditions (\ref{BC}) ($P_\xi \equiv$ 0, $\forall {\bfxi}\in \partial \Xi$) and the isotropy hypothesis, $K$ and $G$ are written as
\bea
\begin{array}{l@{\hspace{+0.0cm}}l}
\ds K =  u^2 \int_\Xi P_\xi \frac{\partial f}{\partial {\bfxi}} \cdot \dot{\bfxi}  
\ d\Xi = u^2 \int_\Xi P_\xi \frac{\partial f}{\partial \xi} \frac{\bfxi}{\xi} \cdot \dot{\bfxi}  
\ d\Xi, \\\\
\ds G = \theta^2 \int_\Xi P_\xi \frac{\partial f_\theta}{\partial {\bfxi}} \cdot \dot{\bfxi}  
 \ d\Xi = \theta^2 \int_\Xi P_\xi \frac{\partial f_\theta}{\partial \xi} \frac{\bfxi}{\xi} \cdot \dot{\bfxi}  
 \ d\Xi,
\end{array}
\label{KG3}
\eea
Now, the Lyapunov theory provides $\dot{\bfxi}$=$\tilde{\lambda}{\bfxi}+{\bfomega_E \times {\bfxi}}$,
and in isotropic homogeneous turbulence $P_\xi =\delta(\vert{\bfxi}\vert-r)/4\pi r^2$, thus $K$ and $G$ are
\bea
\begin{array}{l@{\hspace{+0.0cm}}l}
\ds K =  u^2 \int_\Xi P_\xi \frac{\partial f}{\partial \xi} \xi \tilde{\lambda} 
\ d\Xi =u^2 \frac{\partial f}{\partial r} r \left\langle \tilde{\lambda}\right\rangle_\xi, \\\\
\ds G = \theta^2 \int_\Xi P_\xi \frac{\partial f_\theta}{\partial \xi} \xi \tilde{\lambda} 
 \ d\Xi = \theta^2 \frac{\partial f_\theta}{\partial r} r 
\left\langle \tilde{\lambda} \right\rangle_\xi,
\end{array}
\label{KG5}
\eea
Furthermore, the finite scale Lyapunov theory also gives the relationship between velocity correlation and Lyapunov exponents according to 
\bea
\begin{array}{l@{\hspace{+0.0cm}}l}
\ds \left\langle (u_\xi^*-u_\xi)^2 \right\rangle_\xi = 2 u^2 \left( 1 - f(r) \right) =  \left\langle \tilde{\lambda}^2 \right\rangle_\xi r^2, 
\end{array}
\label{lambda f}
\eea
where $\langle \tilde{\lambda} \rangle_\xi$ and $\langle \tilde{\lambda}^2 \rangle_\xi$ are linked with each other through Eq. (\ref{mPl}), therefore the closure formulas of $K$ and $G$ are in terms of autocorrelations and of their gradients
\bea
\begin{array}{l@{\hspace{+0.0cm}}l}
\ds K(r) = u^3 \sqrt{\frac{1-f}{2}} \frac{\partial f}{\partial r}, \\\\
\ds G(r) = u \theta^2 \sqrt{\frac{1-f}{2}} \frac{\partial f_\theta}{\partial r},
\end{array}
\label{K}
\label{K closure}
\eea
These closure formulas do not include second order derivatives of autocorrelations, thus
Eqs. (\ref{K}) do not correspond to a diffusive model.
The energy cascade expressed by Eqs. (\ref{K closure}) is not based on the eddy viscosity concept, being the result of the trajectories divergence in the continuum fluid.
This cascade phenomenon and Eqs. (\ref{K closure}) are here interpreted as follows: 

\bigskip

1) In fully developed chaos, the Navier--Stokes bifurcations determine a continuous distribution of velocity, temperature and of length scales, where one single bifurcation causes doubling of velocity, temperature, length scale and of all the properties associated with the velocity and temperature fields according to Eqs. (\ref{invr})--(\ref{Dini}).
This leads to nonsmooth spatial variations of velocity field and very frequent kinematic bifurcations. 

2) The huge kinematic bifurcations rate generates in turn continuous distributions of $\tilde{\lambda}$ and $\bfxi$, while fluid incompressibility and the mentioned alignment property of $\bfxi$ make $\tilde{\lambda}$ unsymmetrically distributed with $\bar{\lambda}(r)$ $\equiv$ $\left\langle \tilde{\lambda}\right\rangle_\xi>$0, and the relative particles trajectories to be chaotic.

3) The tendency of the material vorticity to follow direction and variations of the Lyapunov vectors gives the phenomenon of the kinetic energy cascade. 

\bigskip

The main asset of Eqs. (\ref{K closure}) with respect to the other models is that Eqs. (\ref{K closure}) are not based on phenomenological assumptions, such as for instance, the eddy viscosity paradigm \cite{Hasselmann58, Millionshtchikov69, Oberlack93, Mellor84, Baev, Antonia2013}, but are obtained through theoretical considerations concerning the statistical independence of $\bfxi$ from $\bf u$, and the Liouville theorem. 
Thanks to their theoretical foundation, Eqs. (\ref{K closure}) do not exhibit free model parameters or empirical constants which have to be identified.
These closure formulas coincide with those just obtained by the author in the previous works \cite{deDivitiis_1, deDivitiis_4} and \cite{deDivitiis_5}.
While Refs. \cite{deDivitiis_1, deDivitiis_4} derive such closures expressing the local fluid act of motion in the finite scale Lyapunov basis and using the frame invariance property of $K$ and $G$, Ref. \cite{deDivitiis_5} achieves the same formulas adopting maximum and average finite scale Lyapunov exponents, properly defined, and the statistical independence of $\bfxi$ and $\bf u$.
Here, unlike Refs.  \cite{deDivitiis_1, deDivitiis_4} and \cite{deDivitiis_5}, 
Eqs. (\ref{K closure}) are determined exploiting the unsymmetric distribution function of $\tilde{\lambda}$ just studied in Ref. \cite{deDivitiis_6}, showing that the assumptions of Refs. \cite{deDivitiis_1, deDivitiis_4} and \cite{deDivitiis_5} are congruent with the present analysis, corroborating the results of the previous works.

Refs. \cite{deDivitiis_1, deDivitiis_4} show that these closures describe adequately the energy cascade phenomenon and the energy spectra. In detail, $K$ reproduces the kinetic energy cascade mechanism following the Kolmogorov law, and $G$ gives the thermal energy cascade in line with the theoretical argumentation of Kolmogorov, Obukhov--Corrsin and Batchelor \cite{Batchelor_2, Batchelor_3, Obukhov}, with experimental results \cite{Gibson, Mydlarski}, and with numerical data \cite{Rogallo, Donzis}.
Moreover, Eq. (\ref{K closure}) allows the calculation of the skewness of $\Delta u_r$ and 
$\partial u_r/ \partial r$ which is directly linked to the energy cascade intensity. 
This is \cite{Batchelor53}
\bea
\ds H_3(r) \equiv 
\frac{\langle (\Delta u_r)^3 \rangle }{\langle (\Delta u_r)^2 \rangle^{3/2}} 
=
\frac{6 k(r)}{(2(1-f(r)))^{3/2}}
\label{H3}
\eea
Then, substituting Eqs. (\ref{K}) in Eq. (\ref{H3}), the skewness of $\partial u_r/ \partial r$ is 
\bea
H_3(0) = -\frac{3}{7}
\eea
This constant quantifies the effect of chaotic relative trajectories on the energy cascade in isotropic turbulence, and agrees with the several results obtained through direct numerical simulation of the Navier--Stokes equations (DNS) \cite{Chen92, Orszag72, Panda89} ($-0.47 \div -0.40$), and by means of Large--eddy simulations (LES) \cite{Anderson99, Carati95, Kang2003} ($-0.42 \div -0.40$). For sake of reader convenience, Table 1 recalls the comparison, just presented in \cite{deDivitiis_5, deDivitiis_6}, between the value of the skewness $H_3(0)$ of this analysis and those achieved by the aforementioned works. It results that the maximum absolute difference between the proposed value and the other ones results to be less than 10 $\%$. 
Therefore, the proposed hypotheses, leading to the distribution function (\ref{Pl}) and to the closures (\ref{K}), seem to be adequate assumptions for estimating turbulent energy cascade and spectra.
\begin{table}[h]
\centering
\caption{Comparison of the results: Skewness of $\partial u_r/ \partial r$ at diverse Taylor--scale Reynolds number $R_T \equiv u \lambda_T/\nu$ following different authors.}
\vspace{2. mm}
\begin{tabular}{cccc} 
\hline
\hline
Reference   &  Simulation  &  $R_T$  & $H_3(0)$  \\
\hline 
Present analysis      &  -  &  -    & -3/7 = -0.428...  \\
\cite{Chen92}       & DNS &    202 & -0.44   \\ 
\cite{Orszag72}      & DNS &    45 & -0.47    \\
\cite{Panda89}      & DNS &    64 & -0.40    \\
\cite{Anderson99}   & LES &    $<$ 71 &  -0.40          \\ 
\cite{Carati95}     & LES &   $\infty$     &  -0.40 \\  
\cite{Kang2003}     & LES &  720        &    -0.42      \\ 
\hline
\hline
 \end{tabular}
\label{table1}
\end{table} 

{
%\color{blue} 
{\bf Remark.} At this stage of the present analysis, it is worth remarking the importance of the hypothesis of statistical independence of $\bf u$ and $\bfxi$ expressed by Eq. (\ref{indep stat}).
This latter represents the hypothesis of fully developed turbulence of the present analysis, and leads to the analytical expressions of $K$ and $G$ separating the effects of the trajectories divergence in the physical space from those of the velocity field fluctuations in the Navier--Stokes phase space. Without such hypothesis, the energy cascade effect can not be expressed through the term $- F \partial(P_\xi \dot{\bfxi})/\partial{\bfxi}$ and using Eqs. (\ref{KG1}), thus the proposed closures (\ref{K}) can not be determined.}

We conclude this section by observing the limits of the proposed closures (\ref{K}).
These limits directly derive from the hypotheses under which Eqs. (\ref{K}) are obtained: 
Eqs. (\ref{K}) are valid only in regime of fully developed chaos where the turbulence
exhibit homogeneity and isotropy. Otherwise, during the transition through intermediate stages of turbulence, or in more complex situations with particular boundary conditions, for instance in the presence of wall, Eqs. (\ref{K}) cannot be applied.

\bigskip

\section{Properties of the proposed closures}

Here, some of the properties of the proposed closures (\ref{K}) are renewed, with particular reference to the evolution times of the developed velocity and temperature autocorrelations. In detail, we will show that these correlations reach their developed shape in finite times which depend on the initial condition, and that, after this period, the hypothesis of statistical independence could be not more verified. 
This result is just given in Ref. \cite{deDivitiis_5}, where the author adopts a specific Lyapunov analysis using two exponents properly defined. Unlike  Ref. \cite{deDivitiis_5}, such result is here achieved through the previously obtained local finite scale Lyapunov exponent distribution (\ref{Pl}). 
To analyze this, the evolution equations of $u$, $\theta$, $\lambda_T$ and $\lambda_\theta$ are first obtained taking the coefficients of order $r^0$ and $r^2$ of Eqs. (\ref{vk-h}) arising from the Taylor series expansion of even powers of $f$ and $f_\theta$ \cite{Karman38}, \cite{Corrsin_1, Corrsin_2} 
\bea
\begin{array}{l@{\hspace{+0.0cm}}l}
\ds f= 1-\frac{1}{2}\left( \frac{r}{\lambda_T}\right)^2 + ..., \\\\
\ds f_\theta= 1-\left( \frac{r}{\lambda_\theta}\right)^2 + ...,
\end{array}
\label{f f_theta}
\eea
This leads to the following equations
\bea
\begin{array}{l@{\hspace{+0.2cm}}l}
\ds \frac{d u^2}{d t} = - \frac{10 \nu}{\lambda_T^2} u^2, \\\\
\ds \frac{d \theta^2}{d t} = - \frac{12 \chi}{\lambda_\theta^2} \theta^2, 
\end{array}
\label{u2dot_thetadot}
\eea
\bea
\begin{array}{l@{\hspace{+0.0cm}}l}
\ds  \frac{d \lambda_T}{dt} = -\frac{u}{2} + \frac{\nu}{\lambda_T}
\left( \frac{7}{3} f^{IV}(0) \lambda_T^4-5\right), \\\\
\ds  \frac{d \lambda_\theta}{dt} = -\frac{u}{2}\frac{\lambda_\theta}{\lambda_T} + \frac{\chi}{\lambda_\theta}
\left( \frac{5}{6} f^{IV}_\theta(0) \lambda_\theta^4-6\right)
\end{array}
\label{lambda_dot}
\eea
While Eqs. (\ref{u2dot_thetadot}) do not depend on the particular adopted closures \cite{Karman38}, \cite{Corrsin_1, Corrsin_2}, Eqs. (\ref{lambda_dot}) are obtained using the 
proposed closures (\ref{K closure}).
On the other hand, it is useful to consider the fluctuations of the classical Lyapunov exponent, defined as 
\bea
\begin{array}{l@{\hspace{+0.0cm}}l}
\ds \tilde{\Lambda} = \lim_{r \rightarrow 0} \tilde{\lambda} 
= \lim_{r \rightarrow 0} \frac{d}{dt} \ln \varrho, \\\\
\varrho=\vert {\bfxi} \vert
\end{array}
\eea
which are related to $f$ through Eqs. (\ref{f f_theta}) and (\ref{lambda f}) in such a way that
\bea
\ds \Lambda = \sqrt{\left\langle \tilde{\Lambda}^2 \right\rangle } =
\ds \frac{u}{\lambda_T} \propto \lim_{r \rightarrow 0} \frac{d}{dt} \left\langle \ln \varrho \right\rangle_\xi \approx \vert \frac{d \ln \lambda_T}{ dt} \vert.
\eea
being $\Lambda$ the root mean square of $\tilde{\Lambda}$.

Following Eqs. (\ref{lambda_dot}), the time variations of $\lambda_T$, $\lambda_\theta$ and $\Lambda$ are now discussed.
The first terms at the R.H.S. of Eqs. (\ref{lambda_dot}) provide the turbulent energy cascade, 
whereas the other ones arise from the fluid diffusivities. While these latter give contributions to increase both the correlation lengths, the energy cascade mechanism tends to reduce these scales, and if such mechanism is sufficiently stronger than diffusivities, then $d \lambda_T/dt<$ 0 and $d \lambda_\theta/dt<$ 0. 

\begin{figure}[h]
\centering
\includegraphics[width=8.0cm, height=12.2cm]{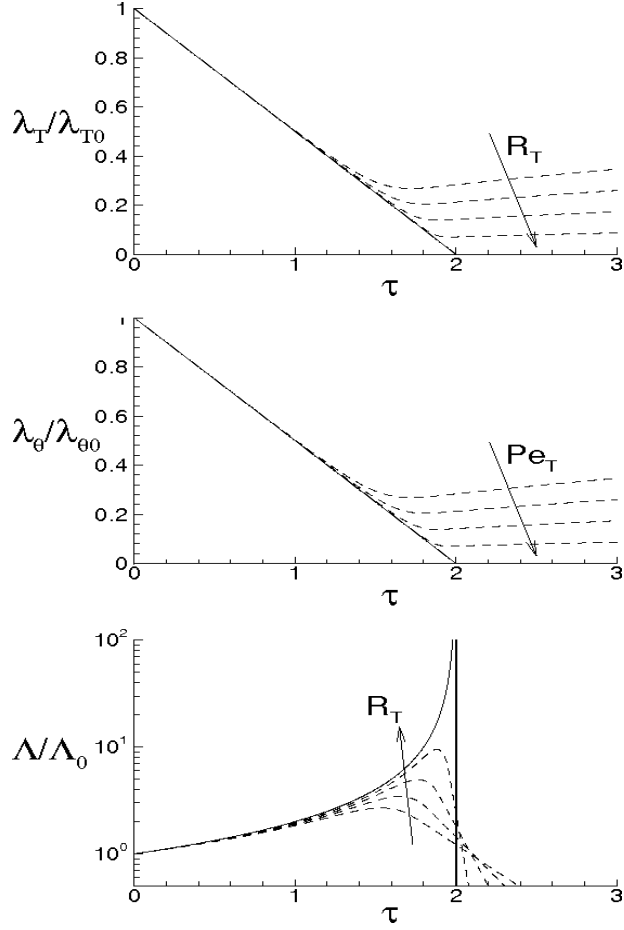}
\caption{Taylor and Corrsin microscales, and root mean square of classical Lyapunov exponent in function of the dimensionless time.}
\label{figura_3}
\end{figure}

For sake of our convenience, the condition $\nu=0$, $\chi=0$ is first studied. In this case, 
$u$ and $\theta$ are both constants, whereas $\lambda_T$, $\lambda_\theta$ and $\Lambda$ vary with t. In detail, $\lambda_T$ and $\lambda_\theta$ are proportional with each other, and vary linearly with the time according to  
\bea
\begin{array}{l@{\hspace{+0.0cm}}l}
\ds \frac{\lambda_T(t)}{\lambda_T(0)} \equiv 
\ds \frac{\lambda_\theta(t)}{\lambda_\theta(0)}=
1- \frac{\tau}{2}, \\\\
\ds \frac{\Lambda(t)}{\Lambda(0)}=
\frac{1}{1-\tau/2}, \\\\
\ds \tau = t \ \Lambda(0),
\end{array}
\eea
while $\Lambda$ monotonically rises and goes to infinity in a finite time,
being $\tau$ the dimensionless time. When $\nu=\chi=0$, the energy cascade provides that both the   microscales decrease until to $\tau \rightarrow 2$, where both the correlations are considered to be fully developed, $\lambda_T \rightarrow$0, $\lambda_\theta \rightarrow$0 and $\Lambda \rightarrow \infty$ (see solid lines of Fig. \ref{figura_3} ).
Thus, the two correlations will exhibit developed shapes in finite times whose values depend on
the initial condition $\Lambda(0)$. The meaning that both the microscales are decreasing functions of $\tau$ is that kinetic and thermal energies are continuously transferred from large to small scales following the previous scheme. 
Next, as $\tau \rightarrow$2, $\Lambda \rightarrow +\infty$ and this means that the velocity gradient diverges in a finite time depending on $\Lambda(0)$ and that contiguous particles trajectories diverge with a growth rate infinitely faster than velocity and temperature fields.

For $\nu>$0, $\chi>$0, then $du/dt<$0 and $d\theta/dt<$0 in any case, and $f$ and $f_\theta$ are here supposed to be fully developed as soon as $d \lambda_T/dt$=0 and $d\lambda_\theta/dt$=0, respectively. These situations are qualitatively shown in the figure by the dashed lines for different values of $R_T$ and $Pe$, where $R_T$=$\lambda_T u/\nu$, $Pe$=$Pr$ $R_T$ and $Pr$=$\nu/\chi$ are, respectively, Reynolds number and P\'eclet number, both referred to the Taylor microscale, and the Prandtl number.
When the initial microscales are relatively large, the diffusivities effects are
quite smaller than the convective terms, the energy cascade is initially stronger than the diffusivities effects and both the microscales exhibit about the same trend just discussed for $\nu$=$\chi$=0. According to Eqs. (\ref{u2dot_thetadot}) and (\ref{lambda_dot}), the interval where $\tau$ ranges can be splitted in two subregions for both $f$ and $f_\theta$. 
The first ones correspond to values of $\tau \in (0,2)$ such that $d \lambda_T/dt<$0 and $d\lambda_\theta/dt<$0, which are upper bounded by the endpoints $\tau_1<$2, $\tau_2<$2 where $d\lambda_T/dt(\tau_1)$= 0 and $d\lambda_\theta/dt(\tau_2)$ =0 (dashed lines), respectively, being in general $\tau_1$ $\ne$ $\tau_2$. There, kinetic and thermal of energy cascade are momentarily balanced by viscosity and thermal diffusivity, respectively, and both the autocorrelations can be considered to be fully developed. For both the correlations, such momentary balance happens in finite times $\tau<$2 which depend on the initial condition.
As far as $\Lambda$ is concerned, this initially coincides about with that obtained for $\nu=$0, then reaches its maximum for $\tau \lesssim$2 and thereafter diminishes due to viscosity. 
When $\Lambda$ achieves its maximum, $d\Lambda/dt$=0, chaos and mixing reach their maximum levels, the correlations are about fully developed, thus relative kinematics and fluid strain change much more rapidly than velocity field. Thereafter, we observe regions where $d\Lambda/dt<$0. There, due to the relatively smaller values of the microscales, the dissipation is stronger than the energy cascade, and both the correlation lengths tend to rise according to Eq. (\ref{lambda_dot}). Such region, which occurs immediately after the condition $d\Lambda/dt=$0, corresponds to the regime of  decaying turbulence.

Observe that the proposed closures (\ref{K}) are expected to be verified where $d\Lambda/dt>$0, in which the Navier-–Stokes bifurcations generate the regime of fully developed turbulence. On the contrary, in regime of decaying turbulence --$d\Lambda/dt<$0--, after a certain time, say $\tau^+>\tau_1 \approx2$, it results $\Lambda/\Lambda(0)<1$. In such situations, the relative kinematic and fluid strain could be not faster than velocity field, thus the statistical independence hypothesis (\ref{indep stat}) could be not satisfied and Eqs. (\ref{K}) will be not defined. Therefore, the condition $\tau \approx 2$ or $\Lambda/\Lambda(0)<1$ provides a further limit of validity for the proposed closure formulas.

\bigskip

\section{$^*$Self--similarity and developed correlations of the proposed closures}

This section analyzes self--similarity and developed shape of $f$ and $f_\theta$ produced by the proposed closures. 
{ %\color{blue} 
The new result with respect to the previous works consists in to remark that the proposed closures generate correlations self--similarity in proper ranges of $r$, which is directly related to the fluid trajectories divergence.}
To study this question, observe that a given function of $t$ and $r$, say $\psi=\psi(t,r)$, which completely exhibits self--similarity with respect to $r$ as $t$ changes, is a function of the kind  
\bea
\begin{array}{l@{\hspace{+0.0cm}}l}
\ds \psi(t, r) = \psi\left( \frac{r}{\hat{L}(t)}\right)
\end{array}
\eea
and exactly satisfies the equation 
\bea
\begin{array}{l@{\hspace{+0.0cm}}l}
\ds \frac{\partial \psi}{\partial t} = -\frac{\partial \psi}{\partial r} \frac{r}{\hat{L}} \frac{d\hat{L}}{dt} \equiv C(t) r \frac{\partial \psi}{\partial r}, \\\\
\ds C(t)= \frac{d \ln \hat{L}}{dt} 
\end{array}
\label{similar eq}
\eea
wherein $\hat{L}$ is the characteristic length associated with the specific problem.
From such equation, the self--similarity of $\psi$ is linked to the variation rate 
$d \ln \hat{L}(t)/dt$.
Now, thanks to the mathematical structures of the proposed closures (\ref{K}), 
and taking into account that $f$ and $f_\theta$ are both even functions of $r$ which near the origin behave like Eqs. (\ref{f f_theta}), $K$ and $G$ can be expressed through even power series of $f$ as follows
\bea
\begin{array}{l@{\hspace{+0.0cm}}l}
\ds K = u^3 \sqrt{\frac{1-f}{2}} \frac{\partial f}{\partial r} 
= \frac{u^3}{2} \frac{r}{\lambda_T} \frac{\partial f}{\partial r}+ ... 
= \frac{u^2}{2} \Lambda r \frac{\partial f}{\partial r}+ ... \\\\
%\equiv C_u(t) r \frac{\partial f}{\partial r}+ ..., \\\\
\ds G = \theta^2 u \sqrt{\frac{1-f}{2}} \frac{\partial f_\theta}{\partial r} 
= \frac{\theta^2 u}{2} \frac{r}{\lambda_T} \frac{\partial f_\theta}{\partial r}+ ... 
=\frac{\theta^2}{2} \Lambda r \frac{\partial f_\theta}{\partial r}+ ..., \\\\
\ds \Lambda \propto \frac{d}{dt} \left\langle \ln \varrho \right\rangle_\xi
\approx  \vert \frac{d \ln \lambda_T}{d t} \vert 
\end{array}
\label{expansion}
\eea
thus, the evolution equations of both the autocorrelations can be written in the following way
\bea
\begin{array}{l@{\hspace{+0.0cm}}l}
\ds \frac{\partial f}{\partial t} = u \sqrt{\frac{1-f}{2}} \frac{\partial f}{\partial r} + ...
= \frac{u}{2 \lambda_T} r \frac{\partial f}{\partial r}+ ... 
= \frac{\Lambda}{2} r \frac{\partial f}{\partial r}+ ...\\\\
 \ds \frac{\partial f_\theta}{\partial t} = u \sqrt{\frac{1-f}{2}} \frac{\partial 
f_\theta}{\partial r} + ...
= \frac{u}{2 \lambda_T} r \frac{\partial f_\theta}{\partial r}+ ...
= \frac{\Lambda}{2} r \frac{\partial f_\theta}{\partial r}+ ..., \\\\
\ds \Lambda \propto \frac{d}{dt} \left\langle \ln \varrho \right\rangle_\xi
\approx  \vert \frac{d \ln \lambda_T}{d t} \vert 
\end{array}
\label{ff}
\eea 
Comparing Eqs. (\ref{ff}) and (\ref{similar eq}), it follows that the proposed closures
(\ref{K}) generate self--similarity in a range of variation of $r$ where  
$\Lambda/2 r \partial f/\partial r$ and $\Lambda/2 r \partial f_\theta/\partial r$ are dominant with respect to the other terms. As the result, such self--similarity is directly caused by the continuous fluid trajectory divergence --quantified by $\Lambda$-- which happens thank to very frequent kinematic bifurcations.
In such these intervals, the correlations will exhibit self--similarity during their time evolution, thus $f$ and $f_\theta$ can be expressed there as follows
\bea
\begin{array}{l@{\hspace{+0.0cm}}l}
\ds f(t, r) \simeq f\left( \frac{r}{\lambda_T(t)}\right),  
\\\\ 
\ds f_\theta(t, r) \simeq f_\theta \left( \frac{r}{\lambda_T(t)}\right), 
\end{array}
\label{self sim}
\eea
In such regions, the energy cascade is intensive and much stronger than the diffusivities  effects, thus following Eq. (\ref{lambda_dot}), $\lambda_\theta(t)$ is expected to be proportional to $\lambda_T(t)$
\bea
\begin{array}{l@{\hspace{+0.0cm}}l}
\ds \frac{\lambda_\theta(t)}{\lambda_\theta(0)} \simeq\frac{\lambda_T(t)}{\lambda_T(0)},
\end{array}
\label{fdt l}
\eea 
Next, as $\vartheta$ is a passive scalar, energy cascade and fluid diffusivities act on $u$ and $\theta$ in such a way that their increments are proportional with each other. Therefore, far from the initial condition, we expect that
\bea
\begin{array}{l@{\hspace{+0.0cm}}l}
\ds \frac{\theta(t)}{\theta(0)} \simeq \frac{u(t)}{u(0)}, 
\end{array}
\label{fdt}
\eea 
Now, Eq. (\ref{fdt l})  provides a link between the correlation scales and $Pr$. In fact,
substituting Eq. (\ref{fdt l}) in Eq. (\ref{u2dot_thetadot}), we obtain
\bea
\begin{array}{l@{\hspace{+0.0cm}}l}
\ds \frac{\lambda_\theta}{\lambda_T} = \sqrt{\frac{6}{5}\frac{1}{Pr}}
\end{array}
\label{fdt ll}
\eea
Furthermore, from Eqs. (\ref{lambda_dot}), also $f^{IV}(0)$ and $f^{IV}_\theta(0)$
are related to the Prandtl number 
\bea
\begin{array}{l@{\hspace{+0.0cm}}l}
\ds \frac{f^{IV}_\theta(0)}{f^{IV}(0)} = \frac{7}{3} Pr^2
\end{array}
\label{fdt f4}
\eea
Hence, the developed autocorrelations can be estimated searching for the solutions of the 
closed von K\'arm\'an--Howarth and Corrsin equations in the self--similar form (\ref{self sim}) 
when $d \lambda_T/dt$=$d\lambda_\theta/dt$=0. This leads to the following ordinary 
differential equations system
\bea
\begin{array}{l@{\hspace{+0.0cm}}l}
\ds \sqrt{\frac{1-f}{2}} \frac{d f}{d \hat{r}} +
\ds \frac{2}{R_T} \left( \frac{d^2f}{d\hat{r}^2} +\frac{4}{\hat{r}} \frac{d f}{d\hat{r}} \right) +
\ds  \frac{10}{R_T} f=0, \\\\
\ds \sqrt{\frac{1-f}{2}} \frac{d f_\theta}{d \hat{r}} +
\ds \frac{2}{R_T Pr} \left( \frac{d^2f_\theta}{d\hat{r}^2} +\frac{2}{\hat{r}} \frac{d f_\theta}{d\hat{r}} \right) +
\ds  \frac{12}{R_T Pr} \left( \frac{\lambda_T}{\lambda_\theta}\right)^2 f_\theta=0, \\\\
\ds \hat{r}=\frac{r}{\lambda_T}.
\end{array}
\label{sseq}
\eea
Several solutions of these equations were numerically obtained in \cite{deDivitiis_2} and \cite{deDivitiis_4}, where the author shows that velocity and temperature correlations agree with the Kolmogorov law, with the theoretical arguments of Obukhov-–Corrsin and Batchelor and with the numerical simulations and experiments known from the literature \cite{Corrsin_1, Batchelor_2, Batchelor_3, Obukhov, Mydlarski, Rogallo, Donzis, Gibson}. 
\begin{figure}[t]
\centering
\vspace{-0.mm}
\hspace{-0.0mm}
\includegraphics[width=6.5cm, height=5.5cm]{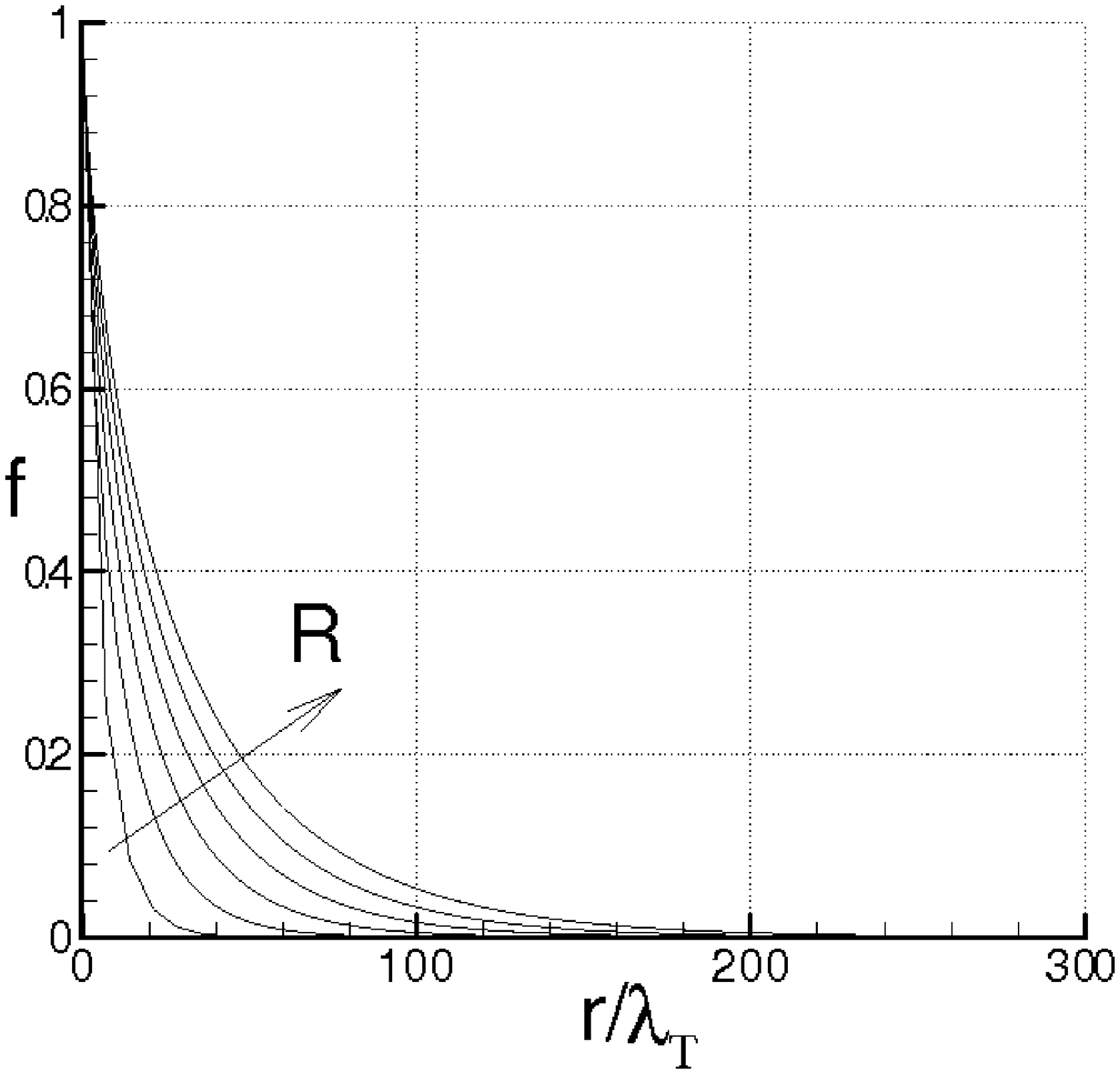} 
\includegraphics[width=6.5cm, height=5.5cm]{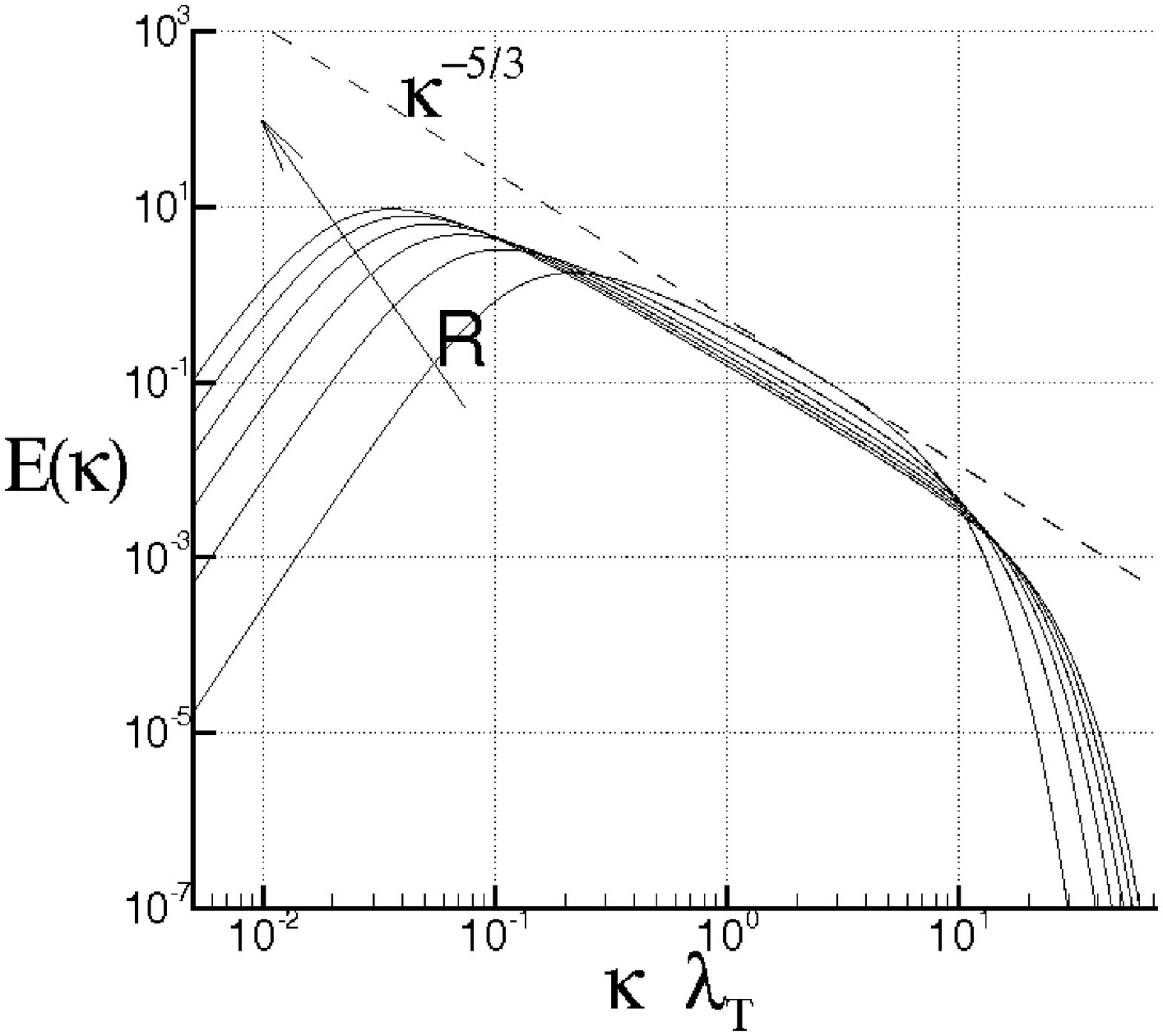}
\caption{%\color{blue} 
Longitudinal velocity correlations (left) and energy spectra (right) at different Taylor scale Reynolds numbers $R_T$=100, 200, 300, 400, 500, 600.
}
\label{figura_r1}
\end{figure}
\begin{figure}[h]
\centering
\vspace{-0.mm}
\hspace{-0.0mm}
\includegraphics[width=6.50cm, height=5.50cm]{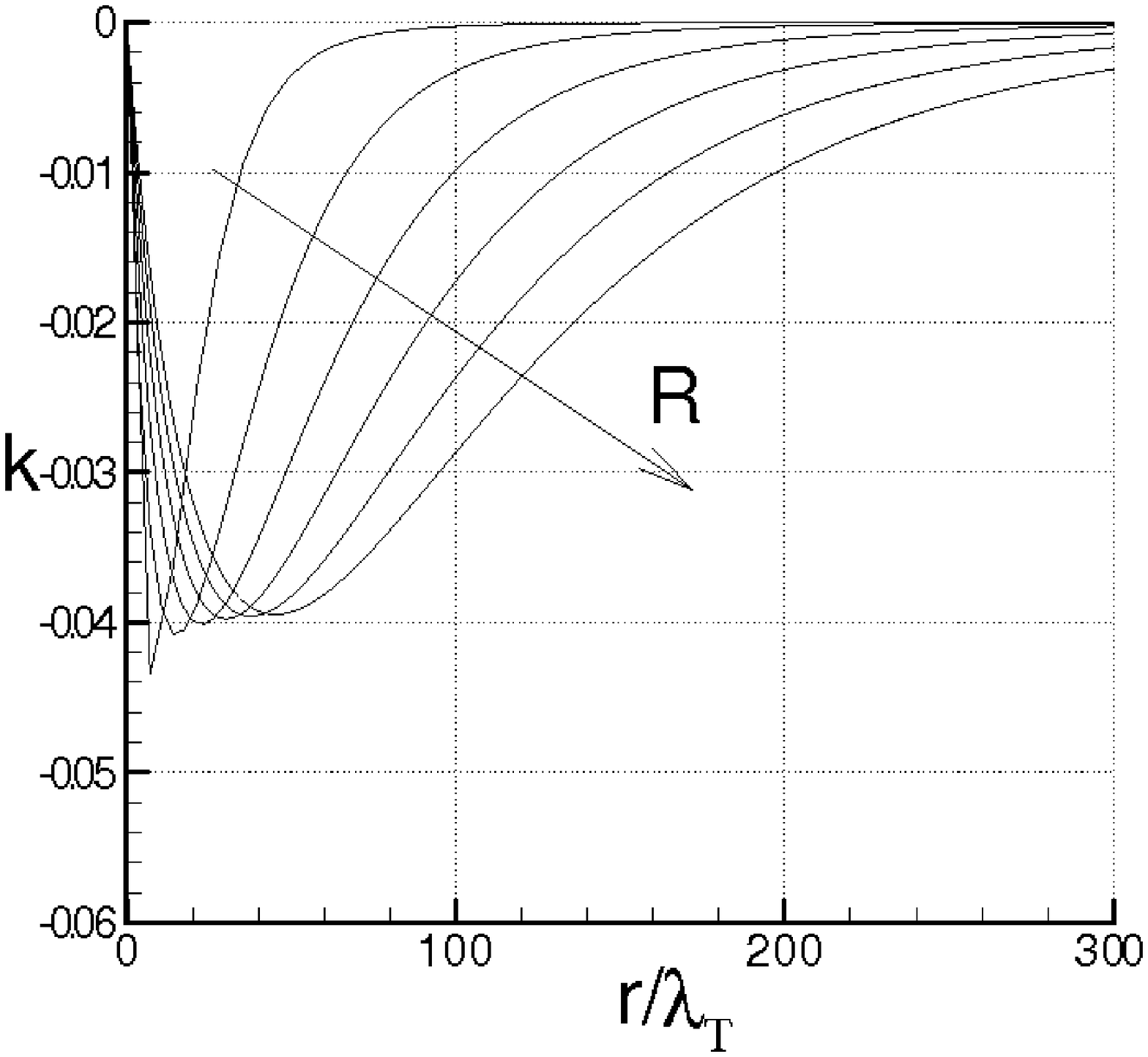} 
\includegraphics[width=6.50cm, height=5.50cm]{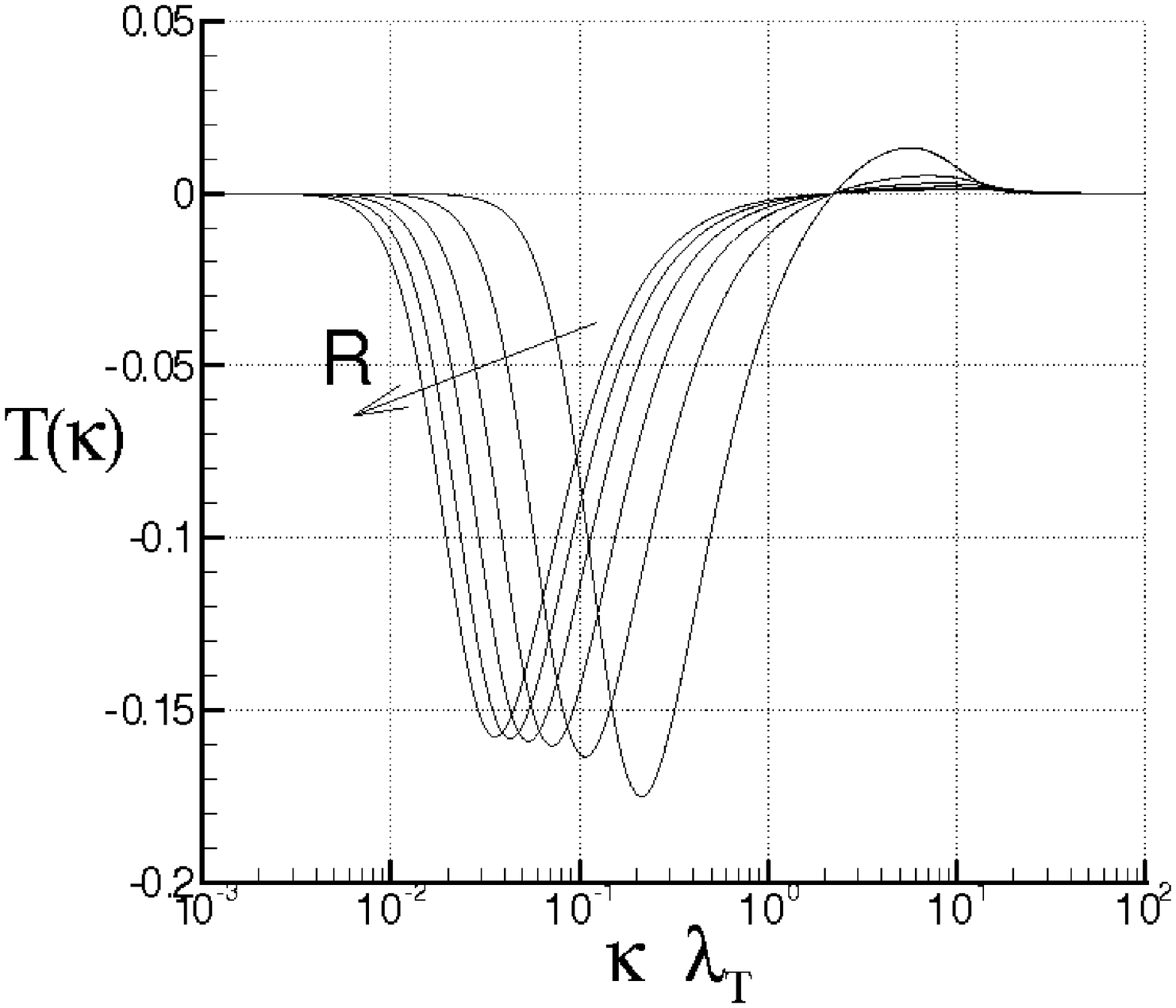}
\caption{%\color{blue} 
Triple longitudinal velocity correlations (left) and the corresponding  spectra (right) at different Taylor scale Reynolds numbers $R_T$=100, 200, 300, 400, 500, 600.
}
\label{figura_r2}
\end{figure}
{%\color{blue} 
For sake of reader convenience, Figs. \ref{figura_r1} and \ref{figura_r2} report the velocity correlations and the corresponding spectra $E(\kappa)$, $T(\kappa)$ numerically calculated with the first equation of Eq. (\ref{sseq}) for $R_T$=100, 200, 300, 400, 500, 600, being
\bea
\left[\begin{array}{c}
\ds E(\kappa) \\\\
\ds T(\kappa)
\end{array}\right]  
= 
 \frac{1}{\pi} 
 \int_0^{\infty} 
\left[\begin{array}{c}
 \ds  u^2 f(r) \\\\
 \ds K(r)
\end{array}\right]  \kappa^2 r^2 
\left( \frac{\sin \kappa r }{\kappa r} - \cos \kappa r  \right) d r 
\label{Ek}
\eea
where all these cases correspond to the same level of average kinetic energy.
The integral correlation scale of $f$ results to be a rising function of $R_T$, 
while the triple longitudinal velocity correlation $k$ maintains negative with a minimum of about -0.04 whose value is achieved for values of $r/\lambda_T$ which rise with the Reynolds number. For what concerns the spectra, observe that increasing $\kappa$, the kinetic energy spectra behave like $E(\kappa) \approx \kappa^4$ near the origin, then exhibit a maximum and thereafter are about parallel to the dashed line $\kappa^{-5/3}$ in a given interval of the wave--numbers. The size of this latter, which defines the inertial range of Kolmogorov, rises as $R_T$ increases. For higher values of $\kappa$, which correspond to scales less than the Kolmogorov length, $E(\kappa)$ decreases more rapidly than in the inertial range.
As $K$ does not modify the kinetic energy, the proposed closure gives $\int_0^\infty T(\kappa) d \kappa \equiv 0$.
\begin{table}[h]
\centering
  \begin{tabular}{cc} 
\hline
\color{blue}$R_T$ \       &  \ \color{blue}$C$ \\[2pt] 
\hline
\hline \color{blue}
\color{blue}100 \       & \ \color{blue}1.8860      \\
\color{blue}200 \       & \ \color{blue}1.9451      \\
\color{blue}300 \       & \ \color{blue}1.9704      \\
\color{blue}400 \       & \ \color{blue}1.9847      \\
\color{blue}500 \       & \ \color{blue}1.9940    \\
\color{blue}600 \       & \ \color{blue}2.0005    \\
\hline
 \end{tabular}
\caption{%\color{blue}
Kolmogorov constant for different Taylor-Scale Reynolds number.}
\label{table2}
\end{table} 
From these solutions, the Kolmogorov constant $C$, here calculated as
\bea
C = \max_{\kappa \in (0, \infty)} \frac{E(\kappa) \kappa^{5/3}}{\varepsilon^{2/3}} 
\eea
is shown in table \ref{table2} in function of the Reynolds number, where $\ds \varepsilon = - 3/2 \ d u^2/ dt$. The obtained values of $C \approx 2$ are in good agreement with the corresponding values known from the literature.

Next, Fig. \ref{figura_r3} shows the temperature spectra $\Theta(\kappa)$ 
and the temperature transfer function $\Gamma(\kappa)$ calculated as follows 
\cite{Ogura}
\bea
\left[\begin{array}{c}
\ds \Theta(\kappa) \\\\
\ds \Gamma(\kappa)
\end{array}\right]  
= 
 \frac{2}{\pi} 
 \int_0^{\infty} 
\left[\begin{array}{c}
 \ds  \theta^2 f_\theta(r) \\\\
 \ds G(r)
\end{array}\right]  
\kappa r \sin \kappa r \ dr 
\label{Tk}
\eea
in such a way that  
\bea
\int_0^\infty \Theta (\kappa)  \ d\kappa = \theta^2, \ \ \
\int_0^\infty \Gamma (\kappa)  \ d\kappa = 0
\label{Tk0}
\eea
The variations of $\Theta(\kappa)$ with $R_T$ and $Pr$ are quite peculiar and 
consistent with previous studies according to which there are
regions where $\Theta(\kappa)$ exhibits different scaling laws $\Theta(\kappa) \approx \kappa^n$.
\begin{figure}[h]
\centering
\vspace{-0.mm}
\hspace{-50.0mm}
\includegraphics[width=11.50cm, height=9.0cm]{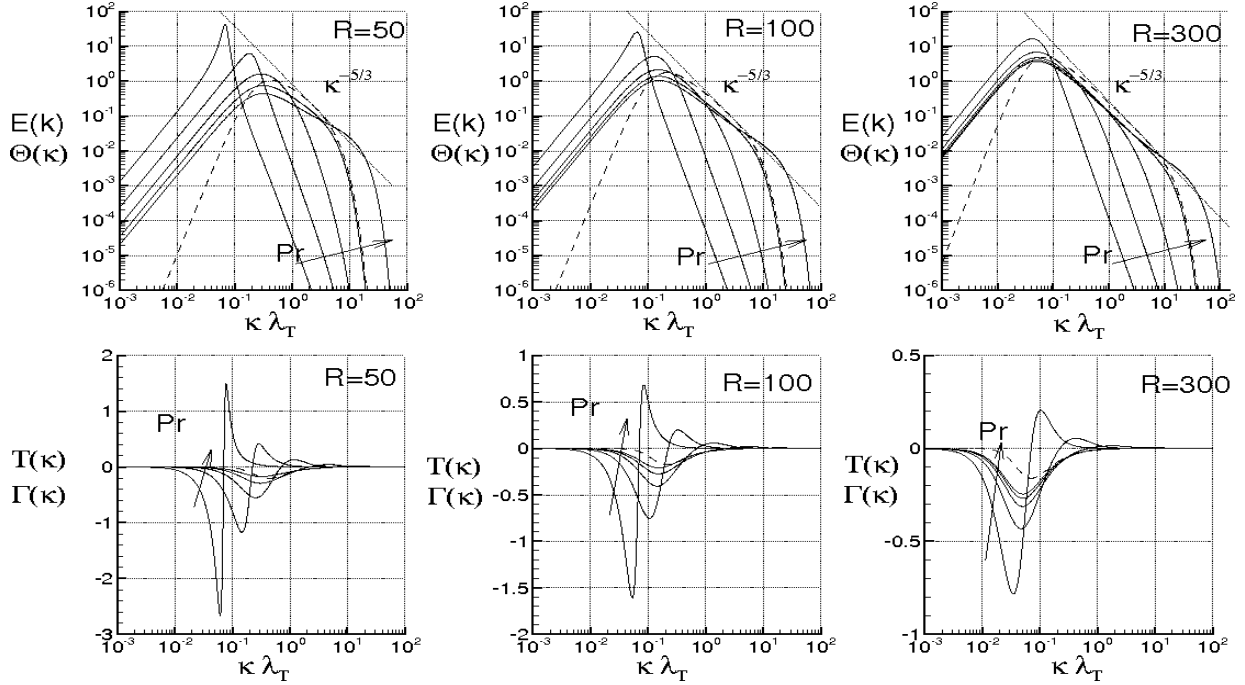}  
\caption{%\color{blue} 
Spectra for Pr= 10$^{-3}$, 10$^{-2}$, 0.1, 1.0 and 10, at different Reynolds numbers. 
Top: kinetic energy spectrum $E(\kappa)$ (dashed line) and temperature spectra 
$\Theta(\kappa)$ (solid lines). Bottom: velocity transfer function $T(\kappa)$ (dashed line) and temperature transfer function $\Gamma(\kappa)$ (solid line).
}
\label{figura_r3}
\end{figure}
Following the proposed closures, $n \simeq$ 2 when $\kappa \rightarrow$ 0 in any case.
For $Pr=$ 0.001, when $R_T$ ranges from 50 to 300, the temperature spectrum essentially exhibits 
two regions: 
one in proximity of the origin where $n \simeq 2$, and the other one, at higher values of $\kappa$, where $-17/3 < n < -11/3$,  (value very close to $-13/3$). 
The value of $n \approx -13/3$, here obtained in an interval around to 
$\hat{r} \approx$1, is in between the exponent proposed by \cite{Batchelor_3} ($-17/3$) and the value determined by  \cite{Rogallo} ($-11/3$) by means of numerical simulations.
Increasing $\kappa$, $n$ significantly diminishes, and $\Theta(\kappa)$ does not show scaling law.
When $Pr=$0.01, an interval near $\hat{r} \approx 1$ where $-17/3 < n < -13/3$ appears, and this
is in agreement with \cite{Batchelor_3}.
Next, for $Pr$ =0.1, the previous scaling law vanishes, whereas for $R_T=$ 50 and 100, $n$ changes with $\kappa$, and $\Theta(\kappa)$ does not show clear scaling laws.
When $R=300$, the birth of a small region is observed, where $n \approx -5/3$ has an inflection point. For $Pr =$ 0.7 and 1, with $R_T =$ 300, the width of this region is increased, whereas
at $Pr$ = 10, and $R =$ 300, we observe two regions:
one interval where $n$ has a local minimum with $n \simeq -5/3$, and the other 
one where $n$ exhibits a relative maximum, with $n \simeq -1$.
For larger $\kappa$, $n$ diminishes and the scaling laws disappear.
The presence of the scaling law $n \simeq -5/3$ agrees with the theoretical
arguments of \cite{Corrsin_2, Obukhov} (see also \cite{Mydlarski, Donzis} and references therein).
Figure \ref{figura_r3} also reports (on the bottom) the spectra 
$\Gamma(\kappa)$ (solid lines) and $T(\kappa)$ (dashed lines) which
describe the energy cascade mechanism. 
}

\bigskip

\section{$^*$Bifurcation analysis of closed von K\'arm\'an--Howarth equation: from fully developed turbulence toward non--chaotic regimes.
}

Starting from non--chaotic regimes, the transition toward the fully developed turbulence happens through intermidiate stages \cite{Ruelle71, Feigenbaum78, Pomeau80, Eckmann81} which correspond to bifurcations where the relative Reynolds numbers show the same order of magnitude.
This section presents a specific bifurcation analysis, which, unlike the classical route toward the chaos \cite{Ruelle71, Feigenbaum78, Pomeau80, Eckmann81}, analyzes the inverse route: the starting condition is represented by the fully developed homogeneous isotropic turbulence, and the route followed is that towards the non--chaotic regime. Such route corresponds to the path 
$B \rightarrow A$ of Figs. \ref{figura_1} (f), (g) and Fig. \ref{figura_2}. Along the line
$B \rightarrow A$, $R_T$ gradually diminishes and the bifurcations of the closed von K\'arm\'an--Howarth equation, properly defined, will be here studied. 
This analysis estimates $R_T^*$ through the closures (\ref{K closure}) and their previously seen properties, where $R_T^*$ defines the minimum value of $R_T$ for which the turbulence maintains fully developed, homogeneous and isotropic. This provides the order of maginitude of $Re$ at the transition, indicating a further limit of the proposed closures.

In order to formulate a bifurcation analysis for the velocity correlation equation,
consider now the various coefficients of the closed von K\'arm\'an--Howarth equation which arise from the even Taylor series expansion of $f(t, r)=\Sum_k f^{(k)}_0 r^k/k!$. Each of such these coefficients corresponds to one of the following equations
\bea
\left\lbrace 
\begin{array}{l@{\hspace{+0.0cm}}l}
\ds  \frac{du}{dt} = -5 \nu \frac{u}{\lambda_T^2}, \\\\
\ds  \frac{d \lambda_T}{dt} = -\frac{u}{2} + \frac{\nu}{\lambda_T}
\left( \frac{7}{3} f^{IV}_0 \lambda_T^4-5\right), \\\\
\ds \frac{d f^{IV}_0}{dt} = ..., \\
\ds ...  \\
\ds \frac{d f^{(n)}_0}{dt} = ..., \\
\ds ... 
\end{array}\right.
\label{dec vkh}
\eea
Such equations can be written by introducing the infinite dimensional state vector 
\bea
\ds {\bf Y} \equiv \left( u, \lambda_T, f_0^{IV},....f_0^{(n)},...\right).
\eea
which represents the state of the longitudinal velocity correlation. 
Therefore, Eqs. (\ref{dec vkh}), formally written as
\bea
\ds \dot{\bf Y} ={\bf F}({\bf Y}, \nu)
\label{kh bif}  
\eea
are equivalent to the closed von K\'arm\'an--Howarth equation. 
Equation (\ref{kh bif}) defines a bifurcation problem where $\nu$ plays the role of control parameter. Thus, this bifurcation analysis studies the variations of $\bf Y$ caused by $\nu$ according to  
\bea
\ds {\bf F}({\bf Y}, \nu)={\bf F}({\bf Y}_0, \nu_0)
\eea
For $\nu > \nu_0$, $\bf Y$ is formally calculated through the implicit functions inversion theorem 
\bea
\ds {\bf Y} = {\bf G}({\bf Y}_0, \nu_0, \nu) \equiv {\bf Y}_0 - \int_{\nu_0}^{\nu} 
\left( \nabla_{\bf Y} {\bf F} \right)^{-1} \frac{\partial {\bf F}}{\partial \nu} \ d \nu
\label{bif Y}
\eea
where $\nabla_{\bf Y} {\bf F}$ is the jacobian $\partial {\bf F}/\partial {\bf y}$. 
A bifurcation of Eq. (\ref{kh bif}) happens when this jacobian is singular, i.e.
\bea
\ds \det \left( \nabla_{\bf Y} {\bf F} \right) =0
\label{bif F}
\eea
If $\nu_0$ is quite small ($R_T$ properly large), the energy cascade is dominant 
with respect to the viscosity effects and $\nabla_{\bf Y} {\bf F}$ is expected to be nonsingular.
Increasing $\nu$, then  $\bf Y$ smoothly varies according to Eq. (\ref{bif Y}), and thereafter the dissipation gradually becomes stronger than the energy cascade until to reach the first bifurcation where condition (\ref{bif F}) occurs. With reference to Fig. \ref{figura_2}, this corresponds to the path $B \rightarrow A$ until to reach $A$. There, a hard loss of stability is expected for the fully developed turbulence toward non--chaotic regimes \cite{Arnold92}. Therefore, $R_T^*$ is calculated as that value of $R_T$ at bifurcation which gives the maximum of the largest real part of the eigenvalues of $\nabla_{\bf Y}{\bf F}$ \cite{Arnold13, Arnold92} compatible with the current value of the average kinetic energy $u^2$, i.e.
\bea
\begin{array}{l@{\hspace{+0.0cm}}l}
\ds R_T^*  \  \ \vert  \ \ \sup_k\left\lbrace\Re(l_k) \right\rbrace=\mbox{max}, \\\\
\ds \det \left( \nabla_{\bf Y} {\bf F} \right) =0, \\\\
\ds  u^2=\mbox{given}
\end{array}
\eea
where $l_k$, $k$=1, 2,... are the eigenvalues of $\nabla_{\bf Y}{\bf F}$.

On the other hand, as previously seen, far from the initial condition, the energy cascade acts keeping $f$ similar in the time in a given interval of variation of $r$. There, the evolution of $f$ is expected to be described --at least in first approximation-- by Eq. (\ref{self sim}) and this suggests that --under such approximation-- the knowledge of $u$ and $\lambda_T$ can be considered to be sufficient to describe the evolution of $f$. Hence, only the first two components of the state vector $\bf Y$ are taken which correspond to the coefficients of the order of $r^0$ and $r^2$ of Eqs. (\ref{dec vkh}). Thus, thank to the self--similarity, the infinite dimensional space where $\bf Y$ lies is replaced by a finite dimensional manifold, and the state vector is reduced to  
\bea
\ds {\bf Y} \equiv \left( u, \lambda_T\right),
\eea
$f_0^{IV}$ plays the role of a parameter which characterizes the
velocity correlation, and  the jacobian $\nabla_{\bf Y} {\bf F}$ reads as  
\bea
\begin{array}{l@{\hspace{+0.0cm}}l}
\nabla_{\bf Y}{\bf F}
=
\left(\begin{array}{cc}
\ds  \frac{\partial \dot{u}}{\partial u} & \ds \frac{\partial \dot{u}}{\partial \lambda_T} \\\\
\ds  \frac{\partial \dot{\lambda_T}}{\partial u} & \ds \frac{\partial \dot{\lambda_T}}{\partial \lambda_T} 
\end{array}\right)
\end{array}
\eea
whose determinant is
\bea
\ds \det\left( \nabla_{\bf Y}{\bf F}\right) = -\frac{5 \nu^2}{\lambda_T^2} 
\left(7 f_0^{IV} \lambda_T^2  + \frac{10}{\lambda_T^2}\right) +
5 \nu \frac{u}{\lambda_T^3} 
\label{det grafF}
\eea
From Eq. (\ref{det grafF}), as long as $\nu>0$ is properly small, $\det\left( \nabla_{\bf Y}{\bf F}\right)>0$. In order that a bifurcation happen, $\det\left( \nabla_{\bf Y}{\bf F}\right)$ must vanish for a certain value of $\nu$ and this implies that $f^{IV}_0 \lambda_T^4>-10/7$. Thus, increasing $\nu$, $\det\left( \nabla_{\bf Y}{\bf F}\right)/\nu$ diminishes and there exists a value of $\nu$ where this jacobian determinant vanishes.
To determine $R_T^*$, $f_0^{IV}$ is eliminated through the bifurcation condition ($\det \left( \nabla_{\bf Y} {\bf F}\right)=0$) and Eq. (\ref{det grafF}), i.e.
\bea
\ds f_0^{IV} = \frac{1}{7 \lambda_T^2 \nu} \left( \frac{u}{\lambda_T}-10\frac{\nu}{\lambda_T^2}\right) 
\eea
Therefore, the singular jacobian is 
\bea
\begin{array}{l@{\hspace{+0.0cm}}l}
\nabla_{\bf Y}{\bf F}
=
\left(\begin{array}{cc}
\ds  -5 \nu/\lambda_T^2 & \ds 10 \nu u/\lambda_T^3  \\\\
\ds  -1/2 & \ds u/\lambda_T
\end{array}\right)
\end{array}
\eea
and admits the following eigenvalues and eigenvectors $l_1$, $l_2$, $\bfy_1$ and $\bfy_2$, respectively 
\bea
\begin{array}{l@{\hspace{+0.0cm}}l}
\ds l_1 =0, \ \ \ \ {\bfy_1} = \left( u, \frac{\lambda_T}{2} \right)  \\\\
\ds l_2 = \frac{u^2}{\nu}\left( \frac{1}{R_T} - \frac{5}{R_T^2} \right) , \ \ \ \ {\bfy_2} = \left(u, R_T \frac{\lambda_T}{10} \right)
\end{array}
\eea
The eigenvalue $l_2 \in \mathbb{R}$ maintains positive for $R_T>5$ and reaches its maximum $l_{2 max}=5 \nu/\lambda_T^2$ for $R_T$=10.
Accordingly, $R_T^*$ is estimated as
\bea
\begin{array}{l@{\hspace{+0.0cm}}l}
R_T^* = 10
\end{array}
\label{RT min}
\eea
which corresponds to $f^{IV}_0$=0.

Another characteristic value of $R_T$ is obtained in the case where both the eigenvalues vanish. This is $R_T$=5 and is expected to represent the onset of the decaying turbulence regime.
In fact, in such situation, it is reasonable that $f$ and $\lambda_T$ are 
\bea 
\begin{array}{l@{\hspace{+0.0cm}}l}
\ds \frac{d \lambda_T}{dt} \simeq0, \\\\
\ds f \simeq \exp \left( -\frac{1}{2} \left( \frac{r}{\lambda_T}\right)^2 \right)  
\end{array}
\eea
Hence, $f^{IV}_0 \lambda_T^4 \simeq$3 and $R_T \simeq$4, in agreement with the previous
estimation.

{\bf Remark}: It is worth remarking that $R_T^*$ provides the minimum of $R_T$ in fully developed isotropic homogeneous turbulence, thus this gives the order of magnitude
of $R_T$ at the transition. Of course, the transition toward the chaos consists in intermediate stages (bifurcations of Navier--Stokes equations) where the turbulence is not developed and the velocity statistics does not exhibit, in general, isotropy and homogeneity.
Hence, the obtained results provide the order of magnitude of $R_T$ at the transition.
On the basis of this analysis, during the transition, $R_T$ ranges as  
\bea
\begin{array}{l@{\hspace{+0.0cm}}l}
\ds 4 \lesssim R_T \lesssim 10
\end{array}
\eea
The obtained value of $R_T^*$=10 is in very good agreement with the bifurcations analysis of the turbulent energy cascade \cite{deDivitiis_3}, where the author shows that, in the transition toward
the developed turbulence, if the bifurcations cascade follows the Feigenbaum scenario \cite{Feigenbaum78, Eckmann81}, the critical Taylor scale Reynolds number is about $10.13$ and occurs after three bifurcations.

We conclude this section by remarking the limits under which $R_T^*$ is estimated. 
Such limits derive from the local self--similarity produced by the closures (\ref{K}) which allow to consider only the first two equations of (\ref{dec vkh}).

\bigskip

\section{Velocity and temperature fluctuations}

The purpose of this section is to obtain, by means of the previous Lyapunov analysis, formal expressions of velocity and temperature fluctuations which will be useful for estimating the statistics of these latter.
For sake of our convenience, Navier--Stokes and thermal energy equations are now written in the following dimensionless divergence form
\bea
\begin{array}{l@{\hspace{-0.cm}}l}
\ds \frac{\partial {\bf u}}{\partial t} =
\mbox{div} \
\hat{\bf T}, \\\\
\ds \frac{\partial \vartheta }{\partial t} =
- \mbox{div} \ \hat{\bf q}  
\label{NST_eq_div}
\end{array}
\ \ \ \ \mbox{in which} \ \ \ \
\begin{array}{l@{\hspace{-0.cm}}l}
\ds \hat{\bf T} = {\bf T} -{\bf u} \otimes {\bf u}, \\\\
\ds \hat{\bf q} = {\bf q} +{\bf u} \vartheta
\end{array}
\eea
where ${\bf T}$ and ${\bf q}$ denote, respectively, dimensionless 
stress tensor and heat flux, according to the Navier-Fourier laws 
\bea
\begin{array}{l@{\hspace{-0.cm}}l}
\ds {\bf T} = -{\bf I} p + {\bf T}_v, \\\\
\ds {\bf T}_v = \frac{1}{Re} \left( \nabla_{\bf x} {\bf u} + 
\nabla_{\bf x} {\bf u}^T \right), \\\\ 
\ds {\bf q} = - \frac{1}{Pe}\nabla_{\bf x} \vartheta 
\label{Tq}
\end{array}
\eea
being $\bf I$ the identity tensor, ${\bf T}_v$ the viscous stress tensor, and the pressure $p$ is given according to Eq. (\ref{pressure}).

In order to obtain the analytical forms of velocity and temperature fluctuations, 
Eqs. (\ref{NST_eq_div}) are first expressed in terms of referential coordinate ${\bf x}_0$
\bea
\begin{array}{l@{\hspace{-0.cm}}l}
\ds \frac{\partial u_i}{\partial t} 
= \left( \frac{\partial \hat{T}_{i j}}{\partial x_{0 k}} \right) 
\left( \frac{\partial x_{0 k}}{\partial x_{j}}\right) 
\equiv \left( \frac{\partial \hat{T}_{i j}}{\partial x_{0 k}} \right) \ G^{-1}_{j k} \exp\left(-\tilde{\Lambda} t \right), \ \ i =1, 2, 3 \\\\
 \ds \frac{\partial \vartheta}{\partial t} 
= -\left( \frac{\partial \hat{q}_j}{\partial x_{0 k}} \right) 
\left( \frac{\partial x_{0 k}}{\partial x_{j}}\right) 
\equiv -\left( \frac{\partial \hat{q}_j}{\partial x_{0 k}} \right) \ G^{-1}_{j k} \exp\left(-\tilde{\Lambda} t \right)
\end{array}
\label{rate1}
\eea
where the repeated index denotes the summation convention. 
The adoption of the referential coordinates allows to factorize of $\partial {\bf u}/\partial t$ and $\partial \vartheta/\partial t$ as product of two statistically uncorrelated matrices: 
one depending on velocity and temperature fields, and the other one representing the local fluid deformation. Velocity and temperature fluctuations are here obtained integrating Eqs. (\ref{rate1}) in the set $(t, a)$. 
Due to the alignment property of the Lyapunov vectors \cite{Ott2002},  $\exp(-\tilde{\Lambda} t)$ rapidly goes to zero as $t \rightarrow \infty$ in any case, whereas $\partial \hat{T}_{i j}/\partial x_{0 k}$ and $\partial \hat{q}_j/\partial x_{0 k}$ are functions of slow growth of $t$. Hence, velocity and temperature fluctuations are formally calculated integrating Eqs. (\ref{rate1}) in the set $(t, \infty)$ where $\partial \hat{T}_{i j}/\partial x_{0 k}$ and $\partial \hat{q}_j/\partial x_{0 k}$ are considered to be constant and equal to the corresponding values at the current time. Such fluctuations are then expressed in function of current velocity and temperature fields according to
\bea
\begin{array}{l@{\hspace{-0.cm}}l}
\ds u_i
=  \left( \frac{\partial \hat{T}_{i j}}{\partial x_{0 k}} \right) \ W_{j k}, \ \ i =1, 2, 3 \\\\
 \ds  \vartheta 
=  -\left( \frac{\partial \hat{q}_j}{\partial x_{0 k}} \right) \ W_{j k}
\end{array}
\label{fluct u t}
\eea
being
\bea
W_{j k} = \int_0^\infty  G^{-1}_{j k} \exp\left(-\tilde{\Lambda} t \right) \ dt 
\eea
where $\vert W_{j k} \vert < \infty$ as $G^{-1}_{j k}$ is represented by slow growth functions of $t$.

{%\color{red} 
It is worth to remark that Eqs. (\ref{fluct u t}) are, in general, rough approximations of velocity and temperature fluctuations. Nevertheless, in fully developed turbulence, $d {\bfx}(t)$ is  considered to be much more rapid than ${\bf u}(t, {\bf x})$, thus Eqs. (\ref{fluct u t}) provide one accurate way to express velocity and temperature in terms of referential coordinates by means of the Lyapunov theory.}

\bigskip

\section{$^*$Statistics of velocity and temperature difference}

In developed turbulence, longitudinal velocity and temperature difference, $\Delta u_r$ = $({\bf u}(t, {\bf x}') -{\bf u}(t, {\bf x})) \cdot {\bf r}/r$ and $\Delta \vartheta$ = $\vartheta(t, {\bf x}') -\vartheta(t, {\bf x})$, ${\bf r} = {\bf x}'-{\bf x}$, play a role of paramount importance as these quantities describe energy cascade, intermittency and are linked to dissipation. This section analyzes the statistics of such quantities in fully developed homogeneous isotropic turbulence through the previously seen kinematic Lyapunov analysis and using a proper statistical decomposition of velocity and temperature. 
In order to determine this statistics, the Navier--Stokes bifurcations effect on $\Delta u_r$ and $\Delta \vartheta$ is first analyzed. To this purpose, $\Delta u_r$ and $\Delta \vartheta$ are expressed in function of current velocity and temperature through Eq. (\ref{fluct u t})
\bea
\begin{array}{l@{\hspace{-0.cm}}l}
\ds \Delta u_r
=  \left( \frac{\partial \hat{T}_{i j}}{\partial x_{0 k}} \right)' \ W_{j k}' -
   \left( \frac{\partial \hat{T}_{i j}}{\partial x_{0 k}} \right)  \ W_{j k} 
\\\\
 \ds  \Delta \vartheta 
=  -\left( \frac{\partial \hat{q}_j}{\partial x_{0 k}} \right)' \ W_{j k}' +
\left( \frac{\partial \hat{q}_j}{\partial x_{0 k}} \right) \ W_{j k}
\end{array}
\label{fluct du dt}
\eea
The several bifurcations happening during the fluid motion determine a continuous 
doubling of ${\bf u}$ in several functions, say ${\hat{\bf v}}_k$, $k$=1, 2,..., in the sense that  
each encountered bifurcation introduces new functions ${\hat{\bf v}}_k$ whose characteristics
are independent of the velocity field at previous time. Then, due to bifurcations, $\bf u$ is 
of the form 
\bea
\begin{array}{l@{\hspace{-0.cm}}l}
\ds {\bf u}(t, {\bf x}) \approx \Sum_k {\hat{\bf v}}_k(t, {\bf x}),
\end{array}
\eea
It is worth remarking that, while ${\bf u}(t, {\bf x})$ is solution of the Navier--Stokes equations, the functions $\hat{\bf v}_k$ are not. Therefore, the functions $\hat{\bf v}_k$ are the result of the mathematical segregation due to bifurcations of a fluid state variable which physically only exist in combination, thus each of them is not directly observable. This implies that $\bf u$ will be distributed, in line with the Liouville theorem, according to a classical definite positive distribution function.
On the contrary, each single function $\hat{\bf v}_k$, representing mathematical segregation of the fluid state, will be distributed following extended distribution functions which can exhibit negative values \cite{Feynman87, Burgin2009, Burgin2010} compatible with conditions linked to the specific problem. These conditions mainly arise from a) the Navier--Stokes equations and from b) the isotropic hypothesis. For what concerns a), in order that pressure and inertia forces can cause sizable variations of velocity autocorrelation, each term $\hat{\bf v}_k \equiv \left( \hat{v}_1, \hat{v}_2, \hat{v}_3 \right)$ will be distributed following highly nonsymmetric extended distribution function, for which
\bea
\begin{array}{l@{\hspace{-0.cm}}l}
\ds \frac{\vert \left\langle \hat{v}_{k i}^3 \right\rangle \vert} 
 {\left\langle \hat{v}_{k i}^2 \right\rangle^{3/2}} >>> 1, \ \ \ i = 1, 2, 3
\end{array}
\label{c_xi_1}
\eea
As for b), due to isotropic hypothesis, $\bf u$ would be distributed following a gaussian PDF \cite{Batchelor53}, thus, according to the Navier--Stokes equations, pressure and inertia forces will not give contribution to the time derivative of the third statistical moment of $\bf u$. Accordingly, the absolute value of odd statistical moments of order n of $\hat{\bf v}_k$ is expected to be very high in comparison with the even statistical moments of order n+1, i.e.
\bea
\begin{array}{l@{\hspace{-0.cm}}l}
\ds \frac{\vert \left\langle \hat{v}_{k i}^n \right\rangle \vert} 
 {\left\langle \hat{v}_{k i}^2 \right\rangle^{n/2}} >>>
\ds \frac{\vert \left\langle \hat{v}_{k i}^{n+1} \right\rangle \vert} 
 {\left\langle \hat{v}_{k i}^2 \right\rangle^{(n+1)/2}}, \ \ \ n = 3, 5, 7,... , 
\ \ \ i = 1, 2, 3.
\end{array}
\label{c_xi_2}
\eea
This suggests that $\Delta{\bf u}$ and $\bf u$ can be expressed, through a specific statistical decomposition \cite{Ventsel}, as linear combination of opportune stochastic variables $\xi_k$ which reproduce the doubling bifurcations effect, and whose extended distribution functions satisfy Eqs. (\ref{c_xi_1}) and (\ref{c_xi_2}). Furthermore, as $\vartheta$ is a passive scalar, its fluctuations are the result of $\bf u$ and of thermal diffusivity, thus also $\vartheta$ is written by means of the same decomposition
\bea
\begin{array}{l@{\hspace{-0.cm}}l}
\ds {\bf u} = \Sum_k {\bf U}_k \xi_k, \\\\
\ds \vartheta  = \Sum_k \Theta_k \xi_k
\end{array}
\label{stat decomp}
\eea
where ${\bf U}_k$ and $\Theta_k$�($k$= 1, 2,... ) are coordinate functions of $t$ and ${\bf x}$, being $\nabla_{\bf x} \cdot {\bf U}_k =0, \ \forall k$, and $\xi_k$ ($k$= 1, 2,... ) are dimensionless independent centered stochastic variables such that
\bea
\begin{array}{l@{\hspace{-0.cm}}l}
\left\langle \xi_k \right\rangle=0, \  \ \ 
\left\langle \xi_i \xi_j \right\rangle=\delta_{i j}, \ \ \
\left\langle \xi_i \xi_j \xi_k \right\rangle= \left\lbrace 
\begin{array}{l@{\hspace{-0.cm}}l}
\ds q \ne 0, \ \forall \ i=j=k  \\\\
\ds 0 \ \ \mbox{else} 
\end{array}\right. 
\end{array}
\label{prop1}
\eea
where $q$, providing the skewness of $\xi_k$ k=1, 2..., satisfies to
\bea
\begin{array}{l@{\hspace{-0.cm}}l}
\vert q \vert >>>1, \ \left\langle \xi_i^2 \right\rangle, \ \left\langle \xi_i^4 \right\rangle, \ \  i=1, 2, ...
\end{array}, 
\label{prop2}
\eea
Therefore, the distribution functions of $\xi_k$ can assume negative 
values compatible with  Eqs. (\ref{prop1})--(\ref{prop2}).

Through the decomposition (\ref{stat decomp}), we will show that the negative value of $H^{(3)}_u(r)$ has very important implications for what concerns the statistics of $\Delta u_r$ and $\Delta \vartheta$, with particular reference to the intermittency of these latter which rises as Reynolds number and P\'eclet number increase.
To study this question, consider first the analytical forms of the fluctuations of $u_i$ and $\vartheta$ in terms of $\xi_k$ obtained by substituting Eq. (\ref{stat decomp}) into Eq. (\ref{fluct u t})
\bea
\begin{array}{l@{\hspace{-0.cm}}l}
\ds u_i = \Sum_j \Sum_k A^{(i)}_{j k} \xi_j \xi_k + \frac{1}{R_T} \Sum_k a^{(i)}_k \xi_k, \ i=1, 2, 3 \\\\
\ds \vartheta  = \Sum_j \Sum_k B_{j k} \xi_j \xi_k + \frac{1}{Pe} \Sum_k b_k \xi_k,
\end{array}
\label{fluct u t xi}
\eea
where $\Sum_j \Sum_k A^{(i)}_{j k} \xi_j \xi_k$ and $1/R_T \Sum_k a^{(i)}_k \xi_k$ are the contributions of inertia and pressure forces, and of the fluid viscosity, respectively,
whereas $\Sum_j \Sum_k B_{j k} \xi_j \xi_k$ and $1/Pe \Sum_k b_k \xi_k$ arise from the convective term and fluid conduction.
Because of turbulent isotropy, it is reasonable that $u_i$ and $\vartheta$ are both Gaussian stochastic variables \cite{Batchelor53, Ventsel, Lehmann99}, thus the various terms of Eq. (\ref{fluct u t xi}) satisfy the Lindeberg condition, a very general, necessary, and sufficient condition for satisfying the central limit theorem  \cite{Ventsel, Lehmann99}. Such theorem does not apply to $\Delta u_i$ and $\Delta \vartheta$ as these latter are the difference between two correlated Gaussian variables, thus their PDF are expected to be very different with respect to Gaussian distributions.
To study the statistics of $\Delta u_r$ and $\Delta \vartheta$, the fluctuations of these latter are first expressed in terms of $\xi_k$ 𝜉
\bea
\begin{array}{l@{\hspace{-0.cm}}l}
\ds \Delta u_r({\bf r}) = \Sum_j \Sum_k \Delta A_{j k} \xi_j \xi_k + \frac{1}{R_T} \Sum_k \Delta a_k \xi_k, \\\\
\ds \Delta \vartheta({\bf r})  = \Sum_j \Sum_k \Delta B_{j k} \xi_j \xi_k + \frac{1}{Pe} \Sum_k \Delta b_k \xi_k,
\end{array}
\label{fluct du dt xi}
\eea
being
\bea
\begin{array}{l@{\hspace{-0.cm}}l}
\ds \Delta A_{j k}=\Sum_{i=1}^3\left( A^{(i)}_{j k}({\bf x} +{\bf r})-A^{(i)}_{j k}({\bf x})\right) \frac{r_i}{r} \equiv S_{u j k} + \Omega_{u j k}, \\\\
\ds \Delta a_k= \Sum_{i=1}^3\left( a^{(i)}_k({\bf x} +{\bf r})-a^{(i)}_k({\bf x})\right) \frac{r_i}{r}, \\\\
\ds \Delta B_{j k}= B_{j k}({\bf x} +{\bf r})-B_{j k}({\bf x})
\equiv S_{\theta j k} + \Omega_{\theta j k}, \\\\
\ds  \Delta b_k= b_k({\bf x} +{\bf r})-b_k({\bf x}),
\end{array}
\label{dec0}
\eea
In Eq. (\ref{dec0}), the matrices $\Delta A_{j k}$ and $\Delta B_{j k}$ are decomposed following their symmetric and antisymmetric parts, respectively $S_{u j k}$, $S_{\theta j k}$ and $\Omega_{u j k}$, $\Omega_{\theta j k}$.
These last ones give null contribution in Eqs. (\ref{fluct du dt xi}), whereas the terms 
arising from $S_{u j k}$ and $S_{\theta j k}$ are expressed as 
\bea
\begin{array}{l@{\hspace{-0.cm}}l}
\ds \Sum_{j} \Sum_{k} S_{X j k} \xi_j \xi_k = \Sum_{i} S_{X i i} \xi_i^2  + 
\ds \Sum_{j \ne k} S_{X j k} \xi_j \xi_k, \\\\
X= u, \theta
\end{array}
\label{dec1}
\eea
in which the first term of Eq. (\ref{dec1}) is decomposed in the following manner 
\bea
\begin{array}{l@{\hspace{-0.cm}}l}
\ds \Sum_{i} S_{X i i} \xi_i^2 = 
S_X^+ \left(\eta_X^2 - \Sum_{j\ne k}^{+} \xi_j \xi_k \right)+
\Sum_{i}^+ \left( S_{X i i} - S_X^+\right) \xi_i^2 +
S_X^- \left(\zeta_X^2 - \Sum_{j\ne k}^-  \xi_j \xi_k \right)+
\Sum_{i}^- \left( S_{X i i} - S_X^-\right) \xi_i^2, \\\\
X= u, \theta
\end{array}
\label{dec2}
\eea
being 
\bea
\begin{array}{l@{\hspace{-0.cm}}l}
\ds \eta_X  = \Sum_i^+ \xi_i, \\\\
\ds \zeta_X = \Sum_j^- \xi_j, \\\\ 
\ds S_X^+ = \frac{1}{n_X^+}\Sum_i^+ S_{i i} >0, \\\\
\ds S_X^- = \frac{1}{n_X^-}\Sum_i^- S_{i i}< 0, \\\\
\ds X= u, \theta, 
\end{array}
\label{eta zeta}
\label{xi}
\eea
and
\bea
\begin{array}{l@{\hspace{-0.cm}}l}
\ds \xi_X = -S_X^+ \Sum_{j \ne k}^+ \xi_j \xi_k + 
\Sum_i^+ \left( S_{X i i} - S_X^+\right) \xi_i^2 -
S_X^- \Sum_{j \ne k}^-  \xi_j \xi_k +
\Sum_i^- \left( S_{X i i} - S_X^-\right) \xi_i^2 +
 \Sum_{j \ne k} S_{X j k} \xi_j \xi_k +
\Sum_k \Delta a_{X k} \xi_k \\\\
\ds \equiv \Sum_{i j} M_{X i j} \xi_i \xi_j +
 \Sum_k g_{X k} \xi_k, \\\\
\ds g_{u k} =\frac{\Delta a_k}{R_T}, \ \ g_{\theta k} = \frac{\Delta b_k}{Pe}, \ \ k=1, 2, ... \\\\
\ds X= u, \theta, 
\end{array}
\label{eta zeta}
\label{xib}
\eea
where $\Sum_{}^+$ and $\Sum_{}^-$ denote summations for $(S_{X j j}>0, S_{X k k}>0)$ and $(S_{X j j}\leq0, S_{X k k}\leq0)$, and $n_X^+$ and $n_X^-$ are the corresponding numbers of terms of such summations, {%\color{red} 
whereas $\Sum_{j\ne k}^+$ and $\Sum_{j \ne k}^-$ indicate the sums of addends calculated for $j \ne k$ corresponding to  $S_{X j j}>0$, $S_{X k k}>0$ and  $S_{X j j}<0$, $S_{X k k}<0$, respectively.}
The decomposition (\ref{dec1})--(\ref{dec2}) and the definitions (\ref{xi}) lead to the following expression of velocity and temperature difference fluctuations
\bea
\begin{array}{l@{\hspace{-0.cm}}l}
\Delta u_r = \xi_u + S^+_u \eta_u^2 +S^-_u \zeta_u^2, \\\\
\Delta \vartheta = \xi_\theta + S^+_\theta \eta_\theta^2 +S^-_\theta\zeta_\theta^2,
\end{array}
\eea
Now, we show that $\xi_X$, $\eta_X$ and $\zeta_X$, $X=u, \theta$ tend to uncorrelated gaussian variables. 
In fact, from Eq. (\ref{xi}), $\eta_X$ and $\zeta_X$, $X=u, \theta$ are sums of random terms belonging to two different sets of uncorrelated stochastic variables (i.e. the sets for which $S_{X ii}<0$ and $S_{X ii}>0$), therefore $\eta_X$ and $\zeta_X$, are two uncorrelated stochastic variables such that $\langle \eta_X \rangle$=$\langle \zeta_X\rangle$=0, $X$=$u,\theta$. Furthermore, as $\xi_k$ are statistically independent with each other, the central limit theorem applied to Eq. (\ref{xi}) guarantees that both $\eta_X$ and $\zeta_X$ tend to two uncorrelated centered gaussian random variables.
As for $\xi_X$,  $X$=$u,\theta$, the following should be considered:
due to the analytical structure of Eq. (\ref{xib}), 
each term of $\xi_X$ is a centered variable, thus $\langle \xi_X \rangle$=0. 
Next, in Eq. (\ref{xib}), the following terms $-S^+ \Sum_{j\ne k}^+ \xi_j \xi_k + \Sum_i^+ \left( S_{X i i} - S_X^+\right) \xi_i^2$ and $-S_X^-\Sum_{j \ne k}^- \xi_j \xi_k +\Sum_i^- \left( S_{X i i} - S_X^-\right) \xi_i^2$ are mutually uncorrelated, as each of these is sum of random variables belonging to two different uncorrelated sets. Moreover, $\Sum_{i \ne j} \xi_i \xi_j$ includes several weakly correlated terms, whereas 
$\Sum_k g_{X k} \xi_k$ is the sum of independent variables. On the other hand, due to hypothesis of fully developed chaos, the energy cascade, here represented by Eqs. (\ref{fluct du dt xi}), (\ref{prop1})--(\ref{prop2}), will generate a strong mixing on the several terms of 
Eq. (\ref{fluct du dt xi}), thus a proper variant of the central limit theorem can be applied to $\xi_X$ whose several terms are weakly dependent with each other \cite{Lehmann99}. As the result, 
$\xi_X$, $X$=$u,\theta$ will tend to centered gaussian variables statistically independent of $\eta_X$ and $\zeta_X$.

Hence, the statistics of $\Delta u_r$  and $\Delta \vartheta$ is represented by the following structure functions of the independent centered gaussian stochastic variables $\xi_X$, $\eta_X$ and $\zeta_X$ for which
$\langle \xi_X^2 \rangle$=$\langle \eta_X^2 \rangle$=$\langle\zeta_X^2\rangle$=1.
\bea
\begin{array}{l@{\hspace{-0.cm}}l}
\Delta u_r = L_u \xi_u + S^+_u (\eta_u^2-1) -S^-_u(\zeta_u^2-1), \\\\
\Delta \vartheta = L_\theta \xi_\theta + S^+_\theta (\eta_\theta^2-1) -S^-_\theta(\zeta_\theta^2-1),
\end{array}
\eea
{%\color{red} 
where $L_u$ and $L_\theta$ are now introduced to take into account that $\xi_X$, $\eta_X$ and $\zeta_X$ have standard deviation equal to unity. }Thus 
\bea 
\begin{array}{l@{\hspace{-0.cm}}l}
\ds L_u \xi_u = \Sum_{i j} M_{u i j}  \xi_i \xi_j + \frac{1}{R_T}\Sum_k \Delta a_{u k} \xi_k, \\\\ 
\ds L_\theta \xi_\theta = \Sum_{i j} M_{\theta i j}  \xi_i \xi_j + \frac{1}{Pe}\Sum_k \Delta a_{\theta k} \xi_k,
\end{array}
\label{Lx}
\eea
and $L_X$, $S^-_X$ and $S^+_X$ are parameters depending upon $r$ which have to be determined.
To this regard, it worth remarking that, in regime of fully developed isotropic turbulence in infinite domain, the numbers of parameters necessary to describe the statistics of $\Delta u_r$ and
$\Delta \vartheta$ should be minimum compatible with assigned quantities which define the current state of fluid motion, such as average kinetic energy, temperature standard deviation and correlation functions. On the other hand, the evolution equation of $f$ \cite{Karman38} requires the knowledge of the correlations of the third order $k$ to be solved. Therefore, in fully developed homogeneous isotropic turbulence, the sole knowledge of $f$ and $k$ is here considered to be the necessary and sufficient information for determining the  statistics of $\Delta u_r$. This implies that 
%for small values of $r$, $L_u$ and $L_\theta$ tend to be proportional with each 
%other in such a way that
%\bea
%\ds \lim_{r \rightarrow 0} \left( \frac{L_u}{L_\theta}\right)^2 
%=\frac{u^2}{\theta^2},
%\eea
%and that 
$S_u^+$ is proportional to $S_u^-$ through a proper quantity which does not depend on $r$, i.e.
\bea
\begin{array}{l@{\hspace{-0.cm}}l}
\ds S^+_u(r) = \chi S^-_u(r) \equiv \chi S_u(r)
\end{array}
\eea
where $\chi$ $<$1 is a function of $R_T$ giving the skewness of $\Delta u_r$, which has to be identified. Accordingly, $S_u$ and $L_u$ will be determined in function of $f$ and $k$ as soon as $\chi=\chi(Re)$ is known. For what concerns the temperature difference, observe that, due to turbulence isotropy,
the skewness of $\Delta \vartheta$ should be equal to zero and this gives
\bea
\ds S^+_\theta(r) = S^-_\theta(r) \equiv S_\theta(r)
\eea
Therefore, the structure functions of $\Delta u_r$ and $\Delta \vartheta$ read as
\bea
\begin{array}{l@{\hspace{-0.cm}}l}
\Delta u_r = L_u \xi_u + S_u \left( \chi \left( \eta_u^2-1\right)  -\left( \zeta_u^2-1\right) \right) , \\\\
\Delta \vartheta= L_\theta \xi_\theta + S_\theta \left( \eta_\theta^2 - \zeta_\theta^2\right) ,
%\Delta u_r = \xi_u + S^+_u \eta_u^2 +S^-_u \zeta_u^2, \\\\
%\Delta \vartheta = \xi_\theta + S^+_\theta \eta_\theta^2 +S^-_\theta\zeta_\theta^2,
\end{array}
\label{du dt}
\eea
Furthermore, again following the parameters minimum number, the ratio $\Psi_\theta(r) \equiv S_\theta/L_\theta$ would be proportional to $\Psi_u(r) \equiv S_u/L_u$ through a proper coefficient depending upon the Prandtl number alone, that is
\bea
\begin{array}{l@{\hspace{-0.cm}}l}
\ds \Psi_\theta(r) = \sigma(Pr) \Psi_u(r)
\end{array}
\eea
where $\sigma$ is a function of the Prandtl number which has to be determined.

At this stage of the present analysis, we show that, in fully developed turbulence, 
$L_u$ and $L_\theta$ are, respectively, functions of $R_T$ and $Pe$, 
resulting in $L_u \propto R_T^{-1/2}$ and $L_\theta \propto Pe^{-1/2}$. 
In fact, from Eq. (\ref{Lx}) we obtain
\bea
\begin{array}{l@{\hspace{-0.cm}}l}
\ds L_u^2 = 
\Sum_{i j k l}  M_{u i j}  M_{u k l} \left\langle \xi_i \xi_j \xi_k \xi_l \right\rangle + 
\frac{2}{R_T} \Sum_k M_{u k k} \Delta a_{u k} \left\langle \xi_k^3 \right\rangle +
\frac{1}{R_T^2} \Sum_k \Delta a^2_{u k}, \\\\
\ds L_\theta^2 = 
\Sum_{i j k l}  M_{\theta i j}  M_{\theta k l} \left\langle \xi_i \xi_j \xi_k \xi_l \right\rangle + 
\frac{2}{Pe} \Sum_k M_{\theta k k} \Delta a_{\theta k} \left\langle \xi_k^3 \right\rangle +
\frac{1}{Pe^2} \Sum_k \Delta a^2_{\theta k},
\end{array}
\label{Lx2}
\eea
As $\vert \left\langle \xi_k^3 \right\rangle \vert >>>1, \left\langle \xi_i \xi_j \xi_k \xi_l \right\rangle$, first and third addend of Eq. (\ref{Lx2}) are negligible with respect to second one, thus $L_u$ and $L_\theta$ tend to functions of the kind
\bea
\begin{array}{l@{\hspace{-0.cm}}l}
\ds L_u = \frac{F_u(r)}{\sqrt{R_T}}, \\\\
\ds L_\theta = \frac{F_\theta(r)}{\sqrt{Pe}}.
\end{array}
\eea
where $F_u(r)$ and $F_\theta(r)$ are functions of $r$ which do not directly depend on $R_T$ and $Pe$.
Hence, the dimensionless $\Delta u_r$ and $\Delta \vartheta$, normalized with respect to the corresponding standard deviations, are expressed in function of $R_T$ and $Pe$
\bea
\begin{array}{l@{\hspace{-0.cm}}l}
\ds \frac{\Delta u_r}{\sqrt{ \langle (\Delta u_r)^2 \rangle}} = 
\frac{\xi_u + \Psi_u (\chi(\eta_u^2-1)-(\zeta_u^2-1))}{\sqrt{1+2 \Psi_u^2(1+ \chi^2) }}, 
\ \ \ \  \Psi_u (r) = \frac{S_u(r)}{L_u(r)}=\Phi(r) \sqrt{R_T}, \\\\
\ds \frac{\Delta \vartheta}{\sqrt{ \langle (\Delta \vartheta)^2 \rangle}} = 
\frac{\xi_\theta + \Psi_\theta (\eta_\theta^2-\zeta_\theta^2)}{\sqrt{1+4 \Psi_\theta^2 }}, \ \ \ \
\Psi_\theta(r) = \frac{S_\theta(r)}{L_\theta(r)}=\Phi(r) \sqrt{Pe} 
%\ds \Psi_u (r) = \frac{S_u(r)}{L_u(r)}=\Phi_u(r) \sqrt{R_T}, \\\\
%\ds \Psi_\theta(r) = \frac{S_\theta(r)}{L_\theta(r)}=
%\Phi_\theta(r) \sqrt{Pe}, 
%\ds \Psi_X (r) = \frac{S_X(r)}{L_X(r)}, \ \ \ \ \ds X = u, \theta
\end{array}
\label{struct funs}
\eea
and this identifies $\sigma=\sqrt{Pr}$.
Equations (\ref{struct funs}) provide peculiar structure functions giving the statistics of $\Delta u_r$ and $\Delta \vartheta$. 

Now, if  $\chi=\chi(R_T)$ is considered to be known, $L_u$ and $S_u$ can be expressed in function of  $\langle \Delta u_r^2 \rangle$ and  $\langle \Delta u_r^3 \rangle$, where this latter
is calculated adopting the proposed closure (\ref{K}).
In fact, $L_u$ and $S_u$ are related to $\langle \Delta u_r^2 \rangle$ and  
$\langle \Delta u_r^3 \rangle$ through Eq. (\ref{du dt})
\bea
\begin{array}{l@{\hspace{-0.cm}}l}
\ds \left\langle (\Delta u_r)^3 \right\rangle = 6 u^3 k = 8 S_u^3 (\chi^3-1), \\\\
\ds \left\langle (\Delta u_r)^2 \right\rangle = 2 u^2 (1-f) = L_u^2+ 2 S_u^2 (\chi^2+1),
\end{array}
\eea
thus, $L_u$,  $S_u$ and $\Phi$ are expressed in function of $f(r)$ and $k(r)$ as
\bea
\begin{array}{l@{\hspace{-0.cm}}l}
\ds S_u(r) =  \left(  \frac{3/4}{\chi^3-1}\right)^{1/3} u \ k(r)^{1/3}, \\\\
\ds L_u(r)=\sqrt{2} \ u  \sqrt{1-f(r)-(1+\chi^2) \left( \frac{3/4}{\chi^3-1}\right)^{2/3} \ k(r)^{2/3}}, \\\\
\ds \Phi = \frac{S_u}{L_u} \frac{1}{\sqrt{R_T}}
\end{array}
\label{S_u L_u}
\eea
In the expression of $L_u(r)$ of Eqs. (\ref{S_u L_u}), the argument of the square root must be
greater than zero, and this leads to the following implicit condition for $\chi$ 
\bea
\begin{array}{l@{\hspace{-0.cm}}l}
\ds \frac{1+ \chi^2}{\left( \chi^3-1\right)^{2/3} } \le \frac{1}{2} \left( \frac{56}{3} \right)^{2/3} 
\end{array}
\label{chi}
\eea
where the proposed closure (\ref{K}) is taken into account.
Inequality (\ref{chi}), solved with respect to $\chi$, gives the upper limit for $\chi$
\bea
\begin{array}{l@{\hspace{-0.cm}}l}
\ds \chi \le \chi_\infty = 0.8659...
\end{array}
\eea
As far as the temperature difference is concerned, we have 
\bea
\ds \frac{\left\langle \left( \Delta u_r\right)^2 \right\rangle} {\left\langle \left( \Delta \vartheta \right)^2 \right\rangle} \equiv 
\frac{u^2}{\theta^2} \ \frac{1-f}{1-f_\theta} =
\frac{L_u^2}{L_\theta^2} \ \frac{1+2\Psi_u^2(1+\chi^2)}{1+4\Psi_\theta^2}
\label{phi_theta}
\eea
thus Eq. (\ref{phi_theta}) allows to calculate $L_\theta$ in terms of the other quantities
\bea
\begin{array}{l@{\hspace{-0.cm}}l}
\ds L_\theta= L_u \frac{\theta}{u} \sqrt{\frac{1-f_\theta}{1-f}}
\sqrt{\frac{1+2 \Phi^2 R_T (1+\chi^2)}{1+4 \Phi^2 Pe}}
\end{array}
\label{phi_theta b}
\eea
In Eqs. (\ref{phi_theta b}) and (\ref{S_u L_u}), the function $\chi$=$\chi(R_T)$ has to be identified, and
$\Phi(r)$ depends on the specific shape of $f(r)$, where, due to the constancy of $H^{(3)}_u(0)$, $\Phi(0)$ is assumed to be constant, independent of $R_T$.

The distribution functions of $\Delta u_r$ and $\Delta \vartheta$ are formally calculated through the Frobenius--Perron equation \cite{Nicolis95}, taking into account that $\xi_X$, $\eta_X$ and $\zeta_X$ are independent identically distributed centered gaussian variables such that $\langle \xi_X^2 \rangle$=$\langle \eta_X^2 \rangle$=$\langle \zeta_X^2 \rangle$=1, $X=u, \theta$
\bea
\begin{array}{l@{\hspace{-0.cm}}l}
\ds F_{u}(\Delta u_r') = \int_\xi \int_\eta \int_\zeta P(\xi, \eta, \zeta) \
\delta (\Delta u_r'-\Delta u_r(\xi, \eta, \zeta)) \ d\xi \ d\eta \ d\zeta, \\\\
\ds F_{\theta}(\Delta \vartheta') = \int_\xi \int_\eta \int_\zeta P(\xi, \eta, \zeta) \
\delta (\Delta \vartheta'-\Delta \vartheta(\xi, \eta, \zeta)) \ d\xi \ d\eta \ d\zeta,
\end{array}
\label{Frobenius-Perron}
\eea
where $\delta$ is the Dirac delta, $P(\xi, \eta, \zeta)$ is the 3D gaussian PDF
\bea
\ds P(\xi, \eta, \zeta) = \frac{1}{\sqrt{(2 \pi)^3}} \exp\left(-\frac{\xi^2+\eta^2+\zeta^2}{2}\right),
\label{gaussian}
\eea
and $\Delta u_r(\xi, \eta, \zeta))$ and $\vartheta(\xi, \eta, \zeta))$
are determined by Eqs. (\ref{struct funs}).

\begin{figure}
\centering
\includegraphics[scale=.4, angle=0]{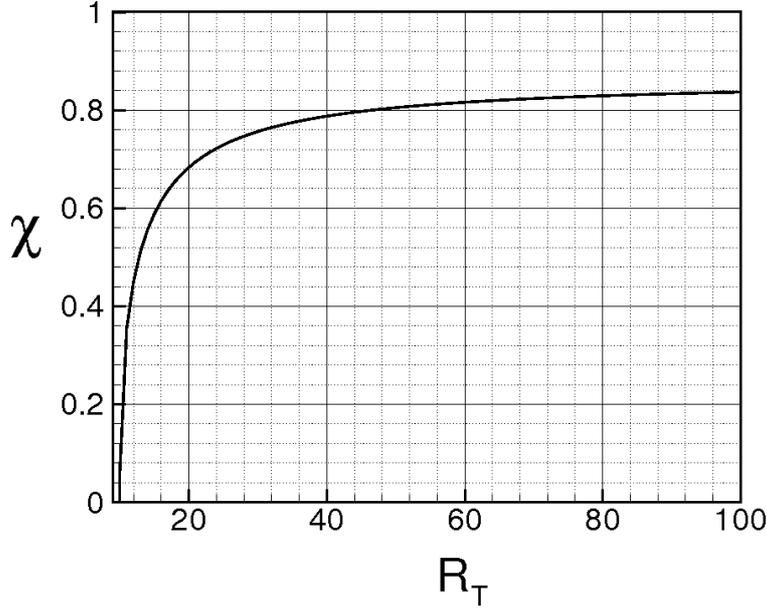}
\caption{Characteristic Function $\chi$=$\chi(R_T)$}
\label{fig_chi_re}
\end{figure}

In other words, the statistics of $\Delta u_r$ and $\Delta \vartheta$ can be inferred looking at the proposed statistical decomposition (\ref{stat decomp}) which includes the bifurcations effects in isotropic turbulence. This is a non--Gaussian statistics, where the absolute value of the dimensionless statistical moments increases with  $R_T$ and $Pe$.
In detail, the dimensionless statistical moments of $\Delta u_r$ and $\Delta \vartheta$ 
are easily calculated in function of $\chi$, $\Psi_u$ and $\Psi_\theta$
\bea
\begin{array}{l@{\hspace{+0.2cm}}l}
\ds H_u^{(n)} \equiv \frac{\left\langle (\Delta u_r )^n \right\rangle}
{\left\langle (\Delta u_r)^2 \right\rangle^{n/2} }
= 
\ds \frac{1} {(1+2(1+\chi^2)  \Psi_u^2)^{n/2}} 
\ds \sum_{k=0}^n 
\left(\begin{array}{c}
n  \\
k
\end{array}\right)  \Psi_u^k
 \langle \xi_u^{n-k} \rangle 
  \langle (\chi (\eta_u^2-1)  - (\zeta_u^2-1) )^k \rangle, \\\\
\ds H_\theta^{(n)} \equiv \frac{\left\langle (\Delta \vartheta )^n \right\rangle}
{\left\langle (\Delta \vartheta)^2 \right\rangle^{n/2} }
= 
\ds \frac{1} {(1+4  \Psi_\theta^2)^{n/2}} 
\ds \sum_{k=0}^n 
\left(\begin{array}{c}
n  \\
k
\end{array}\right)  \Psi_\theta^k
 \langle \xi_\theta^{n-k} \rangle 
  \langle (\eta_\theta^2 - \zeta_\theta^2 )^k \rangle, 
\end{array}
\label{Tm1} 
\eea
where $\Phi(0)$ and $\chi=\chi(R_T)$ have to be identified.
To this end, we first analyze the statistics of $\partial u_r/\partial r$ which, following the proposed Lyapunov analysis, exhibits a constant skewness $H_u^{(3)}(0)$=-3/7.
Then, $H_u^{(3)}(r)$ is first obtained from Eqs. (\ref{Tm1}) 
\bea
\ds H_u^{(3)}(r) = \frac{8 \Psi_u^3(\chi^3-1)}{(1+ 2 \Psi_u^2 (1+\chi^2))^{3/2}} 
\eea
and $H_u^{(3)}(0)$ is calculated for $r \rightarrow 0$
\bea
\ds H_u^{(3)}(0) = \frac{8 \Psi_u^3(0)(\chi^3-1)}{(1+ 2 \Psi_u^2(0) (1+\chi^2))^{3/2}} 
%= - \frac{3}{7}
\label{H30}
\eea
Accordingly, $\chi = \chi(R_T)$ is implicitly expressed in function of $\Phi(0) \sqrt{R_T}$. From Eq. (\ref{H30}), $\chi = \chi(R_T)$ is a monotonic rising function of $R_T$ 
which, for $H_u^{(3)}(0)$=-3/7, admits limit 
\bea
\ds \chi_\infty = \lim_{R_T \rightarrow \infty} \chi (R_T) = 0.8659...
\eea
resulting in $\chi(R_T)<0$ for properly small values of $R_T$.
On the other hand, in fully developed turbulence, the PDF of $\partial u_r/\partial r$ 
exhibits non gaussian behavior (i.e. non gaussian tails) for $\partial u_r/\partial r \rightarrow \pm \infty$, accordingly $\chi$ must be positive.
Hence, the limit condition $\chi=0$ is supposed to be achieved for $R_T=R_T^*$=10 which
represents the minimum value of $R_T$ for which the turbulence is homogeneous isotropic.
This allows to identify $\Phi(0)$  by means of Eq. (\ref{H30})
\bea
\ds \Phi(0) = \frac{1}{\sqrt{R_T^*}} \sqrt{ \frac{{H_{u 0}^{(3)}}^{2/3}}{4-2{H_{u 0}^{(3)}}^{2/3}}}=
0.1409...
\label{phi0}
\eea 
Thus, Eq. (\ref{H30}) gives, in the implicit form, the variation law $\chi=\chi(R_T)$ which is depicted in Fig. \ref{fig_chi_re}.

\bigskip

We conclude this section with the following considerations regarding the proposed analysis, and
summarizing some of the results just obtained in the previous works.

%{\bf Remark.}
For non--isotropic turbulence or in more complex situations with boundary conditions or walls, the velocity will be not distributed following a normal PDF, thus Eq. (\ref{c_xi_2}) will be not verified, and
Eq. (\ref{struct funs}) will change its analytical structure incorporating stronger intermittent terms \cite{Lehmann99} giving the deviation with respect to the isotropic turbulence. Hence, the absolute statistical moments of $\Delta u_r$ will be greater than those calculated through Eq. (\ref{Tm1}), indicating that, in more complex cases than the isotropic turbulence, the intermittency of $\Delta u_r$ can be significantly stronger.

%\bigskip

%{\bf Remark.}
Next, $\Psi_u$ and $\Psi_\theta$ represent the ratios (large scale velocity)-–(small scale velocity) and (large scale temperature)-–(small scale temperature), respectively. In particular,
$
\ds \Psi_u \propto {u}/{u_s} \approx  (u^2/\lambda_T)/(u_s^2/l_s)
$�
being $l_s$ and $u_s$ the characteristic small scale and the corresponding velocity.
This means that �$u/u_s \approx \lambda_T/l_s \approx \sqrt{R_T}$, and that the Reynolds
number relative to $u_s$ and $l_s$ is $u_s l_s/\nu \approx $ 1, that is $l_s$ and $u_s$
identify the Kolmogorov scale and the corresponding velocity.
For what concerns $\Psi_\theta$, $\vartheta$ is a passive scalar, thus $\Psi_\theta$ reads as
$
\ds \Psi_\theta \propto {\theta}/{\theta_s} \approx \theta/\theta_s  (u/\lambda_T)/(u_s/l_s)
$�
and this leads to $u_s l_s/\nu \approx $ 1.

{%\color{blue}
At this stage of the present analysis, we can show that the kinematic bifurcation rate $S_b$, defined by Eq. (\ref{bif rate}), is much larger than the kinematic Lyapunov exponents.
In fact, $S_b$ can be also estimated as the ratio (large scale velocity)--(small scale length),
where large scale velocity and small scale length are given by $u$ and by the Kolmogorov scale, respectively. Taking into account the Kolmogorov scale definition and Eq. (\ref{u2dot_thetadot}), we obtain
\bea
\ds S_b \approx \frac{u}{l_s} = 15^{1/4} R_T^{1/2}  \Lambda 
\eea
confirming the assumption made in the relative section. 
In fully developed turbulence, $S_b >> \Lambda$, and is a rising function of $R_T$.}

%{\bf Remark.}
As show in Ref. \cite{deDivitiis_1}, the statistics given by Eqs. (\ref{Frobenius-Perron}) and (\ref{Tm1}) agrees with the experimental data presented in Refs. \cite{Tabeling96, Tabeling97}. There, in experiments using low temperature helium gas between two counter--rotating cylinders (closed cell), the PDF of $\partial u_r/\partial r$ and its statistical moments are measured. Although the experiments regard wall--bounded flows, the measured PDF of velocity difference are comparable with the present results (Eqs. (\ref{Frobenius-Perron}) and (\ref{Tm1})). Apart from a lightly non--monotonic evolution of $H^{(4)}_u(0)$ and $H^{(6)}_u(0)$ in \cite{Tabeling96, Tabeling97}, the dimensionless statistical moments of 
$\partial u_r/\partial r$ exhibit same trend and same order of magnitude of the corresponding quantities calculated with Eqs. (\ref{Tm1}). In particular, the PDFs of $\partial u_r/\partial r$ obtained with the present analysis show non gaussian tails which coincide with those measured in \cite{Tabeling96, Tabeling97}.
\begin{figure}[h]
\centering
\vspace{-0.mm}
\hspace{-0.0mm}
\includegraphics[width=75.mm, height=11.0cm]{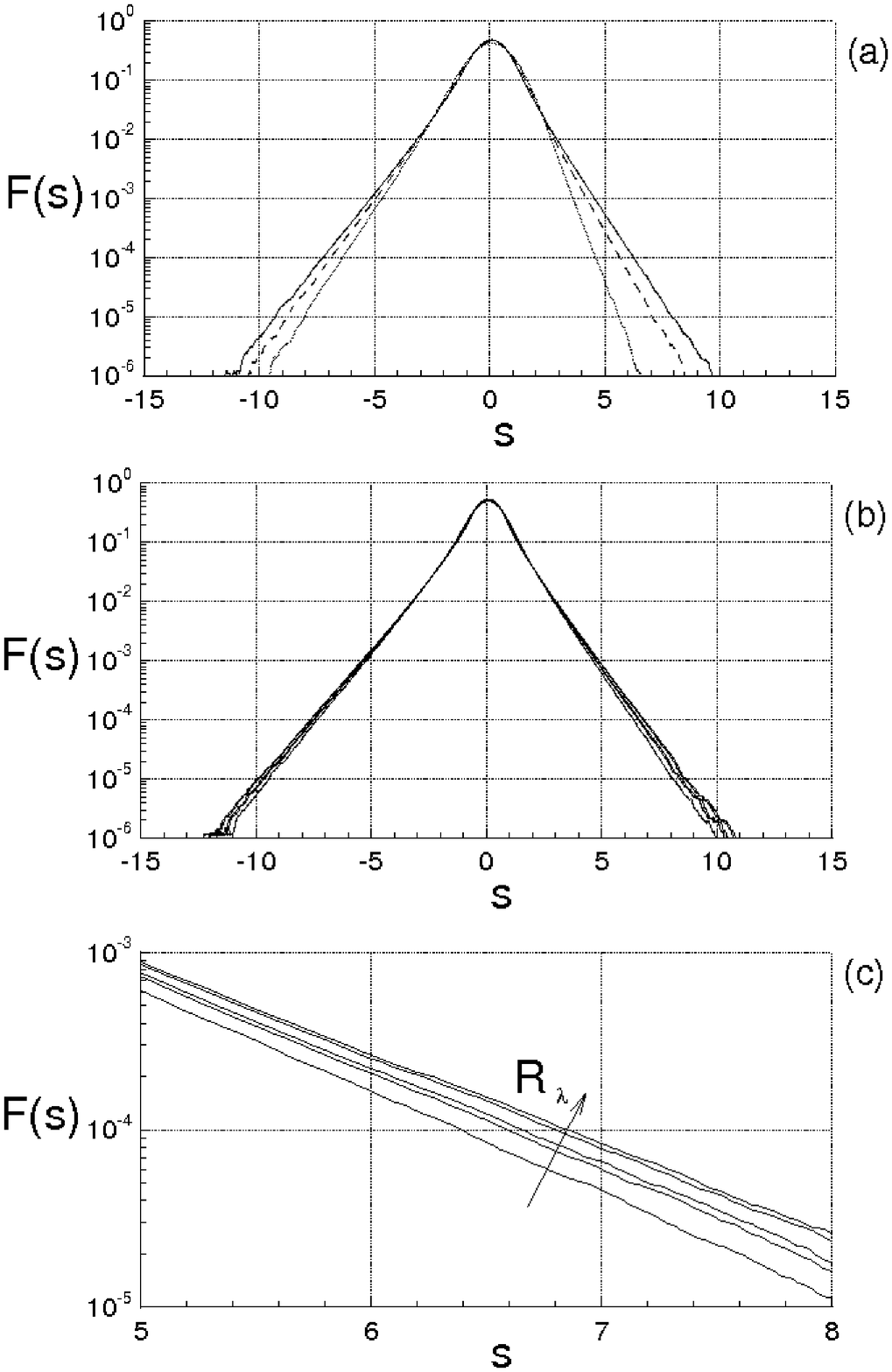} 
%\qquad\qquad
%\includegraphics[width=0.39 \textwidth]{E_k.eps}
\hspace{7.77mm}
\includegraphics[width=70.0mm, height=3.50cm]{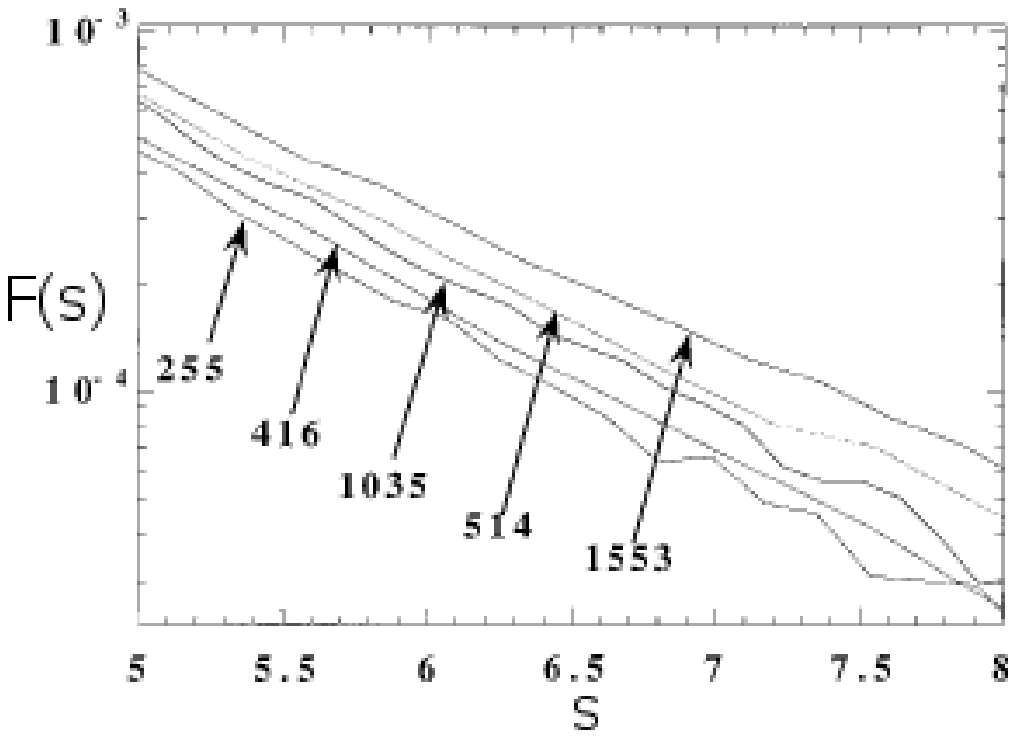}
\caption{%\color{blue} 
Left: PDF of $\partial u_r/\partial r$ for different values of $R_T$. a) Dotted, dash--dotted and continuous lines are for $R_T$ = 15, 30
and 60, respectively. b) and c) PDFs for $R_T$ = 255, 416, 514, 1035 and 1553. c) represents an enlarged part of the diagram (b). Right--bottom: Data from Ref. \cite{Tabeling96}.
}
\label{figura_r5}
\end{figure}
{%\color{blue} 
In Fig. \ref{figura_r5}, the normalized PDFs of $\partial u_r/\partial r$, calculated
with Eqs. (\ref{Frobenius-Perron}) and (\ref{Tm1}), are shown in terms of $s$
\bea
\ds s = \frac{\partial u_r/\partial r}{\sqrt{\left\langle \left( \partial u_r/\partial r\right)^2\right\rangle}}
\eea
in such a way that their standard deviations are equal to the unity.
The results of Fig. \ref{figura_r5}a are performed for $R_T$ = 15, 30 and 60, whereas
Figs. \ref{figura_r5}b and \ref{figura_r5}c report the PDF for $R_T$ = 255, 416, 514, 1035 and 
1553, where Fig. \ref{figura_r5}c represents the enlarged region of Figs. \ref{figura_r5}b, showing the tails of PDF for $5 < s < 8$. 
According to Eqs. (\ref{Frobenius-Perron}) and (\ref{Tm1}), the tails of the PDF rise with the Reynolds number in the interval $10 < R_T < 700$, whereas for $R_T > 700$, smaller variations are observed. On the right--bottom, the results of \cite{Tabeling96} for $R_T$ = 255, 416, 514, 1035 and 1553 are shown. 
Despite the aforementioned non--monotonic trend (see Fig.\ref{figura_r5} (Right--bottom)), 
Fig. \ref{figura_r5}c gives values of the PDFs and of the corresponding average slopes which agree with those obtained in \cite{Tabeling96}, expecially for $5 < s < 8$.
To this regards, it is worth to remark that, in certain conditions, the flow obtained in the experiments of \cite{Tabeling96} could be quite far from the isotropy hypothesis, as such experiments pertain wall--bounded flows, where the walls could significantly influence the fluid velocity in proximity of the probe. 
}
\begin{figure}[h]
\centering
\vspace{-0.mm}
\hspace{-0.0mm}
\includegraphics[width=65.0mm, height=45.0mm]{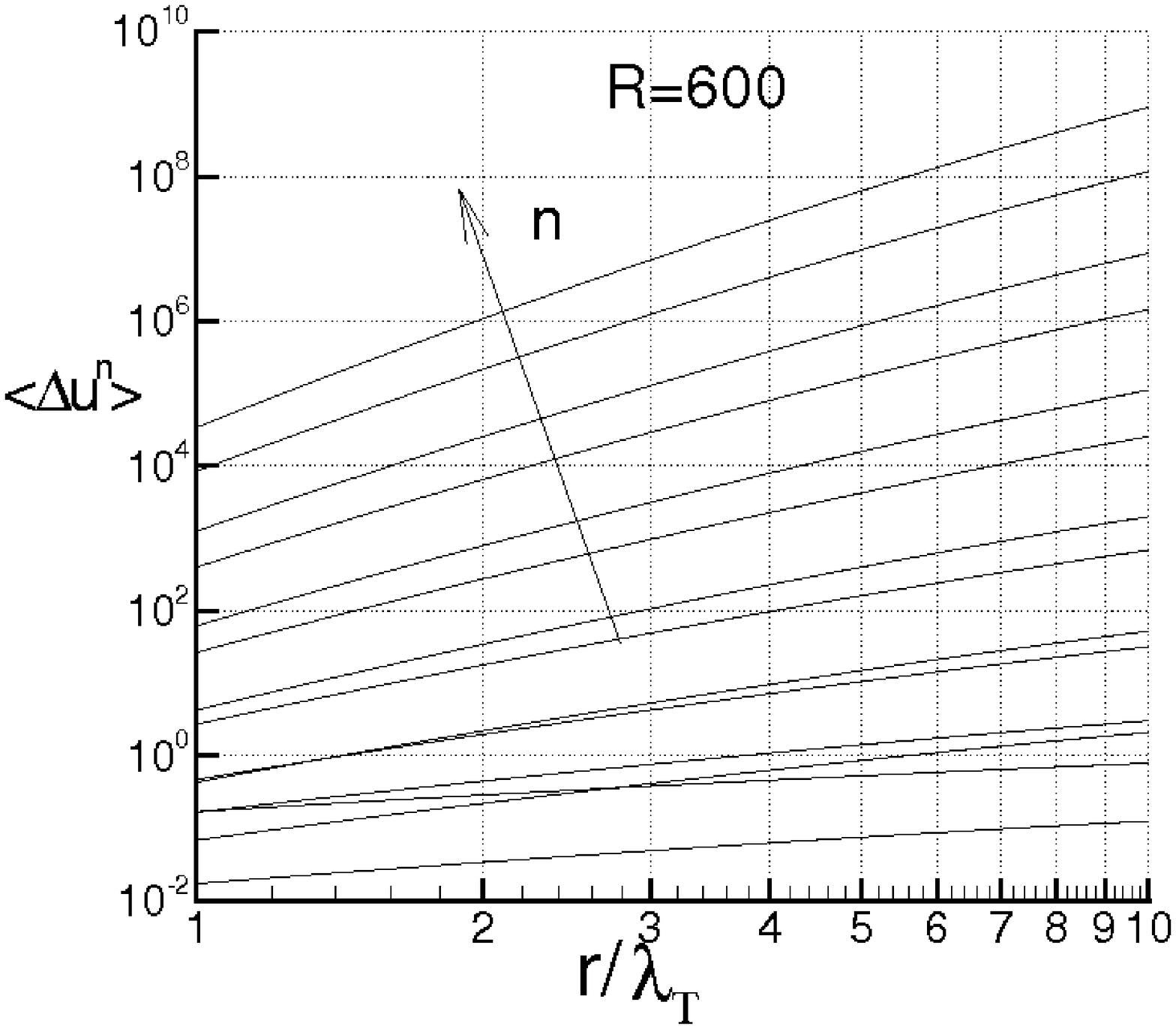} 
%\qquad\qquad
\hspace{7.mm}
\includegraphics[width=65.0mm, height=45.0mm]{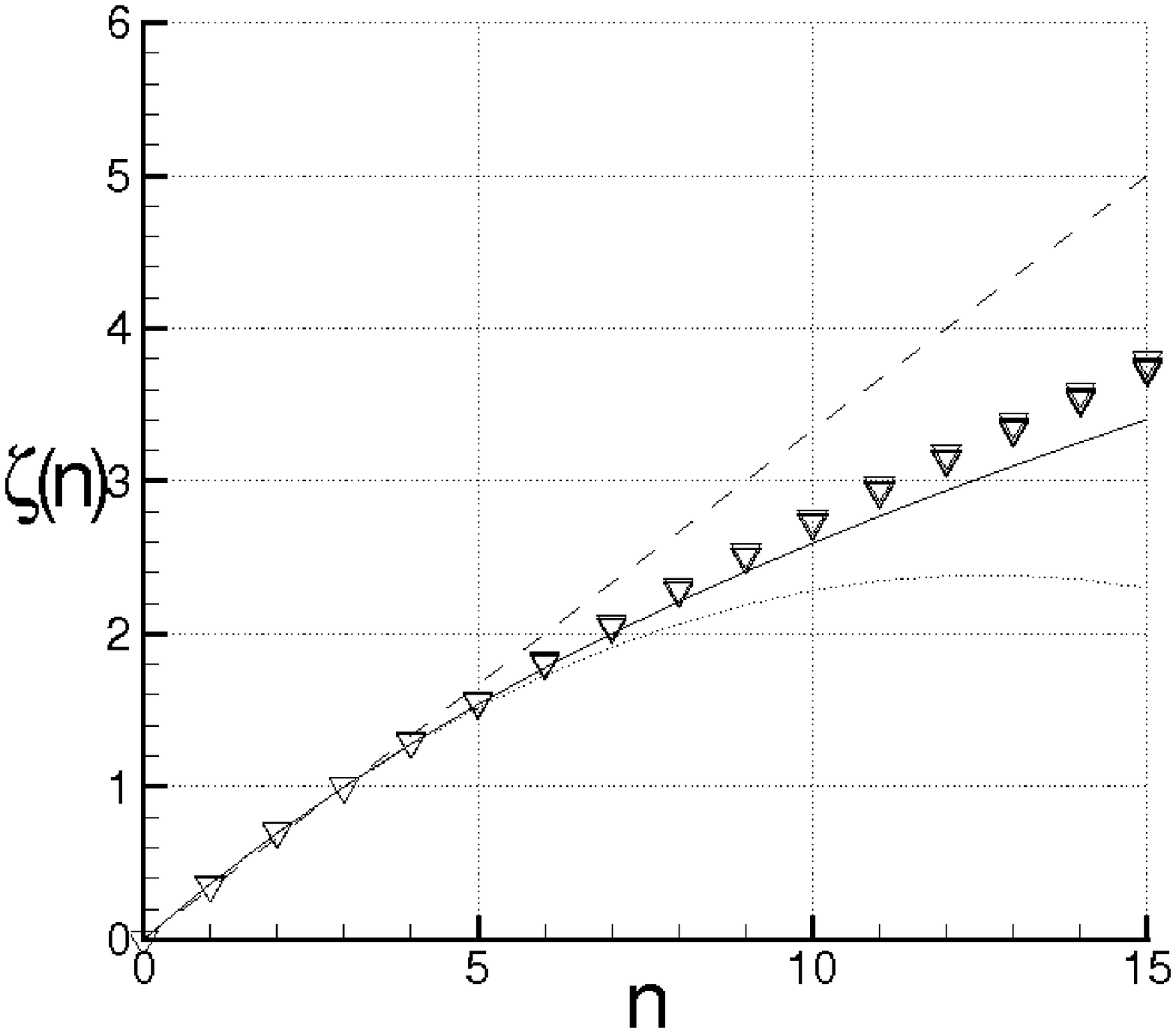}
\caption{%\color{blue} 
Left: Statistical moments of $u_r$ in terms of separation distance, 
for $R_T$=600. Right: Scaling exponents of $\partial u_r/\partial r$ at different $R_T$. 
Solid symbols are for the data calculated with the present analysis. 
Dashed line is for Kolmogorov K41 data \cite{Kolmogorov41}. 
Dotted line is for Kolmogorov K62 data \cite{Kolmogorov62}. 
Continuous line is for She--Leveque data \cite{She94}.
}
\label{figura_r6}
\end{figure}

In Refs. \cite{deDivitiis_1, deDivitiis_2}, and \cite{deDivitiis_4} the scaling exponents $\zeta_V(n)$ associated with the several moments of $\Delta u_r$
\bea
\ds \ \left\langle \left( \Delta u_r \right)^n  \right\rangle \approx A_n r^{\zeta_V(n)}, 
\eea
are calculated with Eq. (\ref{struct funs}) through the following best fitting procedure.
{%\color{blue} 
The statistical moments of $\Delta u_r$ are first calculated in function
of $r$ using Eqs.(\ref{Tm1}) (see Fig. \ref{figura_r6}(Left)). 
Then, the scaling exponents $\zeta_V(n)$ are identified through 
a minimum square method which, for each statistical moment, is applied to the following optimization problem
\bea
\ds J_n(\zeta_V(n), A_n) \hspace{-1.mm} \equiv  
\int_{\hat{r}_1}^{\hat{r}_2} 
\ds ( \langle (\Delta u_r)^n \rangle - A_n r^{\zeta_V(n)} )^2 dr 
 = \mbox{min}, \   n = 1, 2, ...
\eea
where $(\langle (\Delta u_r)^n)\rangle$ are calculated with Eqs. (\ref{Tm1}),
$\hat{r}_1$ is assumed to be equal to 0.1, whereas $\hat{r}_2$ is taken in such a way
that $\zeta_V(3)$ = 1. The so obtained scaling exponents are shown in Fig. \ref{figura_r6} (Right--side) (solid symbols) where these are compared with those given by the Kolmogorov theories K41 \cite{Kolmogorov41} (dashed line) and K62 \cite{Kolmogorov62} (dotted line), and with the exponents calculated by She--Leveque \cite{She94} (continuous curve).
For $n < 4$, $\zeta_V(n) \approx n/3$, and for higher values of $n$, due to nonlinear terms
of Eq. (\ref{struct funs}), $\zeta_V(n)$ shows multiscaling behavior.
The values of $\zeta_V(n)$ here calculated are in good agreement with the She--Leveque data,
and result to be lightly greater than those obtained in \cite{She94} for $n >$ 8.

As far as the temperature difference statistics is concerned, Fig. \ref{figura_r7}(Left) shows the distribution function of $\partial \vartheta/\partial r$ in terms of dimensionless abscissa 
\bea
s = \frac{\partial \vartheta /\partial r}{\sqrt{\left\langle \left( \partial  \vartheta /\partial r\right) ^2 \right\rangle} }
\eea
calculated with Eqs. (\ref{Frobenius-Perron}) and (\ref{struct funs}), 
for different values of $\Psi_\theta$. 
To show the intermittency of such PDF, the flatness $H_\theta^{(4)}$ and the hyperflatness $H_\theta^{(6)}$, defined as
\bea
\ds H_\theta^{(4)} = \frac{\langle s^4 \rangle}{ \langle s^2 \rangle^2}, \ \ \ \
\ds H_\theta^{(6)} = \frac{\langle s^6 \rangle}{ \langle s^2 \rangle^3}
\eea
are plotted in Fig. \ref{figura_r7} (Right) in terms of $\Psi_\theta$.
When $\Psi_\theta=$0, the PDF is gaussian, thus $H_\theta^{(4)}$ = 3 and $H_\theta^{(6)}$ = 15.
Increasing $\Psi_\theta$, the non--linear terms $\eta_\theta$ and $\zeta_\theta$ cause an increment of $H_\theta^{(4)}$ and $H_\theta^{(6)}$, and  when $\Psi_\theta \rightarrow \infty$ $H_\theta^{(4)}\rightarrow$ 9 and $H_\theta^{(6)} \rightarrow$ 225.
\begin{figure}[ht]
	\centering
          \includegraphics[width=60.0mm, height=50.0mm]{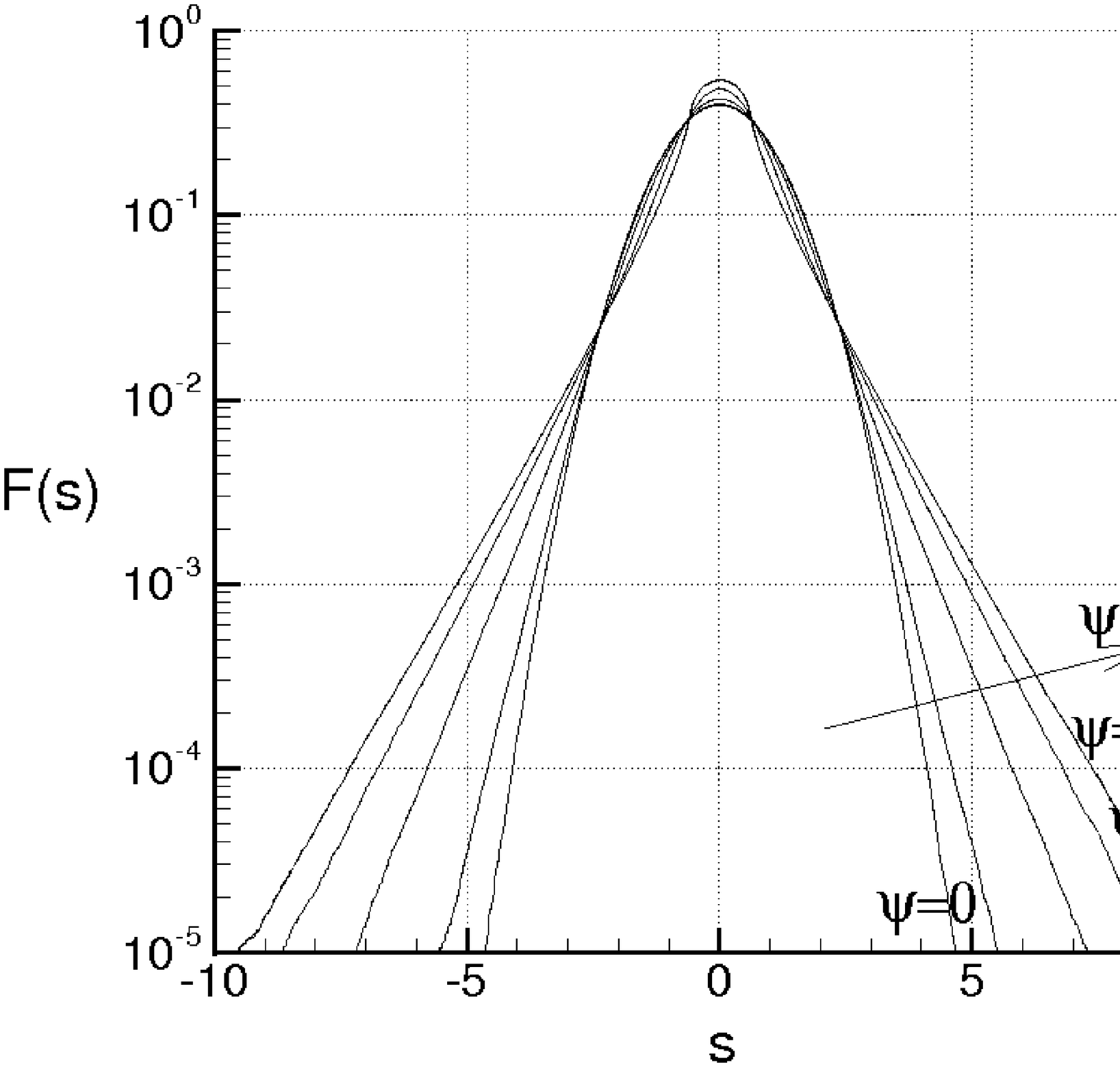}
\hspace{23.mm}
          \includegraphics[width=60.0mm, height=50.0mm]{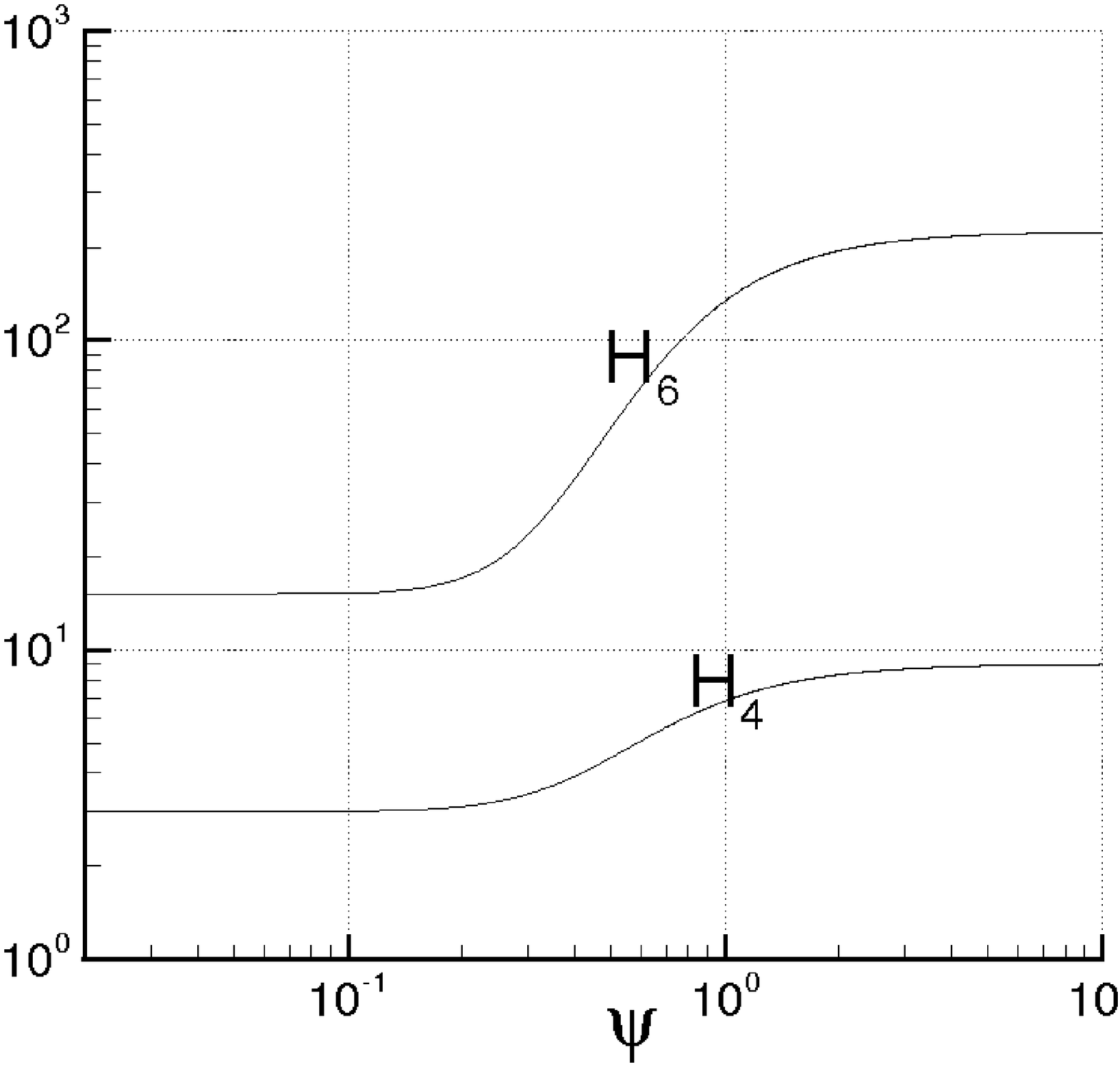}
\caption{%\color{blue} 
Left: Distribution function of the longitudinal temperature derivatives, at different values of $\Psi_\theta$. Right: Dimensionless statistical moments, $H_\theta^{(4)}$ and $H_\theta^{(6)}$ in function of $\Psi_\theta$.}
\label{figura_r7}
\end{figure}
Furthermore, the statistics of the temperature dissipation
\bea
 \varphi = \chi \nabla \vartheta \cdot \nabla \vartheta,
\eea
is analyzed in function of $\Psi_\theta$ with particular reference to its intermittency. 
To this end, the Kurtosis of $\varphi$, $K_4(\varphi)$, is estimated by means of Eq. (\ref{Tm1}), where, thanks to isotropy, the three components of $\nabla \vartheta \equiv ( \vartheta_x, \vartheta_y, \vartheta_z)$ are identically distributed.
Next, $\vartheta_x$, $\vartheta_y$ and $\vartheta_z$ are supposed to be statistically uncorrelated. This last assumption allows to estimate the Kurtosis of $\varphi$ in terms 
of the dimensionless statistical moments of $\partial \vartheta/ \partial r$, according to
\bea
K_4(\varphi)=  \frac{H^{(8)}_\theta - 4 H^{(6)}_\theta +6 H^{(4)}_\theta -3}{3\left( \left( H^{(4)}_\theta\right)^2+1-2 H^{(4)}_\theta\right)} + 2
\eea 
where $H_\theta^{(4)}$, $H_\theta^{(6)}$ and $H_\theta^{(8)}$ are calculated using Eq. (\ref{Tm1}).
Figure \ref{figura_r8} shows $K_4(\varphi)$ in function of $\Psi_\theta$,
\begin{figure}[ht]
	\centering
          \includegraphics[width=7.0cm, height=6.0cm]{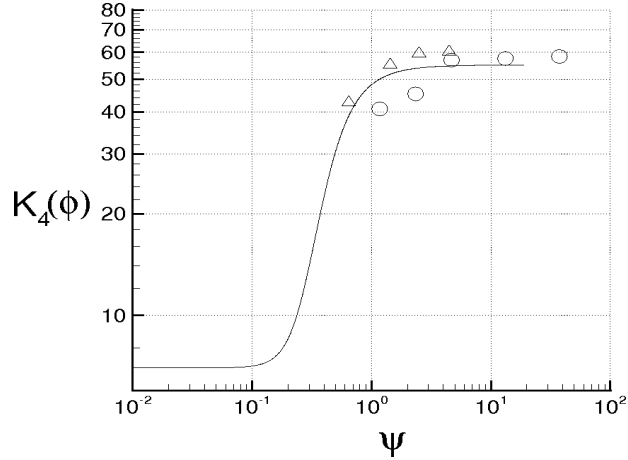}
\caption{%\color{blue} 
Comparison of the results: Kurtosis of temperature dissipation in function of $\Psi_\theta$. The symbols represent the results by \cite{Burton}.}
\label{figura_r8}
\end{figure}
and compares the values calculated with the present theory (solid line), with those obtained by \cite{Burton} through the nonlinear large--eddy simulations (symbols). The comparison shows that the data are in qualitatively good agreement.
More in detail, for $\Psi_\theta \rightarrow \infty$, $K_4 \rightarrow$ 55, whereas the results of \cite{Burton} give a value around to 60. This difference could be due to the fact that the present analysis only considers the isotropic turbulence which tends to bound the values of the dimensionless statistical moments of $\partial \vartheta/ \partial r$ and of $\varphi$, and to the approximation of assuming the components of $\nabla \vartheta$ to be statistically uncorrelated.
\begin{table}[h]
\centering
  \begin{tabular}{cc} 
\hline
\color{blue} Reference  \       &  \ \color{blue}$\Phi(0)$  \\[2pt] 
\hline
\hline \color{blue}
\color{blue} Present Analysis  \ &  \ \ \ \color{blue} 0.1409...      \\
\color{blue}\cite{Tabeling96}  \ & \ \ \ \color{blue} $\simeq$ 0.148     \\
\color{blue}\cite{Sreenivasan} \ & \ \ \ \color{blue} $\simeq$ 0.135   \\
\hline
 \end{tabular}
\caption{%\color{blue} 
Identification of $\Phi(0)$ through elaboration of experimental data
of \cite{Tabeling96} and \cite{Sreenivasan}, and comparison with the present analysis}
\label{table3}
\end{table} 

Finally, observe that the experimental data of \cite{Tabeling96} and \cite{Sreenivasan} allow to identify $\Phi(0)$. Table \ref{table3} reports a comparison between the value of $\Phi(0)$ calculated with the present theory and those obtained through elaboration of the experimental data of \cite{Tabeling96} and \cite{Sreenivasan}. 
Form this comparison, the value of $\Phi(0)$ calculated with Eq. (\ref{phi0}) is in very good agreement with those obtained through the elaboration the data of \cite{Tabeling96} and \cite{Sreenivasan}.
}
\bigskip

\section{Conclusion \label{Conclusion}}

A review of previous theoretical results concerning an original turbulence theory is presented. The theoretical approaches here adopted, different with respect to the other articles, 
confirm and corroborate the results of the previous works.

In separate sections, novel issues regarding the proposed turbulence theory are presented,
and are here summarized.

%$i$) 
{%\color{blue}
-The bifurcation rate of velocity gradient, calculated along fluid particles trajectories
is shown to be much larger than the maximal Lyapunov exponent of the kinematic field.

-On the basis of the previous item, the energy cascade is viewed as
a stretching and folding succession of fluid particles which gradually involves
smaller and smaller scales.
}

%-The energy cascade phenomenon is interpreted as the tendency of the 
%material vorticity to follow direction and variations of the Lyapunov 
%vector, which in turn varies according to the Lyapunov theory. 

%$ii$) 
-The central limit theorem, in the framework of the bifurcation analysis, provides reasonable argumentation that the finite time Lyapunov exponent can be approximated by a gaussian random variable
if $\tau \approx 1/\Lambda$.

%$iii$)
-The closures of von K\'arm\'an--Howarth and Corrsin equations given by this theory determine velocity and temperature correlations which exhibit local self--similarity directly linked to the continuous particles trajectories divergence.

%$iv$)
-The proposed bifurcation analysis of the closed von K\'arm\'an--Howarth equation studies the route from developed turbulence toward non--chaotic regimes, and leads to an estimation of the critical Taylor scale Reynolds number in isotropic turbulence in agreement with the various experiments.

%$v$)
-Finally, a specific statistical decomposition of velocity and temperature is presented. 
This decomposition, adopting random variables distributed following extended distribution functions, leads to the statistics of velocity and temperature difference which agrees with the data of  experiments.

\bigskip 

\section{Acknowledgments}

This work was partially supported by the Italian Ministry for the Universities 
and Scientific and Technological Research (MIUR).

\bigskip

\end{document}